\documentclass[11pt]{article}
\pdfoutput=1
\usepackage{graphicx}
\usepackage{epsfig}
\usepackage{amsfonts}
\usepackage{amsmath}
\usepackage{amssymb}
\usepackage{dsfont}
\usepackage{hyperref}
\usepackage{color}
\usepackage{cite}
\usepackage{pgfplots}
\usepackage{tikz} 
\usepackage{pgfplotstable}
\pgfplotsset{compat=1.8}
    \renewcommand{\arraystretch}{1.5}

\newcommand{\ep}{\epsilon}

\newcommand{\al}{\alpha}
\newcommand{\bt}{\beta}
\newcommand{\g}{\gamma}

\newcommand{\simu}{\sigma^{\mu\nu}}

\newcommand{\Or}{\mathcal O}

\newcommand{\vL}{\ensuremath{\mathcal{L}}}
\newcommand{\vp}{\varphi}
\newcommand{\sq}{^{2}}

\newcommand{\dslash}[1]{#1 \llap{/\kern-0.5pt}}
\newcommand{\Dslash}[1]{#1 \llap{/\kern+1.5pt}}
\newcommand{\DDslash}[1]{#1 \llap{/\kern+2.3pt}}
\newcommand{\dslashh}[1]{#1 \llap{/\kern+1pt}}

\newcommand{\Ex}[1]{\cdot 10^{#1}}
\newcommand{\bea}{\begin{eqnarray}}
\newcommand{\eea}{\end{eqnarray}}
\newcommand{\bma}{\begin{pmatrix}}
\newcommand{\ema}{\end{pmatrix}}
\newcommand{\nn}{\nonumber}

\newcommand\POWHEG{{\tt POWHEG}}
\newcommand\POWHEGBOX{{\tt POWHEG\ BOX}}
\newcommand\POWHEGBOXVTWO{{\tt POWHEG BOX V2}}

\begin{document}
\begin{titlepage}

\begin{flushright}
CERN-TH-2017-043\\
LA-UR-17-21898\\
NIKHEF 2017-014
\end{flushright}

\vspace{2.0cm}

\begin{center}
{\LARGE  \bf 
Right-handed charged currents in the\\[0.3cm] era of the Large Hadron Collider
}
\vspace{2.4cm}

{\large \bf  S. Alioli$^a$, V. Cirigliano$^b$, W. Dekens$^{b,c}$, J. de Vries$^d$, and E. Mereghetti$^b$ } 
\vspace{0.5cm}

{\large 
$^a$ 
{\it CERN Theory Division, CH-1211, Geneva 23, Switzerland}}

\vspace{0.25cm}

{\large 
$^b$ 
{\it Theoretical Division, Los Alamos National Laboratory,
Los Alamos, NM 87545, USA}}

\vspace{0.25cm}
{\large 
$^c$ 
{\it 
New Mexico Consortium, Los Alamos Research Park, Los Alamos, NM 87544, USA
}}

\vspace{0.25cm}
{\large 
$^d$ 
{\it 
Nikhef, Theory Group, Science Park 105, 1098 XG, Amsterdam, The Netherlands
}}

\end{center}

\vspace{1.5cm}

\begin{abstract}

We discuss the phenomenology of right-handed charged currents  in the framework of the Standard Model Effective Field Theory, 
in which they arise due to a single gauge-invariant dimension-six operator.  We study the manifestations of the nine complex couplings  
of the $W$ to right-handed quarks in collider physics, flavor physics, and low-energy precision measurements.  
We first obtain constraints on the couplings under the assumption that the right-handed operator is the dominant correction to the Standard Model 
at observable energies.  We subsequently study the impact of degeneracies with other Beyond-the-Standard-Model effective interactions and identify 
observables, both at colliders and low-energy experiments, that would uniquely point to right-handed charged currents.

\end{abstract}

\vfill
\end{titlepage}

\tableofcontents

\section{Introduction}

The existence of right-handed charged currents (RHCC) is a distinctive signature of left-right symmetric extensions of the Standard Model (SM)~\cite{Mohapatra:1974hk,Senjanovic:1975rk,Senjanovic:1978ev}.  This class of models is quite attractive as it allows parity to be restored at high energies by extending the SM gauge symmetries to $SU(3)_c\times SU(2)_R\times SU(2)_L\times U(1)_{B-L}$.
An explanation for parity violation is then provided by the spontaneous symmetry breaking of this extended gauge group.  Moreover,  TeV scale left-right theories provide an appealing realization of the seesaw mechanism for neutrino masses~\cite{Mohapatra:1979ia}.
In light of the ongoing efforts to search for new physics at the Large Hadron Collider (LHC) and in low-energy precision measurements, 
it is timely to assess the status and prospects of detecting signals of right-handed charged currents over a broad spectrum of probes. 

In this paper, we consider a setup in which  RHCC interactions manifest  themselves  at observable energies, including the scales probed at colliders,  
through a single $SU(3)_c \times SU(2)_L\times U(1)_Y$-invariant  dimension-six operator~\cite{Buchmuller:1985jz,Grzadkowski:2010es}, namely
\bea
\vL_{6,qq\vp \vp } &=&   \frac{2}{v^2} i \tilde{\vp}^{\dagger} D_{\mu} \vp \, \bar{u}^i_R \gamma^\mu \,\xi_{i j} d^j_R 
+  \mathrm{h.c.}\, ,
\label{dim6edms}
\eea
where $D_\mu$ is the covariant derivative, $\vp$ is the Higgs doublet,  $\tilde{\vp}= i \sigma_2  \vp^*$, $v = 246$ GeV is the Higgs vacuum expectation value, $i$ and $j$ are generation indices, and we work in the quark 
mass eigenbasis. 
After electroweak symmetry breaking this operator gives rise to a coupling of the $W^\pm$ boson to a right-handed charged current. In the unitary gauge we have
\bea\label{qqvpvp}
\vL_{6, qq\vp\vp}& =& \frac{g}{\sqrt 2}\bigg[\xi_{ij}\,\bar u^i_R \g^\mu  d^j_R\,W_\mu^+  \bigg]\left(1+\frac{h}{v}\right)^2 +\text{h.c.},
\eea
where $g$ is the $SU(2)_L$ gauge coupling. The operator in Eq.\ \eqref{dim6edms} arises in left-right symmetric models from the mixing between the charged gauge bosons of the $SU(2)_R$ and $SU(2)_L$ gauge groups. In this case 
$\xi_{ij}$ is proportional to a unitary $3 \times 3$ matrix, the right-handed analog of the Cabibbo-Kobayashi-Maskawa (CKM) matrix. Here we do not commit to a specific model, so that $\xi$ is a generic $3 \times 3$ matrix, with 9 independent complex parameters.
The elements $\xi_{ij}$ scale as  $\xi_{ij} \sim \mathcal O(v^2/\Lambda^2)$, where $\Lambda$ is the scale of new physics. 
We work within the framework of the SM effective field theory (SMEFT), in which it is a valid approximation to only consider dimension-six operators as long as there is a gap between the scale of new physics and the largest energy scale in the problem. For low-energy observables the largest energy scale will be the electroweak scale, such that we have the requirement $\Lambda > v$. Instead, due to the larger energies available at colliders, the effects of Eq.\ \eqref{dim6edms} can be investigated in $p p$ collisions at the LHC if one assumes that $\Lambda >  {\rm few} \ {\rm TeV}$. 

Although we do not restrict ourselves to a specific model it is worthwhile to mention how Eq.~\eqref{qqvpvp} can be induced in UV-complete models. For example, in the minimal left-right symmetric model (see e.g. Ref.~\cite{Zhang:2007da}) the effective operator arises due to the mixing between left- and right-handed $W$ bosons. In this set-up, after integrating out the heavy right-handed $W$ boson, we can identify \hbox{$\xi_{ij} \sim (\kappa \kappa'/v_R^2) (V_R)_{ij} e^{i \alpha}$} where $\kappa$ and $\kappa'$ are vacuum expectation values of the order of the electroweak scale, $v_R \gg v $ is the right-handed scale, $\alpha$ is a CP-violating phase arising from the extended Higgs sector of the model, and $V_R$ the right-handed analogue of the CKM matrix. With further assumptions, such as explicit $P$ and/or $C$ symmetry at high energies, $V_R$ can even be calculated in terms of SM quantities, such as quark masses and CKM elements, and the new model parameters $\kappa$, $\kappa'$, and $\alpha$ \cite{Maiezza:2010ic, Senjanovic:2014pva, Senjanovic:2015yea}. In this way, our results can be used in phenomenological analyses of UV-complete scenarios, although care must be taken as other effective operators might be induced at the matching scale.

The operator \eqref{dim6edms} has several interesting manifestations, both at high- and low-energy.  
At colliders, it affects the production and polarization of  $W$ bosons. Furthermore,  the operator \eqref{dim6edms}  affects the production cross section of the Higgs boson, both in associated production with a $W$ boson and in the vector boson fusion (VBF) channel. As we will discuss, invariance under the SM gauge group causes Eq.\ \eqref{qqvpvp} to modify not only the $Wqq'$ vertex, but also the $HWqq'$ interaction.
This latter interaction produces a very different dependence of the Higgs production cross section on kinematic variables such as the partonic center of mass energy or 
the Higgs transverse momentum. This in turn results in a large enhancement of the $\xi$-mediated cross section compared to the SM, especially for $WH$ associated production. As a consequence, we will see that, for the first two rows of the $\xi$ matrix, processes involving the Higgs are already more constraining than single $W$ production.  The third row of the $\xi$ matrix is directly constrained by single-top production, and top decay. In particular, the measurement of the $W$ polarization in top decay allows for a direct access to the Lorentz structure of the $Wtb$ vertex, and to test its left-handed nature.

The operator \eqref{dim6edms} leaves a distinctive trace at low energy as well. Indeed, it is the only dimension-six operator in the SMEFT that induces a
tree-level charged-current coupling of left-handed leptons to right-handed quarks, thus affecting baryon $\beta$ decays, and meson leptonic and semileptonic charged-current decays. 
We will see that, under the assumption that the SM is modified  predominantly  by a RHCC at the high energy scale $\Lambda$, low-energy probes provide very stringent constraints on the first two rows of the $\xi$ matrix. However, the most constraining observables are degenerate enough that,  by introducing 
new physics  beyond the $\xi_{ij}$, the bounds can be  weakened to levels that are comparable to collider sensitivities. Less degenerate observables,
such as decay correlations in the neutron $\beta$ decays, suffer from  comparatively large theoretical uncertainties, so that, once again, they probe 
the $\xi_{ij}$ at levels comparable to collider experiments. 
One of our main findings is that once one tries to remove degeneracies by identifying observables that are sensitive 
primarily to $\xi_{ij}$,  collider searches  and low-energy probes have comparable sensitivity, and it is of great value 
to pursue both.

In addition to observables that are at least in principle directly sensitive to RHCC, we also consider indirect bounds, both at high- and low-energy.
Some of the most stringent indirect limits arise through top-quark loops which induce large corrections, enhanced by $m_t/m_{d_j}$, to the bottom-Yukawa coupling and dipole operators. In turn, the bottom Yukawa induces $h\to b\bar b $, while the dipole operators contribute to $B \rightarrow X_{s,d} \gamma$, the rare decay $K_L \rightarrow \pi^0 e^+ e^-$, and hadronic electric dipole moments (EDMs).
We will see that the constraints from these loop processes are several orders of magnitude stronger than direct constraints from top production and decay. 
Furthermore, RHCC of light quarks induce tree-level contributions to EDMs and direct CP-violation (CPV) in kaon decays \cite{Cirigliano:2016yhc}. The stringent bounds on hadronic EDMs and the experimental value of $\ep^\prime/\ep$ 
can therefore be used to rule out couplings larger that $\textrm{Im}\, \xi_{ud, us} \sim 10^{-6} - 10^{-7}$, suggesting a very high right-handed scale. 
While these indirect probes are certainly more ``degenerate" than  direct observables (i.e. they receive contributions from several other dimension-six operators in the SMEFT), 
the bounds that they imply are nonetheless very significant.
Within the SMEFT these limits put stringent bounds on certain directions in the space of dimension-six Wilson coefficients, thereby imposing non-trivial constraints on new physics scenarios.

This paper is organized as follows. 
We start by investigating the constraints coming from colliders in Section \ref{sec:collider}. We discuss direct $W$ production in Section \ref{sec:W}, 
associated production of a Higgs and a $W$ boson in Section \ref{sec:HW}, and Higgs production via VBF in Section \ref{sec:VBF}. We then consider constraints on the $\xi_{tj}$ elements coming from
top production and decay in Section \ref{sec:top}, and from the decays of the Higgs in Section \ref{sec:hbb}.
To connect to low-energy observables, in Section \ref{sec:LELagr}, we integrate out the heavy SM particles and match onto a low-energy 
effective Lagrangian. In Section \ref{sec:tree-level} we consider the constraints coming from $\beta$ decay, and from leptonic and semileptonic meson decays. 
We then discuss observables sensitive to non-leptonic operators induced by the operator \eqref{dim6edms}, organizing the discussion in $\Delta F=0$ observables (Section \ref{sec:EDM}), which consists of hadronic EDMs,
$\Delta S = 1$ observables, including $\ep^\prime/\ep$ and $K_L \rightarrow \pi^0 e^+ e^-$  (Section \ref{sec:DeltaS1}), and $\Delta B =1$ observables, mainly related to inclusive and exclusive $b \rightarrow s,d \gamma$ transitions (Section \ref{sec:DeltaB1}). In Section \ref{Single}, we obtain limits on the real and imaginary part of $\xi_{ij}$ by 
taking into account all the observables discussed above, in a scenario in which only one of the elements is turned on at the high scale. 
In Section \ref{Disentangling} we discuss strategies to unambiguously identify the signal of a RHCC, both at low energies, and in associated production of a Higgs and a $W$. Finally, in Section \ref{sec:topanomalous} we consider more in detail the $Wtb$ vertex. We conclude in Section \ref{sec:concl}.

\section{Right-handed charged currents at colliders}\label{sec:collider}

In this section we study the effects of the RHCC operator $\xi$ on several processes of interest at the LHC. We focus on $W$ production, associated production of a $W$
and a Higgs boson, Higgs production via vector boson fusion, and
single-top production, as we expect these processes to be the most sensitive to RHCC.  For all these processes, we include NLO QCD
corrections to both the SM and BSM contributions, and, with the exception of single-top production, we implement the
processes in the \POWHEGBOXVTWO{}~\cite{Nason:2004rx,Frixione:2007vw,Alioli:2010xd}.  We also
consider the effect of the $\xi$ operator on the decays of the top-quark and the Higgs boson.

Throughout this Section we always consider $n_f=5$ massless quarks and
we include the exact top-quark mass dependence when necessary. 
For processes involving light quarks only, this is
justified by the fact that the interference of the right-handed
currents with the SM is suppressed by two insertions of the light
quark masses. The interference is negligible for values of the couplings
$\xi_{ij} \gg y_{u_i} y_{d_j}$, where $y_{u_i,d_j}$ are the SM Yukawa
couplings. Even in the most favorable case, $\xi_{cb}$, neglecting the
charm and bottom masses is reasonable for $\xi_{cb} > 10^{-3}$, a level
that, as we will see, is far from the sensitivity that can be reached by present collider experiments.
In the case of single-top production and top decays, interference terms are important for $\xi_{tb} \sim y_b y_t$. While in this case the corrections are more relevant,
they are still subleading with respect to terms quadratic in $\xi_{tb}$ for the values of the coupling accessible at colliders.

\subsection{$W$ production}\label{sec:W}
\begin{figure}
\center 
\includegraphics[width=0.5\textwidth]{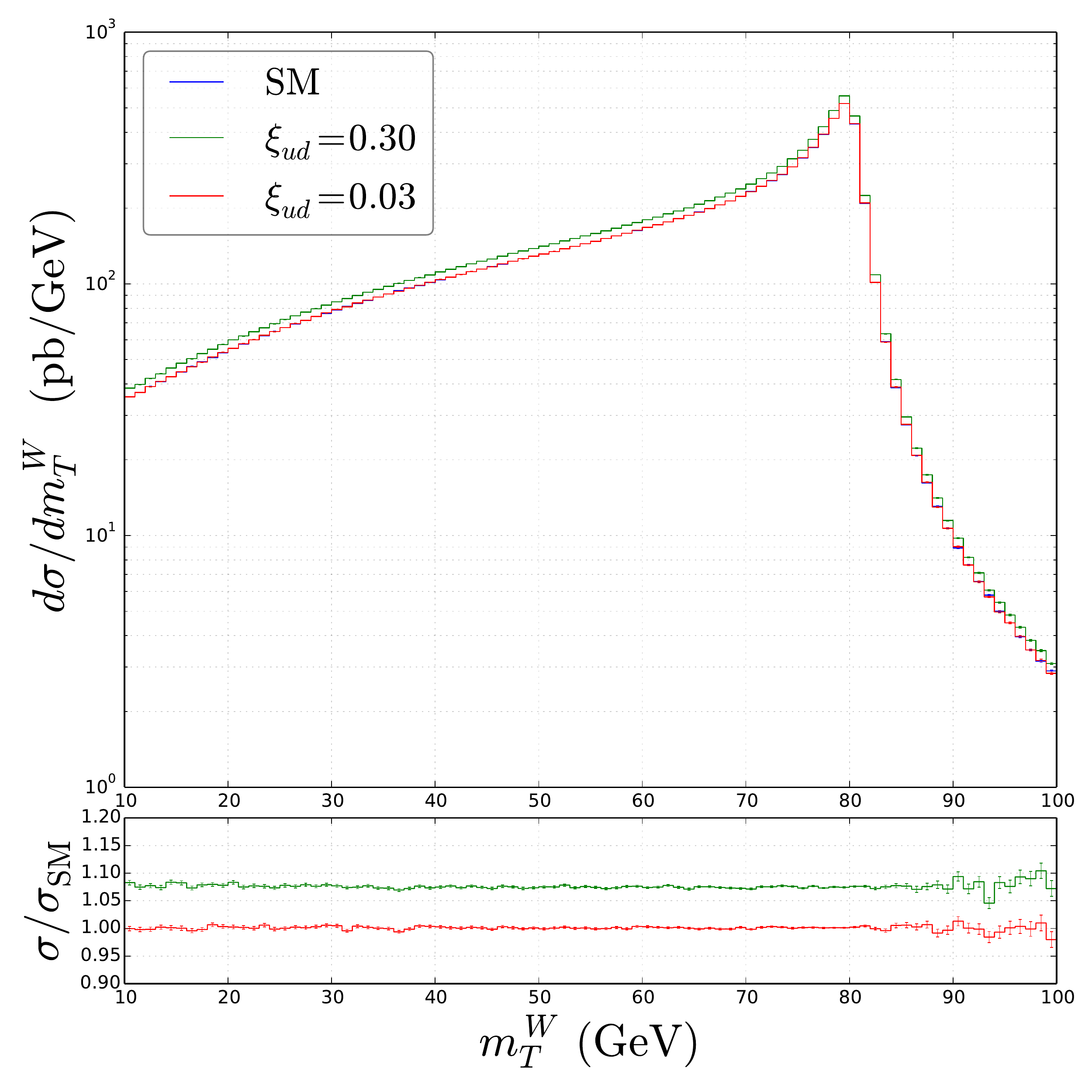}
\caption{Transverse mass  distribution  $m^W_{T}$ in $p p \rightarrow W^+ \rightarrow e^+ \nu$ at $\sqrt{S} = 13$ TeV in the presence of right-handed charged currents.}\label{fig:mW}
\end{figure} 
The first process we analyze to look for manifestations of  RHCC is $W^{\pm}$ production. This process is accurately measured at the LHC, both at the level of the inclusive cross section 
as well as for differential distributions~\cite{Aaboud:2016ylh,Aad:2016naf,Aad:2013ueu,ATLAS:2012au,Aad:2011fp,Khachatryan:2016nbe,Khachatryan:2016pev,Chatrchyan:2014mua,CMS:2011aa,Chatrchyan:2011jz,Khachatryan:2010xn}. 
Precise high-order calculations of the SM background are available up to fixed NNLO QCD corrections~\cite{Anastasiou:2003yy,Melnikov:2006kv,Catani:2009sm} and also include the resummation of the vector-boson transverse momentum~\cite{Bozzi:2010xn,Catani:2015vma}. More recently, the interface of the NNLO predictions with the parton shower has been presented in Ref.~\cite{Karlberg:2014qua}. A careful quantitative assessment of the size of the corrections at different orders, both in QCD and EW, has been presented in Ref.~\cite{Alioli:2016fum}. For this study, we have calculated the NLO QCD corrections to the partonic processes mediated by RHCC and interfaced with the parton shower according to the \POWHEG{} method, extending the original work of Ref.~\cite{Alioli:2008gx}.

The contribution of the RHCC operator to $W^\pm$-production observables 
that are symmetric under the exchange of the
charged lepton and the neutrino momenta, is identical to the SM contribution after
the replacement of the CKM elements $V_{ ij}$ by
$\xi_{ij}$. For example, in Fig.\ \ref{fig:mW} we show the differential distribution with respect to the 
$W$ transverse mass which is defined as \cite{Ellis:1991qj}
\begin{equation}
m^W_{T} =  \sqrt{2 |p_{T\,l}| |p_{T\,\nu}| (1 - \cos \Delta \phi_{l\nu})}\, .
\end{equation}
Here $p_{T\, l}$ and $p_{T\, \nu}$ are the charged-lepton and the
neutrino transverse momenta, respectively,  and $\Delta \phi_{l\nu}$ is their
azimuthal separation.  We evaluate the cross section at $\sqrt{S} =13$ TeV for the SM (blue curve), $\xi_{ud}
= 0.3$ (green curve), and $\xi_{ud} = 0.03$ (red curve). The
effect of the RHCC amounts to a rescaling of the
cross section. Since the correction is quadratic in $\xi_{ij}$,
choosing $\xi_{ud} = 0.3$ gives approximately a 10\% correction to the SM
prediction, while $\xi_{ud} = 0.03$ gives sub-permille corrections.
Presently, the $W^{\pm}$ cross section at 13 TeV is known with roughly $3\%$ experimental uncertainty and a similar theoretical
uncertainty \cite{Aad:2016naf}, implying that the bound on $\xi_{ud}$ that can be
extracted from the $W^{\pm}$ cross section is at best around 
$|\xi_{ud}|\lesssim 0.2$. Due to PDF suppression, the bounds on the other elements of the $\xi_{ij}$ matrix are even weaker.  As we discuss in
Section \ref{sec:HW},  with current sensitivity, the associated production of a Higgs and a $W$
boson is already more constraining than $W$
production. For this reason, we do not list the bounds coming from $W$ production on the various $\xi_{ij}$ elements.

\begin{figure}
\center
\includegraphics[width=7.5cm]{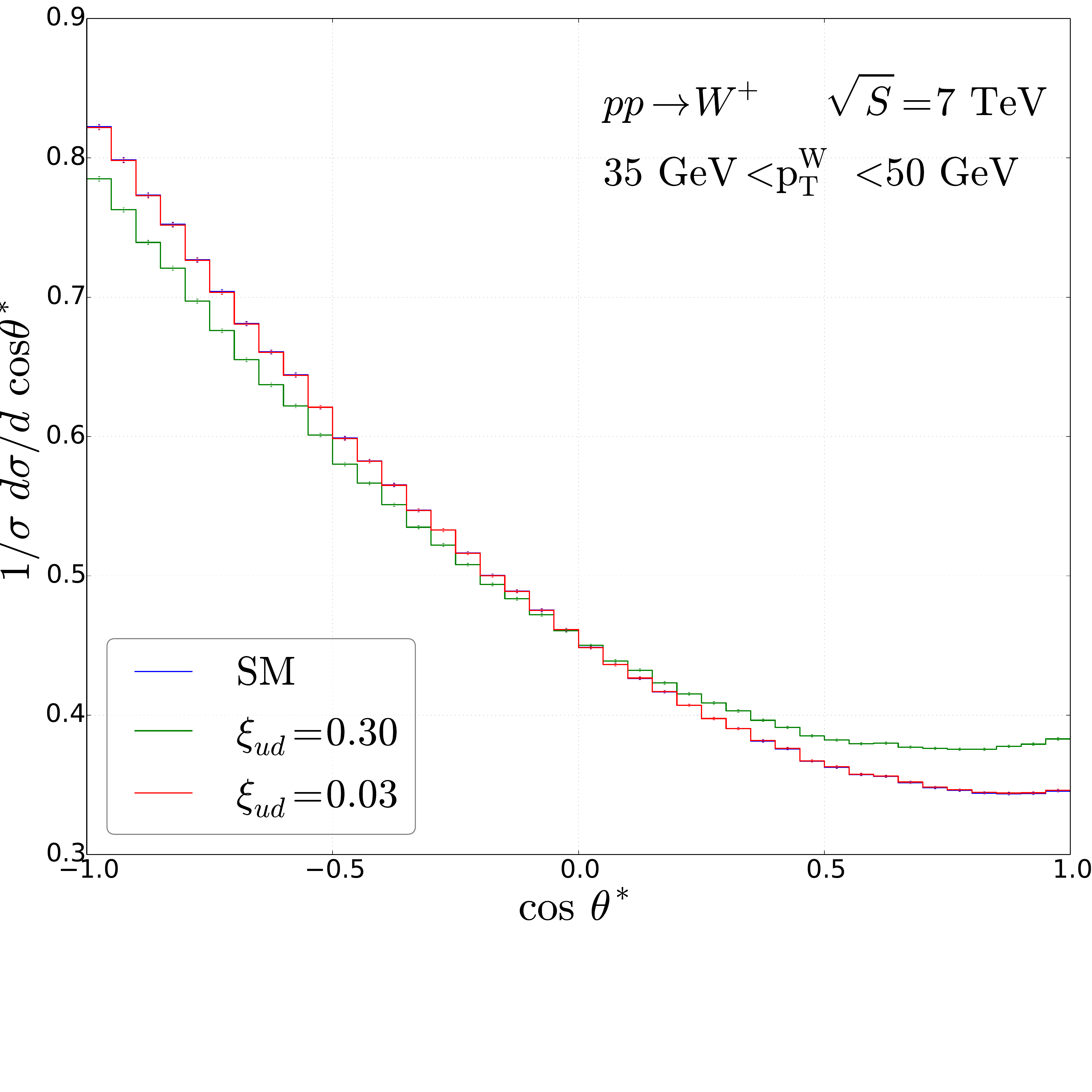}
\includegraphics[width=7.5cm]{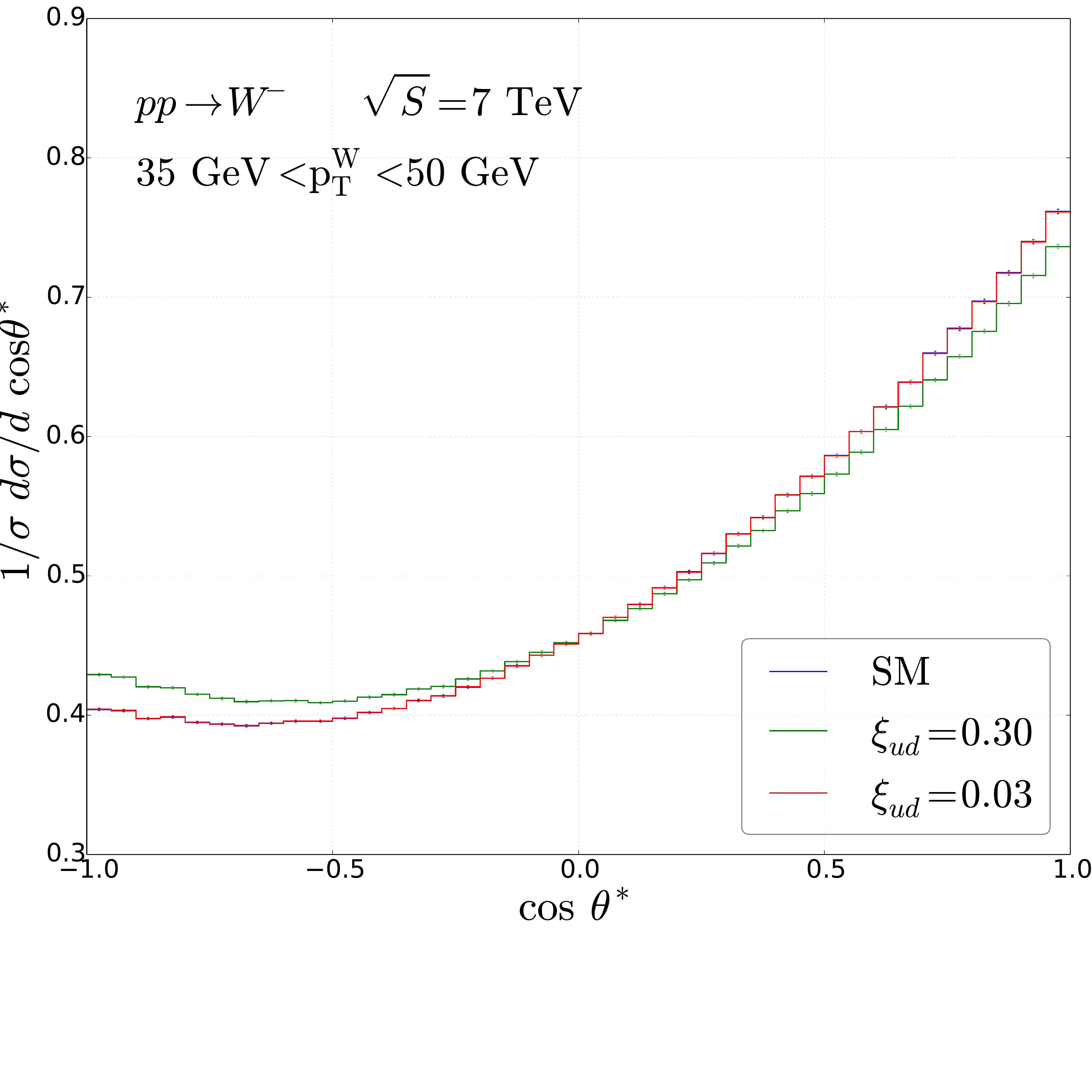}
\vspace{-1cm}\caption{Polar angle $\cos\theta^*$ distribution in $p p \rightarrow W^+ \rightarrow e^+ \nu$ and $p p \rightarrow W^- \rightarrow e^- \bar{\nu}$ at $\sqrt{S} = 7$ TeV, in the SM and in the presence of $\xi_{ud}$. $\theta^*$ is the polar angle of the charged lepton, measured in the $W$ boson rest frame, with the $z$-axis chosen to be oriented along the $W$ direction in the lab frame. }\label{fig:costheta}
\end{figure}

The angular distribution of the charged leptons in $W^{\pm}$ decay is sensitive to the left-handed nature of the $W$ boson in the SM \cite{Bern:2011ie}.  One might therefore expect that angular distributions can provide stronger constraints. In Fig.~\ref{fig:costheta} we plot the differential distribution with respect to $\cos \theta^*$, where $\theta^*$ is the polar angle of the charged lepton in the $W$-boson rest frame,
with the $z$-axis chosen to be in the direction of the $W$-boson momentum in the laboratory frame.  The differential  $\cos \theta^*$ distribution is related to the $W$ boson polarization fractions, $F_{0}$, $F_L$, and $F_R$
by the relations  \cite{Bern:2011ie}
\begin{eqnarray}
F_0 &=& \int d\cos\theta^*  (2 - 5 \cos^2\theta^*) \frac{1}{\sigma} \frac{d \sigma}{d\cos\theta^*}\ , \\ 
 F_L &=& \int d\cos\theta^*  \left(-\frac{1}{2} 
\mp \cos\theta^* 
+ \frac{5}{2} \cos^2 \theta^* \right) \frac{1}{\sigma} \frac{d \sigma}{d\cos\theta^*}\ , \\
F_R &=& \int d\cos\theta^*  \left(-\frac{1}{2} 
\pm \cos\theta^* 
+ \frac{5}{2} \cos^2\theta^* \right) \frac{1}{\sigma} \frac{d \sigma}{d\cos\theta^*}\ ,
\end{eqnarray}
where the upper (lower) sign is for $W^{+ (-)}$. 

Fig.~\ref{fig:costheta} depicts the  $\cos\theta^*$ distribution for the SM (blue curve), for  $\xi_{ud} = 0.3$, and for  $\xi_{ud} = 0.03$. These predictions were evaluated  at $\sqrt{S} = 7$ TeV and we applied the same $p_{T}^{W}$ cuts as used in the analysis of Ref.~\cite{ATLAS:2012au}.  We observe that
$\xi_{ud}$ does not affect the longitudinal fraction $F_0$, while it increases the fraction of right-handed polarized $W$'s, and decreases $F_L  - F_R$.
We find the helicity fractions to be given by
\begin{eqnarray}
F_0 = 0.21, \qquad F_L = 0.54 \frac{1 + 0.3\ \xi_{ud}^2}{1 + 0.8\ \xi_{ud}^2} \qquad  F_R = 0.25 \frac{1 + 1.9\ \xi_{ud}^2}{1 + 0.8\ \xi_{ud}^2}\ , 
\end{eqnarray}
for $35 \, \textrm{GeV} <p_{T}^{W} < 50$ GeV, and 
\begin{eqnarray}
F_0 = 0.19, \qquad F_L = 0.55 \frac{1 + 0.4\ |\xi_{ud}|^2}{1 + 0.8\ |\xi_{ud}|^2} \qquad  F_R = 0.26 \frac{1 + 1.8\ |\xi_{ud}|^2}{1 + 0.8\ |\xi_{ud}|^2} , 
\end{eqnarray}
for $p_{T}^{W} > 50$ GeV.
These helicity fractions were measured in Ref.~\cite{ATLAS:2012au} with uncertainties of roughly 20\% on $F_L - F_R$ and larger uncertainties on $F_0$.
The significant uncertainties  allow for rather large values of $\xi_{ud}$, up to $\xi_{ud} \sim 0.4$.
The contributions to the $W$ helicity fractions involving other elements of the $\xi_{ij}$ matrix are further suppressed by the respective PDF's resulting in even weaker bounds.
We conclude that at the moment the $W$ polarization fractions do not provide strong constraints on the effects of RHCC.

\subsection{Associated production of a Higgs and a $W^\pm$ boson}\label{sec:HW}

As for single W-boson production, $WH$ associated production is available at NNLO in QCD~\cite{Ferrera:2011bk,Ferrera:2013yga,Campbell:2016jau} and is matched to the parton shower up to the NNLO+PS level~\cite{Astill:2016hpa}.

\begin{figure}
\center
\includegraphics[width=12cm]{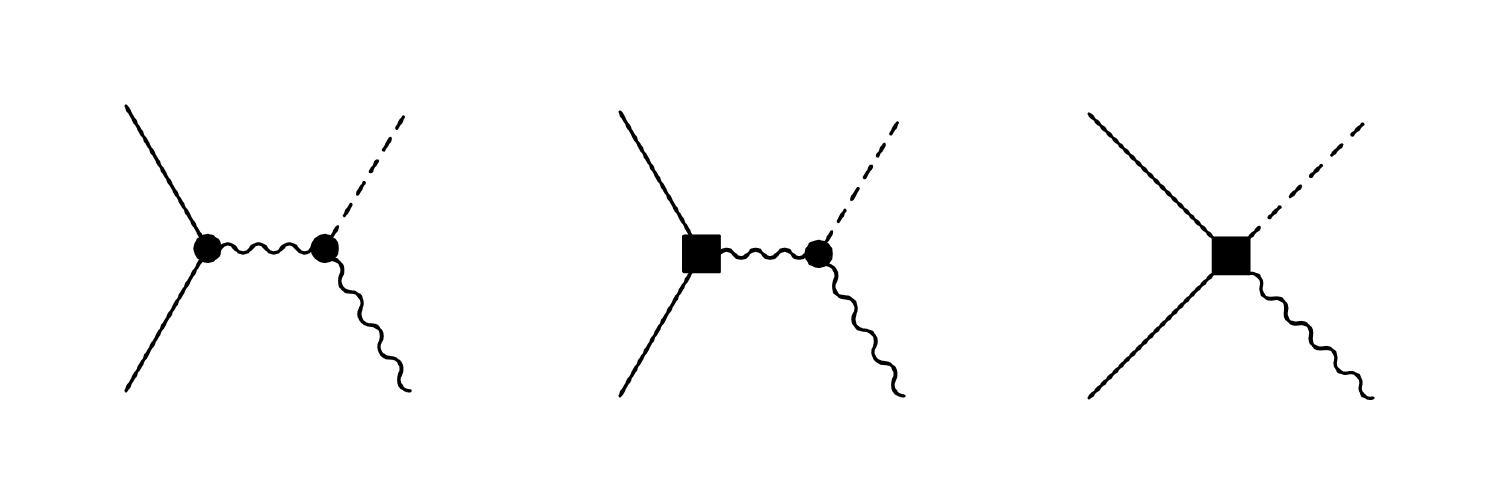}
\vspace{-1cm}\caption{Leading-order contribution to the associated production of a Higgs and a vector boson.
Dots denote SM vertices, while  squares denote an insertion of  $\xi$.}\label{ass1}
\end{figure}

A representative set of tree-level diagrams contributing to associated $WH$ production at LO in presence of RHCC is shown in Fig. \ref{ass1}.
The first diagram denotes the SM contribution. The remaining two diagrams involve an insertion of $\xi$, which is denoted by a square. 
In addition to a contribution similar to the SM (the second diagram), the $\xi$ operator induces a local $(\bar q^\prime  q)_R h W$ interaction (the third diagram), which leads to a large enhancement of the RHCC contribution. The interference of the right-handed currents with the SM is suppressed by two insertions of the light quark masses and is negligible for values of the couplings  that can be probed at the LHC. We therefore focus on the contributions quadratic in $\xi$.

We computed the NLO QCD corrections to the RHCC contributions to the $WH$ cross section, also considering the decay of the $W$ into leptons. This implementation  builds upon the original NLO+PS \POWHEGBOX{} code in~\cite{Luisoni:2013kna}.
The SM NLO cross section at $\sqrt{S} = 8$ TeV is 
\begin{equation}
\sigma_{W^+H} = ( 0.461 \pm 0.021 ) \, \textrm{pb}, \quad \sigma_{W^-H} = (0.264 \pm 0.017) \, \textrm{pb}\ ,
\end{equation}
where we used the NLO PDF sets \texttt{CT10}, \texttt{MSTW08}, and \texttt{NNPDF2.3}. The error is dominated by PDF uncertainties while scale variations are about $(2$-$3)\%$. 
In Tab.~\ref{tabErrors1} we give the cross sections induced by $\xi$ at $\sqrt{S} = 8$ TeV.  The central value is evaluated at the scale $\mu =
m_H + m_W$. As for the SM results, the errors are dominated by PDF variations that are roughly 10\%. As expected, PDF errors are larger for processes involving the strange PDF, which is not known at the same level as the PDFs of the lighter quarks. The cross sections at 13 and 14 TeV have similar theoretical uncertainties.

\begin{table}
\center
\begin{tabular}{|| c | c | c c c c c c||}
\hline
& 				& $ |\xi_{ud}|^2 $  &   $|\xi_{us}|^2$ 	&  $|\xi_{ub}|^2$   & $|\xi_{cd}|^2$    &  $|\xi_{cs}|^2$  & $|\xi_{cb}|^2$  \\ \hline 
&central 			&  230 	            & 158  		& 66.4    		& 12.7 		   	&  7.48                  & 2.84 \\
$\sigma_{W^+ H}$ (pb) &scale	&  $\pm 4$          & $\pm 3$   	        & $^{+0.8}_{-0.4}$      & $^{+0.2}_{-0.1}$    &  $^{+0.08}_{-0.04}$    & $^{+0.08}_{-0.04}$       \\
&PDF				&  20 		    & 36.0   		& 3.2    		& 1.8 		   	&  2.32       	         & 0.48 \\
 \hline \hline 
&central 			&  100 	            & 17.4  		& 6.64    		& 50.4 		   	&  7.72      & 2.84\\
$\sigma_{W^- H}$ (pb) &scale	&  $\pm 2$  	    & $^{+2.8}_{-2.4}$    & $^{+0.08}_{-0.04}$       & $^{+0.8}_{-0.4}$   &  $^{+0.08}_{-0.04}$  & $^{+0.08}_{-0.04}$        \\
&PDF				&  10.4 		        & 4.00   		& 1.12    		& 3.2 		   	&  2.08      & 0.48  \\
\hline
\end{tabular}
\caption{Corrections to the $W^{\pm}H$ cross section induced by the RHCC $\xi_{ij}$ at $\sqrt{S} =  8$ TeV. The total cross section
is $\sigma_{W^\pm H} = \sigma^{SM}_{W^\pm H} + \sum a_{ij} |\xi_{ij}|^2$, where the sum is over all light quarks. The values and theoretical uncertainties of $a_{ij}$ are given in the table. }
\label{tabErrors1}
\end{table}

Using the total cross sections in Tab.~\ref{tabErrors1}, we construct the production signal strength in the presence of the RHCC,
\begin{equation}
\mu^{\mathcal \xi}_{WH} = 1 + \frac{\sigma^{\mathcal \xi}_{W^+ H} + \sigma^{\mathcal \xi}_{W^- H} }{\sigma^{SM}_{W^+ H} + \sigma^{SM}_{W^- H} }\, .
\end{equation}
Our results for the production signal strengths at $\sqrt{S} =$ 8, 13 and 14 TeV are summarized in the upper half of Tab.~\ref{ss1}. 
As discussed in Section \ref{sec:hbb}, the first two rows of the $\xi_{ij}$ matrix do not significantly affect the Higgs 
branching ratios. We can thus safely assume that Higgs decays are SM-like and we simply use the production signal strength in the $WH$ channel  in our analysis. 
The combined result from the  ATLAS and CMS collaborations is \cite{Khachatryan:2016vau}
\begin{equation}
\mu_{WH} (8\, \textrm{TeV})  = 0.89^{+0.40}_{-0.38}\, .
\end{equation}
In the lower half of Tab. \ref{ss1} we list the 90\% CL bounds on $|\xi_{ij}|$,
with the assumption that the SM is modified only by a RHCC at the high-energy scale $\Lambda \gg v$\footnote{Since the coefficients $\xi_{ij}$ depend very mildly on the initial scale $\Lambda$ 
\cite{Jenkins:2013wua,Alonso:2013hga}, we do not specify its precise value.}.
The constraints are much stronger than those extracted from $W^{\pm}$ production. The ATLAS collaboration also published preliminary results obtained at  13 TeV  \cite{ATLAS-CONF-2016-091}
\begin{equation}
\mu_{WH} (13\, \textrm{TeV})  = 0.33^{+0.95}_{-0.92}\, .
\end{equation}
The larger contribution of the RHCC operators to the cross section is compensated by the larger experimental errors, resulting in similar constraints as obtained from the 8 TeV data.

\begin{table}
\center
\begin{tabular}{|| c | c  | c c c c c c||}
\hline
	       &  $\sqrt{S}$ &  $|\xi_{ud}|^2$  & $|\xi_{us}|^2$     & $|\xi_{ub}|^2$ & $|\xi_{cd}|^2$ &  $|\xi_{cs}|^2$  & $|\xi_{cb}|^2$ \\ \hline 
	       & 8  TeV 			&  $451 \pm 29$ 	     & $250\pm 57$  		  & $101 \pm 7$    		& $87  \pm 7$ 		   	&   $22 \pm 6$     & $8 \pm 2$ \\
$\mu_{WH}$ -1  & 13  TeV 			&  $663 \pm 42$ 	     & $381\pm 80$  		  & $164 \pm 11$    		& $142 \pm 10$ 		   	&   $42 \pm 10$    & $16 \pm 3$ \\
	       & 14  TeV 			&  $703 \pm 45$ 	     & $406\pm 84$  		  & $177 \pm 12$    		& $153 \pm 11$ 		   	&   $46 \pm 11$    & $18 \pm 3$ \\
\hline
		&   	     &  $|\xi_{ud}| $  & $|\xi_{us}| $   & $|\xi_{ub}| $ & $|\xi_{cd}| $ &  $|\xi_{cs}| $  & $|\xi_{cb}| $ \\ \hline 
ATLAS, CMS \cite{Khachatryan:2016vau}    & 8  TeV    &  0.04         & 0.05                & 0.08 		& 0.08  	    &  0.19 		  & 0.30 \\ 
ATLAS  \cite{ATLAS-CONF-2016-091}        & 13  TeV   &  0.04         & 0.06                & 0.08 		& 0.09  	    &  0.18 		  & 0.28 \\ 
future          			 & 14  TeV   &  0.02         & 0.03                & 0.04 		& 0.05  	    &  0.09 		  & 0.13 \\  
\hline 
\end{tabular}
\caption{Signal strength for the $WH$ production channel  at $\sqrt{S} = 8$ TeV, 13 TeV, and 14 TeV, and 90\% CL bounds on $\xi_{ij}$. The  projected bounds at 14 TeV assume $20\%$ uncertainties on $\mu_{WH}$.}
\label{ss1}
\end{table}

In Tab.~\ref{ss1} we also show the signal strength and projected bounds at $\sqrt{S} = 14$ TeV. The RHCC contributions to the cross section grow faster than the SM contributions, resulting in a greater sensitivity at the LHC Run 2. The projected bounds in Tab.~\ref{ss1} assume a $20\%$ uncertainty on the $\mu_{WH}$ signal strength. For all coefficients the improvement is roughly a factor of 2.

The pattern of the constraints in Table~\ref{ss1}  can be simply understood in terms of the parton distributions. 
$WH$ production is most sensitive to $\xi_{ud}$, which involves two valence quarks in the proton, followed by the couplings with one valence quark, $\xi_{us}$, $\xi_{ub}$, and $\xi_{cd}$.
The bounds become weaker for $\xi_{cs}$ and $\xi_{cb}$ as the PDFs involve  two sea quarks in these cases.
While the current and projected bounds on $\xi_{ud}$, $\xi_{us}$, $\xi_{cd}$, and $\xi_{cs}$ 
are at least a factor of 10 smaller than the corresponding CKM matrix element, the $WH$ cross section allows for values of $\xi_{ub}$ and $\xi_{cb}$
that are much larger than $V_{ub}$ and $V_{cb}$. As we will discuss, this possibility is excluded by inclusive and exclusive $B$ decays.

\begin{figure}
\center
\includegraphics[width=8cm]{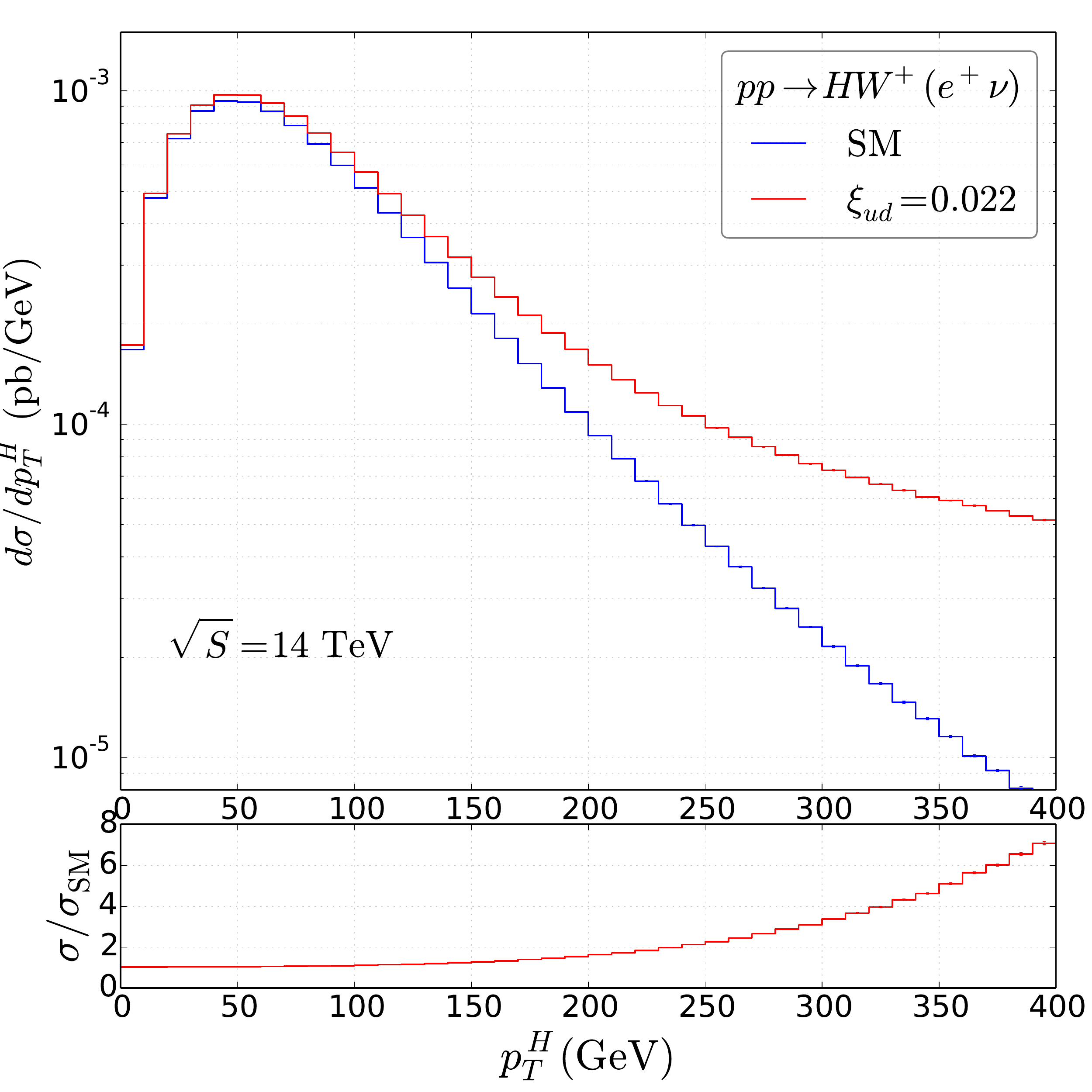}
\caption{Cross section for the production of $H W^{+}$, with $W^{+} \rightarrow \nu e^+$, at $\sqrt{S} = 14$ TeV, differential with respect to the Higgs $p_T$.}
\label{Ass14}
\end{figure}

We have focused so far on the signal strengths alone. Differential distributions can provide additional valuable information. In particular, as shown in Fig.~\ref{Ass14}, the RHCC contribution is enhanced for large values of the Higgs transverse momentum, or of the $WH$ invariant mass, so that the study of differential distributions at the current and future LHC runs could further improve the collider bounds on $\xi$. In Section~\ref{Disentangling} we will explore  in more detail how to construct observables that can discriminate between RHCC and corrections to the $WH$ cross section from other dimension-six operators in the SMEFT.

\subsection{Vector boson fusion}\label{sec:VBF}

\begin{figure}
\center
\includegraphics[width=12cm]{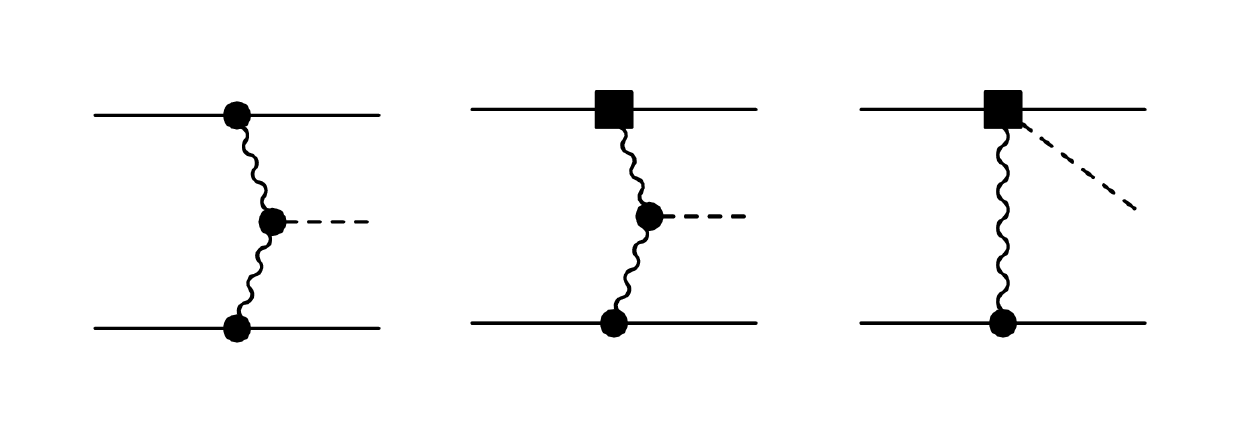}
\vspace{-1cm}\caption{Leading order contribution to Higgs production via vector boson fusion.
Dots denote SM vertices, while  squares denotes an insertion of  $\xi$.}\label{vbf1}
\end{figure}

The production rate for an Higgs boson via vector boson fusion (VBF) is also sensitive to the RHCC.
The tree-level diagrams contributing to VBF Higgs production are shown in Fig.\ \ref{vbf1}. As for $WH$ associated production, one topology is identical to the SM with the CKM matrix replaced by $\xi_{ij}$. In addition,  the $\xi$ operator induces a $\bar q^\prime q h W$ vertex depicted in the last diagram of Fig. \ref{vbf1}. 
This topology again leads to an enhancement with respect to the SM, albeit not as numerically significant as for $WH$ associated production.

The total cross section for Higgs production through VBF has been recently computed in the SM at N${}^3$LO in QCD~\cite{Dreyer:2016oyx}. Fully-differential distributions are available up to NNLO~\cite{Cacciari:2015jma} and the interface with parton showering is available at NLO+PS accuracy~\cite{Nason:2009ai}. For this study, we computed the NLO QCD corrections to both the SM and the RHCC VBF cross sections, building upon the \POWHEG{} implementation presented in Ref.~\cite{Nason:2009ai}.
We find that the total VBF cross section is
\begin{eqnarray}\label{VBF.1}
\frac{\sigma_{\textrm{VBF}} (8 \, \textrm{TeV}) }{\textrm{pb}} = 1.6 \left(1  + 16.4 |\xi_{ud}|^2  + 9.1 |\xi_{us}|^2  + 8.5 |\xi_{ub}|^2  + 6.2 |\xi_{cd}|^2 + 2.6 |\xi_{cs}|^2  + 1.1 |\xi_{cb}|^2 \right)\, ,
\end{eqnarray}
where we used $\mu = m_W $ as the renormalization and factorization scale, and we computed the cross section with the \texttt{CT10nlo} PDF set. PDF and scale variations indicate that the theory uncertainties are below the 10\% level.

In order to disentangle VBF from gluon-fusion contributions, the final state is required to have at least two well-separated jets. The invariant mass of the two jets has to be $m_{j j} > 600$ GeV,  while the rapidity separation has to be $|y_{j_1} - y_{j_2}|  >4.2$.
Furthermore, all jets are required to have $p_{Tj} > 20$ GeV and $|y_j| < 5$. Within these cuts, the NLO cross section becomes
\begin{eqnarray}\label{VBF.2}
\frac{\sigma_{\textrm{VBF}} (8 \, \textrm{TeV})}{\textrm{pb}} = 0.30 \left( 1 + 12.0 |\xi_{ud}|^2 + 7.3 |\xi_{us}|^2 + 6.7 |\xi_{ub}|^2 + 5.7 |\xi_{cd}|^2  + 1.3 |\xi_{cs}|^2 
+ 0.3 |\xi_{cb}|^2 \right)\, .
\end{eqnarray}
From Eqs. \eqref{VBF.1} and \eqref{VBF.2} we see that  the VBF channel is sensitive to RHCC, but the modifications to the signal strength are much smaller than for $WH$ associated production. The combined ATLAS and CMS signal strength from Run I is \cite{Khachatryan:2016vau} 
\begin{equation}
\mu_{VBF} = 1.18 \pm 0.25\, .
\end{equation}
This implies that from the signal strength alone we get constraints on, for example, $|\xi_{ud}| < 0.2$, $|\xi_{ub}| < 0.3$, and $|\xi_{cb}| \sim 1$. These bounds are considerably weaker than those given in Table~\ref{ss1} arising from associated production.

At 14 TeV, the cross section within the aforementioned VBF cuts is 
\begin{eqnarray}\label{VBF.3}
\frac{\sigma_{\textrm{VBF}} (14 \, \textrm{TeV})}{\textrm{pb}} = 1.1 \left( 1 + 23.2 |\xi_{ud}|^2 + 12.4 |\xi_{us}|^2 + 11.8 |\xi_{ub}|^2 + 8.7 |\xi_{cd}|^2  + 2.4 |\xi_{cs}|^2 
+ 1.0 |\xi_{cb}|^2 \right)\,,
\end{eqnarray}
which again signals a smaller sensitivity to $\xi_{ij}$ compared to $WH$ associated production. 

One advantage of using VBF is the presence of jets in the final state, which gives an additional handle on the flavor structure of the RHCC. 
For example, by requiring that at least one of the jets in the final state is $b$ tagged, we can enhance the contributions of the couplings $\xi_{ub}$ and $\xi_{cb}$ at the price of a reduced cross section,
\begin{equation}
\left. \frac{\sigma_{\textrm{VBF}} (14 \, \textrm{TeV})}{\textrm{pb}}  \right|_{\textrm{btag}} = 0.05  \, \left(1  + 235 |\xi_{ub}|^2 + 19.3 |\xi_{cb}|^2  \right)\, .
\end{equation}
After $b$ tagging, the enhancement of the $\xi_{ub}$ and $\xi_{cb}$ contributions to VBF is slightly better than the one quoted in Table \ref{ss1} for  $WH$. Since the flavor tag allows for more direct access to specific elements of the $\xi$ matrix, it can be interesting to pursue such observables at the Run II.

\subsection{Single-top production and $W$-boson helicity fractions in top decays}\label{sec:top}

The processes considered so far ($W$ production, $WH$ production, and Higgs production via VBF) do not constrain the $\xi_{t j}$ elements of the right-handed current matrix. In this Section, we consider single-top production, which is sensitive to $W t q$ couplings, and the $W$-boson helicity fractions in top decay, which probe the $W t b$ coupling, and are sensitive to its left-handed nature.

\begin{table}
\center
\begin{tabular}{|| c |c |c| c |c c||}
\hline
 $\sqrt{S} $ (TeV) & $\sigma_t$ (pb) & $\sigma_{\bar t}$ (pb) & $\sigma_{t +\bar t}$ (pb) & \multicolumn{2}{c||}{ Experiment}  \\
\hline
7   & $46 \pm 6$&   $23 \pm 3$ &  $68 \pm 8$ & ATLAS &  \cite{Aad:2014fwa}\\
    & --        & -- & $67 \pm 7 $  & CMS &  \cite{Chatrchyan:2012ep}\\
    \hline
8   & $57 \pm 4$        & $33 \pm 3$ &   $90 \pm 5$            & ATLAS & \cite{Aaboud:2017pdi} \\
    &  $54  \pm 5$&  $28 \pm 4$ & $84 \pm 8 $ & CMS & \cite{Khachatryan:2014iya}\\ \hline
13  & $156 \pm 28$& $91 \pm 18$ & $247 \pm 46$ & ATLAS &  \cite{Aaboud:2016ymp}\\
    & $150 \pm 22$& $82 \pm 16$ & $232 \pm 31$ & CMS & 		\cite{Sirunyan:2016cdg}\\
\hline
\end{tabular}
\caption{$t$-channel single-top total cross sections measured by ATLAS and CMS.}\label{TabTop}
\end{table}

The largest SM contribution to single-top production at the LHC is via the $t$-channel exchange of a $W$ boson. Smaller contributions arise from the associated production of a top and a $W$ boson and by $s$-channel $W$ exchange.
ATLAS published the measurement of the inclusive and differential  cross sections  at $\sqrt{S} =$ 7, 8, and 13 TeV, with luminosities of, respectively, 5 fb$^{-1}$, 20.2 fb$^{-1}$  
and 3.2 fb$^{-1}$~\cite{Aad:2014fwa,Aaboud:2016ymp,Aaboud:2017pdi}.
CMS published results at $\sqrt{S} = 7$, $8$, and 13 TeV, with luminosities of 1.56, 20, and 2.3 fb$^{-1}$, respectively \cite{Chatrchyan:2012ep,Khachatryan:2014iya,Sirunyan:2016cdg}.
The observed cross sections are summarized in Table \ref{TabTop}. The associated production of a top and a $W$ boson, and $s$-channel single-top production have also been observed  \cite{Aad:2015eto,Chatrchyan:2014tua,CDF:2014uma}. In our analysis, however, we include only $t$-channel production as this gives the most stringent constraints.

The production of a top-quark in the $t$-channel can be described both in the 5-flavor scheme, in which the $b$ quark is considered massless and appears in the initial state, and in the 4-flavor scheme, which
keeps into account $m_b$ effects.
The total and differential SM cross sections are known at NNLO in QCD~\cite{Brucherseifer:2014ama} only in the 5-flavor scheme.  Top-quark decay is also available at NNLO \cite{Brucherseifer:2013iv,Gao:2012ja}.  A combination of production and decay in the top-quark narrow-width approximation has been presented in Ref.~\cite{Berger:2016oht}.  The calculation in the 4-flavor scheme has been presented at NLO in QCD in Ref.~\cite{Campbell:2009ss} and the interface to the parton shower is discussed in Ref.~\cite{Frederix:2012dh}. More recently, this process was interfaced with the parton shower at the NLO+PS level, including off-shell and interference effects~\cite{Jezo:2015aia,Papanastasiou:2013dta,Frederix:2016rdc}.

We computed the corrections of the operators $\xi_{t j}$ to the $t$-channel single top cross section in the 5-flavor scheme,
including NLO QCD effects building upon the POWHEG implementation in Ref.~\cite{Alioli:2009je}.  The right-handed current operators interfere with the SM through terms proportional to the mass of the light down-type quark. For $\xi_{td}$ and $\xi_{ts}$ operators it is safe to neglect these terms considering the current collider sensitivity.  For $\xi_{tb}$, the interference is more important but can also be neglected at present.

In Table~\ref{SingleTop} we give the SM cross section at $\sqrt{S} = 7$, $8$, and $13$ TeV, and the corrections induced by the RHCC operators. 
The cross sections 
are evaluated at the factorization and renormalization scale $\mu = m_t$. 
The scale uncertainty was estimated by varying the factorization and renormalization scales between $\mu = m_t/2$ and $\mu = 2 m_t$. 
The PDF and $\alpha_s$ uncertainties were conservatively estimated following the original PDF4LHC recipe~\cite{Alekhin:2011sk} using the three PDF sets CT10 \cite{Lai:2010vv}, MSTW08 \cite{Martin:2009iq}, and NNPDF2.3 \cite{Ball:2012cx}. PDF uncertainties turn out to dominate the theoretical uncertainty.

In addition to the single-top production cross section, modifications of the $W t q$ vertex strongly affect the decay of the top quark.
An observable particularly sensitive to RHCC is the ratio of the top decay width into a $W$ and a $b$ quark 
and the width into all down-type quarks. This ratio is measured to be \cite{Olive:2016xmw}
\begin{equation} 
\frac{\Gamma(t \rightarrow W b)}{\sum_{q=d,s,b} \Gamma(t \rightarrow W q)} =  0.957 \pm 0.034\, .
\end{equation}
In the presence of $\xi_{tj}$, neglecting  terms proportional to  $m_b$, this becomes
\begin{equation} \label{eq:TopDecayWidth}
\frac{\Gamma(t \rightarrow W b)}{\sum_{q=d,s,b} \Gamma(t \rightarrow W q)} =  \frac{ |V_{tb}|^2 + |\xi_{tb}|^2}{\sum_{q=d,s,b} \left( |V_{t q}|^2 +   |\xi_{t q}|^2 \right)}\, .
\end{equation}

\begin{table}
\center
\begin{tabular}{|| c| c |c   || c  | c  |  c || }
\hline
 $\sqrt{S}$ (TeV) &    	    & SM 		    & $|\xi_{td}|^2$    & $|\xi_{ts}|^2$  & $|\xi_{tb}|^2$ \\
\hline
7 & $\sigma_t$ (pb)         & $41.9^{+1.8}_{-0.9}$  & $309  \pm 3$    	&  $85.6 \pm 6.4$ &  $36.0 \pm 1.2$ \\
  & $\sigma_{\bar t}$ (pb)  & $22.7^{+0.9}_{-1.0}$  & $89.2 \pm 2.4$ 	&  $58.0 \pm 4.8$ &  $24.0 \pm 0.8$\\ 
\hline
8 & $\sigma_t$ (pb) 	    & $56.4^{+2.4}_{-1.1}$  & $375 \pm 4$ 	& $110  \pm 8$ 	  & $48 \pm 1.6$\\
  & $\sigma_{\bar t}$ (pb)  & $30.7^{+1.1}_{-1.3}$  & $115 \pm 3$ 	& $76 \pm 6.4$	  & $32 \pm 1.0$\\ 
\hline
13 & $\sigma_t$ (pb) 	    & $136\pm 5$ 	    & $720 \pm 12$ 	& $259 \pm 18$	  & $126 \pm 12$\\
   & $\sigma_{\bar t}$ (pb) & $81.0^{+4.1}_{-3.6}$  & $265 \pm  11$ 	& $200 \pm 14$	  & $88 \pm 2.4$ \\ 
\hline
\end{tabular}
\caption{$t$-channel single-top cross section in the presence of  $\xi_{tj}$. }\label{SingleTop}
\end{table}

Additional information is carried by the helicity fractions of $W$ bosons produced from top quark decays, which are mostly sensitive to $\xi_{tb}$.
The helicity fractions have been measured at the Tevatron~\cite{Aaltonen:2012rz} and at the LHC \cite{Aad:2012ky,Chatrchyan:2013jna,Aad:2015yem,Khachatryan:2014vma,Aaboud:2016hsq}. In Table~\ref{Tab:HelFrac} we summarize the results used in our analysis. The experimental uncertainty is obtained by combining in quadrature the statistical and systematic errors reported by the experimental collaborations,
and considering the correlations between $F_0$ and $F_{L,R}$ in the  determination of the $\chi^2$ function.

The SM helicity fractions have been computed at NNLO in QCD \cite{Czarnecki:2010gb}. They can be expressed as functions of the ratio $x = m_W/m_t$. Choosing the top pole mass $m_t = 173$ GeV and $m_W = 80.4$ GeV,  
the SM helicity fractions at NNLO are $F_0 = 0.687$, $F_L = 0.311$, and $F_R = 0.0017$. The theoretical uncertainty is very small, only at the permille level, and is negligible compared to the experimental uncertainty.  
The corrections to the helicity fractions induced by a right-handed $W t b$ vertex have been computed at NLO in QCD in Ref. \cite{Drobnak:2010ej},
and they can be obtained from the SM by exchanging $F_R$ and $F_L$.

Combining the information from single-top production, top decay, and $W$ boson helicity fractions, we obtain the following 90\% C.L. bounds
\begin{eqnarray}
|\xi_{td} | < 0.13, \qquad  |\xi_{ts} | < 0.22, \qquad |\xi_{tb} | < 0.16\, ,
\end{eqnarray}
where, as in the previous Sections, we assumed that the SM is modified only by a RHCC.

\subsection{Higgs branching ratios}\label{sec:hbb}

RHCC  affect  Higgs decays at tree level, by modifying the $h \rightarrow W W^*$ channel, and, more importantly,  
contributing at one loop to the Yukawa couplings of the Higgs boson to fermions. In particular, $\xi_{tb}$ induces large corrections to the bottom Yukawa coupling $y_b$. The corrections are proportional to the top Yukawa coupling $y_t$ and  alter the $h\to b\bar b $ width and thereby the total Higgs width as well. 
Defining the quark Yukawa couplings as 

\begin{table}
\center
\begin{tabular}{||c|c|c|c c||}
\hline
\hline
$F_0$            	  & $F_L$  & $F_R$  &  \multicolumn{2}{c||}{experiment} \\
\hline
$0.72 \pm 0.08$    &  $0.31 \pm 0.09$          & $-0.03 \pm 0.04$  	& CDF \& D0 &\cite{Aaltonen:2012rz}\\
$0.67 \pm 0.07 $   &  $0.32 \pm 0.04$          & $ 0.01 \pm 0.04$  	& ATLAS &\cite{Aad:2012ky} \\
$0.68 \pm 0.04$    &  $0.31 \pm 0.03$	       & $ 0.01 \pm 0.01$  	& CMS &\cite{Chatrchyan:2013jna} \\
$0.72 \pm 0.06 $   &  $0.30 \pm 0.04$	       & $-0.02 \pm 0.02 $ 	& CMS &\cite{Khachatryan:2014vma} \\
$0.709 \pm0.019$   &   $0.299 \pm 0.015$       &    $0.008 \pm 0.014$              &  ATLAS     & \cite{Aaboud:2016hsq}  \\  
\hline
\end{tabular}
\caption{$W$ helicity fractions measured at CDF, D0, ATLAS, and CMS. }\label{Tab:HelFrac}
\end{table}

\bea
\vL \supset -y_q \, \bar q_L q_R\, h+{\rm h.c.}\,\, ,
\eea
the running of the bottom Yukawa and mass are modified by $\xi_{tb}$ as follows \cite{Jenkins:2013zja,Jenkins:2013wua,Alonso:2013hga},
\bea \label{YbRGE}
\frac{d m_b}{d \ln\mu} &=& \g_m \frac{\al_s}{4\pi}m_b + \frac{g\sq}{(4\pi)\sq}m_t(x_t-3)V_{tb}^*\xi_{tb}\, ,\nn\\
\frac{d y_b}{d \ln\mu} &=& \g_m \frac{\al_s}{4\pi}y_b + \frac{g\sq}{(4\pi)\sq}\frac{m_t}{v}(x_h+3x_t-9)V_{tb}^*\xi_{tb}\, ,
\eea
where $x_i=m_i\sq/m_W\sq$. The anomalous dimension  $\g_m = -6C_F$ determines the usual one-loop evolution from QCD effects, while the $\xi_{tb}$ terms alter the evolution between the scale of new physics, $\Lambda$, and $\mu=M_W$. We use the boundary condition $y_b(\Lambda)=m_b(\Lambda)/v $, which follows from our assumption that the $\xi$ operators are the only dimension-six terms present at $\mu = \Lambda$.  In this case the SM Yukawa couplings are not modified at this scale.

Apart from these RG effects, the right-handed current operators affect the $h\to b\bar  b$ process through finite loop contributions. The effective coupling that is probed in Higgs decays, is then given by
\bea
y_b^{\rm (eff)}& =& y_b(\mu) - \frac{g^2}{(4\pi)^2} \frac{m_t}{v} \frac{\xi_{t b}V^*_{t b}}{2} \bigg[( x_h + 3 x_t-9) \log \frac{\mu^2}{m_W^2}  - 3\frac{x_t (x_t  -3)}{x_t -1} \log x_t 
+  x_t \beta_t \log\left(\frac{\beta_t -1}{\beta_t + 1} \right) \nn \\
& &+  (x_h - 2)  \beta_w \log\left(\frac{\beta_W -1}{\beta_W + 1}\right) +  (2 x_h + x_t (x_t - 7)) f_1(x_h,x_t) \nn \\
& & - (4 + (2 - x_h) x_t) f_2(x_h,x_t) + (-5 + 2 x_h + 4 x_t)\, ,
\bigg],
\eea
where $y_b(\mu)$ is given by the solution of Eq.\ \eqref{YbRGE}, $\beta_i = \sqrt{1 - 4 m_i^2/m_h^2}$, and the two functions $f_{1,2}$ are given by 
\begin{eqnarray}
f_{1}(x_h,x_t) = \int_0^1 dz \frac{1}{1 - x_t + x_h z } \log\left(\frac{x_t - x_h (1 - z) z}{1 + (x_t -1) z}\right)\, , \nn \\
f_{2}(x_h,x_t) = \int_0^1 dz \frac{1}{-1 + x_t + x_h z } \log\left(\frac{1 - x_h (1 - z) z}{x_t  + z (1-x_t)}\right)\, .
\end{eqnarray}
Setting $\mu=m_H$ and $\Lambda = 1$ TeV, we find numerically $y_b^{\rm (eff)} \simeq  0.012-0.019\, \xi_{tb}V_{tb}^*$.
We then compute the  $h\to b \bar b$ decay rate using expressions in Ref.\ \cite{Spira:1997dg}. 
Using the ATLAS and CMS combination \cite{Khachatryan:2016vau} for the $\g\g,\, WW,\, ZZ,\, \tau\tau,\, b\bar{b}$, and $\mu\mu$ signal strengths, we obtain,
\bea\label{hbbbound}
{\rm Re}\, \xi_{tb} \in [-0.01,\, 0.13]\, ,\qquad {\rm Im}\, \xi_{tb}  \in [-0.22,\, 0.22]\, ,
\eea
where we turned on only one parameter, the real or imaginary part, at the time, and the bounds assume $\Lambda =1$ TeV. 
The limits are only mildly sensitive to the choice of initial scale $\Lambda$. Setting, for example, $\Lambda = 10$ TeV results in  ${\rm Re}\, \xi_{tb} \in [-0.005,\, 0.10]$, and ${\rm Im}\, \xi_{tb}  \in [-0.15,\, 0.15]$.
Notice that the bounds from $h \rightarrow b\bar b$ are already competitive with the direct bounds from single-top production and top decay. 

Analogously to $\xi_{tb}$, the  $\xi_{td}$ and $\xi_{ts}$ elements give   contributions to the down- and strange-quark Yukawas that are enhanced by $y_t$, with, in this case, some suppression from the small CKM elements $V_{td}$ and $V_{ts}$.   
Since, at the moment, $y_d$ and $y_s$ are constrained at the same level as $y_b$ \cite{Perez:2015aoa,Soreq:2016rae}, the bounds on  $\xi_{td}$ and $\xi_{ts}$ are weaker than Eq. \eqref{hbbbound}
and not competitive with single top production and top decay. 
The contributions of the first two rows of the $\xi_{ij}$ matrix to the quark Yukawa couplings are not enhanced by the top Yukawa. 
In this case, tree level corrections to $h \rightarrow W W^*$ are more  important.
Neglecting quark and lepton masses,  the decay rate at LO in QCD is equal to
\begin{equation}
\Gamma(h \rightarrow W W^*) = \frac{3 m_H m_W^4}{32 \pi^3 v^4}  \left(  R(x_W) +  \sum_{i,j}  \left( |V_{ij}|^2  R(x_W)  +     |\xi_{ij}|^2 R_\xi(x_W) \right)  \right)\ ,   
\end{equation}
where $x_W = m_W^2/m_H^2$, the sum over $i$ and $j$ extends over all light quarks, and the functions $R$ and $R_\xi$ are \cite{Spira:1997dg}
\begin{eqnarray}
R(x) &=& -\frac{1 - x}{6 x} (2 - 13 x + 47 x^2) - \frac{1}{2} (1 - 6 x + 4 x^2) \log x  + \frac{ (1 - 8 x + 20 x^2)}{\sqrt{4 x - 1}} \arccos\left( \frac{3 x -1}{2 x^{3/2}} \right) \nn \\
R_\xi(x) &=& -\frac{1 - x}{36 x^2} (-3 + 53 x - 541 x^2 + 407 x^3 ) + \frac{1}{6 x} (2 + 3 x + 114 x^2 - 12 x^3) \log x  \nn \\ & &+ \frac{-2 + 19 x - 80 x^2 + 156 x^3}{3 x \sqrt{4 x - 1}} \arccos\left( \frac{3 x -1}{2 x^{3/2}} \right)\ .
\end{eqnarray}
For $m_H = 125$ GeV, $R_\xi(x_W)/R(x_W) = 0.03$, such that 
\begin{equation}
\Gamma(h \rightarrow W W^*) = \Gamma_{\textrm{SM}}(h \rightarrow W W^*) \left(1 + 0.01 \sum_{i j} |\xi_{ij}|^2 \right)\, ,
\end{equation}
and very large couplings, $\xi_{ij} > 1$, are needed to significantly affect the Higgs branching ratios.

\subsection{Summary of collider bounds}
To summarize, the 90\% C.L.\ bounds on the $\xi_{ij}$ matrix are 
 \begin{equation}
\left( \begin{array}{ccc}
|\xi_{ud}| 	& |\xi_{us}|    & |\xi_{ub}| \\
|\xi_{cd}|  	& |\xi_{cs}|    & |\xi_{cb}| \\
|\xi_{td}|   	& |\xi_{ts}|    & |\xi_{tb}| \\
\end{array}
\right)  \le  \left( \begin{array}{ccc}
0.04  & 0.05 & 0.08 \\
0.08  & 0.19 & 0.30 \\
0.13  & 0.22 &  0.16 \\
\end{array}
\right)\, . \label{sumcoll}
\end{equation}
The bounds on the first two rows are dominated by $WH$ production, while the bounds on $\xi_{tj}$ 
are determined by single-top production and top decay.
Including the constraint from $h \rightarrow b \bar b$ changes the bounds on $\xi_{tb}$ into $\textrm{Re}\, \xi_{tb} \in [-0.01,0.13]$ and $\textrm{Im}\, \xi_{tb} \in [-0.15,0.15]$.
The above collider constraints still leave room for BSM physics around, or even slightly below, the TeV scale.

\section{Low-energy effective Lagrangian}\label{sec:LELagr}
The effects of the RHCC operator $\xi$ at low energies are obtained by integrating out the $W$ boson and the other heavy SM degrees of freedom.
We start by analyzing the tree-level contributions to semileptonic and four-quark operators and then discuss the loop-level operators that are relevant for EDMs ($\Delta F =0$) and 
rare flavor-changing neutral-current (FCNC)  processes such as $B \rightarrow X_q \gamma$ and $K_L \rightarrow \pi^0 e^+ e^-$ ($\Delta F =1$).
The matching onto $\Delta F =1$ penguin operators and $\Delta F =2$ operators relevant for meson-antimeson mixing, which involve two insertions of $\xi$, are discussed in Appendix \ref{AppB}.
A similar analysis for SM-EFT operators involving the Z boson was recently reported in ref. \cite{Bobeth:2017xry}.

\subsection{Tree-level effective Lagrangian}
The combination of the SM left-handed charged current and the RHCC generates at low-energy one semileptonic four-fermion 
and two four-quark operators. Including the four-fermion operators induced by the SM at tree level, we have
\begin{eqnarray}\label{LagTree}
\mathcal L &=& -\frac{4 G_F}{\sqrt{2}} \left(  V^*_{ij}\,  \bar d^j \gamma^\mu P_L u^i \, \bar \nu \gamma_\mu P_L l  + 
\xi^*_{ij}\,  \bar d^j \gamma^\mu P_R u^i \, \bar \nu \gamma_\mu P_L l  + \textrm{h.c.}
\right) \nn \\
& &  - \sum^{2}_{a=1} \,\left( C^{ij\, lm}_{a\, LL} \mathcal O^{ij\, lm}_{a\, LL}  + C^{ij\, lm}_{a\, LR} \mathcal O^{ij\, lm}_{a\, LR} + C^{{ij\, lm}\,*}_{a\, LR} \big(\mathcal O^{ij\, lm}_{a\, LR}    \big)^\dagger  \right)\ ,      
\end{eqnarray}
where $P_{L,R} = (1\mp\gamma_5)/2$, $i$-$m$ are flavor indices, and the four-quark operators are defined as
\begin{eqnarray}\label{eq:4q1}
\mathcal O^{ij\, lm}_{1\, LL} = \bar d^m \gamma^\mu P_L u^l \, \bar u^i \gamma_\mu P_L d^j\ , \qquad  \mathcal O^{ij\, lm}_{2\, LL}  =\bar d_\al^m \gamma^\mu P_L u_\bt^l \, \bar u_\bt^i  \gamma_\mu P_L d_\al^j\ , \nn \\
\mathcal O^{ij\, lm}_{1\, LR} = \bar d^m \gamma^\mu P_L u^l \, \bar u^i \gamma_\mu P_R d^j\ , \qquad  \mathcal O^{ij\, lm}_{2\, LR}  =\bar d_\al^m   \gamma^\mu P_L u_\bt^l \, \bar u_\bt^i  \gamma_\mu P_R d_\al^j\ ,
\end{eqnarray}
where  $\al,\,\bt$ are color indices. 
The tree-level  matching coefficients at the scale $m_W$ are given by
\begin{eqnarray}\label{eq:4q2}
C^{ij\, lm}_{1\, LL}(m_W)  = \frac{4 G_F}{\sqrt{2}}  V^*_{lm} V_{i j}\ ,  \qquad C^{ij\, lm}_{1\, LR}(m_W)  =\frac{4 G_F}{\sqrt{2}}  V^*_{lm} \xi_{i j}\ , \qquad C^{ij\, lm}_{2\, AB}(m_W) = 0 \ ,
\end{eqnarray}
where $A,B \in \{L,R\}$ and hermiticity implies $C^{lm\, ij \,*}_{1\, LL}= C^{ij\, lm}_{1\, LL}$. As usual, the SM couplings scale as two inverse powers of the electroweak scale, $C_{i\, LL}\sim 1/v\sq$, while the `left-right' operators induced by the RHCC scale as two inverse powers of the scale of new physics, $C_{i\, LR}\sim \xi/v\sq\sim 1/\Lambda\sq$. We neglect four-quark operators that are quadratic in $\xi$ and are suppressed by $v^2/\Lambda\sq$ with respect to the linear terms.

The operators in Eq. \eqref{LagTree} need to be evolved to lower energies. While the semileptonic operators are not affected by QCD RG evolution, the anomalous dimensions of the four-quark operators are defined by\cite{Buchalla:1995vs,Dekens:2013zca}
\begin{eqnarray}\label{eq:4q3}
 \frac{d}{d\log\mu} \left(
C_{1\, AB}, C_{2\, AB}  \right)^T  = \frac{\alpha_s}{4\pi} \sum_{n=0} \left(\frac{\alpha_s}{4\pi}\right)^n \gamma^{(n)}_{AB}  \left(
C_{1\, AB}, C_{2\, AB}  \right)^T  \,,
\end{eqnarray}
and, at lowest order,  we have
\begin{eqnarray}
\gamma^{(0)}_{LL} =\left( \begin{array}{cc}
-\frac{6}{N_c}& 6\\
6& -\frac{6}{N_c} \\
\end{array} 
\right)\ ,  \qquad \gamma^{(0)}_{LR} =   \left( \begin{array}{cc}
\frac{6}{N_c}& 0\\
-6&-6\frac{N_c\sq-1}{N_c}
\end{array}
\right)\ .
\end{eqnarray}
The solution of the RGE for the operators of interest is given in Tab.  \ref{MQCDXi}.

The semileptonic operators in Eq.\ \eqref{LagTree} affect leptonic and semileptonic decays of mesons and the $\beta$ decay of baryons. 
In particular, $\xi$ is the only dimension-six operator in the SMEFT that induces a tree-level charged-current coupling of  right-handed quarks to left-handed leptons, allowing for clean low-energy tests. 
The coefficients of the four-quark operators $\mathcal O_{i\, LR}$ have an imaginary part which leads to CP violation even if all generation indices are the same. They therefore
induce large tree-level contributions to observables such as EDMs and $\ep^\prime/\ep$.

\subsection{One-loop contributions to $\Delta F =0$ and $\Delta F=1$ operators}

Next we consider loop diagrams, which can induce important contributions to processes such as $B \rightarrow X_q \gamma$, $K_L \rightarrow \pi^0 e^+ e^-$ and to EDMs.
At linear order in $\xi$, the most important operators that are generated are dipole operators and the Weinberg three-gluon operator,
described by 
\begin{eqnarray}\label{eq:dipole1}
\mathcal L &=&   \bigg(- \frac{g_s}{2} \sum_{i j \in \{u,c\}}   m_{u_j}  C^{i j}_{g u}  \bar u^{i}_L \sigma^{\mu \nu} \, G_{\mu\nu}^a t^a u^j_R  
- \frac{g_s}{2} \sum_{i j \in \{d,s,b\}}     m_{d_j}  C_{g d}^{i j}  \bar d^{i}_L \sigma^{\mu \nu} \, G_{\mu\nu}^a t^a d^j_R  \nn   \\ & & 
- \frac{e Q_u}{2} \sum_{i j \in \{u,c\}}     m_{u_j}  C_{\gamma u}^{i j}  \bar u^{i}_L \sigma^{\mu \nu} \, F_{\mu\nu} u^j_R 
- \frac{e Q_d}{2} \sum_{i j \in \{d,s,b\}}    m_{d_j}  C_{\gamma d}^{i j}  \bar d^{i}_L \sigma^{\mu \nu} \, F_{\mu\nu} d^j_R \nn \\ & & 
- \frac{e Q_e}{2} \sum_{i j \in \{e,\mu,\tau\}}     m_{e_j}  C_{\gamma l}^{i j}  \bar e^{i}_L \sigma^{\mu \nu} \, F_{\mu\nu} e^j_R +{\rm h.c.}\bigg)  
 + \frac{g_s }{3} f^{abc} \left(  C_{G} G^{a\,\mu \nu}  + \frac{C_{\tilde G}}{2} \varepsilon^{\mu \nu \alpha \beta} G^a_{\alpha \beta} \right) G^b_{\mu \rho} G^{c\, \rho}_{\nu}\, , \nn\\
\end{eqnarray}
where we chose to factor the mass of the right-handed quark or lepton  out of the definition of the dipole operators. 
The lepton dipoles receive a contribution from $\xi_{tb}$ at two loops, which, neglecting neutrino mass effects, is diagonal in lepton flavor. We discuss this contribution in Appendix \ref{app:eEDM}.   
The operators in Eq. \eqref{eq:dipole1} satisfy the RGEs 
\begin{eqnarray}
\frac{d}{d \log \mu} \left( C^{i j}_{\gamma u}, C^{i j}_{g u}, C_G + i C_{\tilde G} \right)^T  &=& \frac{\alpha_s}{4\pi} \sum_n \left( \frac{\alpha_s}{4\pi} \right)^n \gamma^{(n)}_{\textrm{dipole}}  \left( C^{ij}_{\gamma u}, C^{ij}_{g u}, C_G + i C_{\tilde G} \right)^T \nn \\ & &  +  \frac{\alpha_s}{4\pi} \frac{m_{d_m}}{m_{u_j}} \sum_n \left( \frac{\alpha_s}{4\pi} \right)^n \gamma^{(n)}_{\textrm{dipole, uLR}} \left(C^{j m\, i m \,*}_{1\, LR}, C^{j m\, i m\, *}_{2\, LR}\right)^T  \\
\frac{d}{d \log \mu} \left( C^{i j}_{\gamma d}, C^{i j}_{g d}, C_G + i C_{\tilde G} \right)^T  &=& \frac{\alpha_s}{4\pi} \sum_n \left( \frac{\alpha_s}{4\pi} \right)^n \gamma^{(n)}_{\textrm{dipole}}  \left( C^{{ij}}_{\gamma d}, C^{{ij}}_{g d}, C_G + i C_{\tilde G} \right)^T \nn \\ & &  +  \frac{\alpha_s}{4\pi} \frac{m_{u_m}}{m_{d_j}} \sum_n \left( \frac{\alpha_s}{4\pi} \right)^n \gamma^{(n)}_{\textrm{dipole, dLR}} \left(C^{m j\,m i}_{1\, LR}, C^{m j\, m i}_{2\, LR}\right)^T\, .
\end{eqnarray}
At lowest order $\gamma_{\textrm{dipole}}$ is given by \cite{Weinberg:1989dx,Wilczek:1976ry,BraatenPRL,Dekens:2013zca}
\begin{eqnarray}\label{eq:dipole2}
\gamma^{(0)}_{\textrm{dipole}} &=& \left(
\begin{array}{ccc}
8 C_F 
& -8 C_F 
& 0 \\
0     & (16 C_F - 4N_c) 
& 2  N_c \delta_{i j}\\
0     &    0 	     & N_c + 2 n_f + \beta_0
\end{array}
\right)\, .
\end{eqnarray}
The mixing between the tree-level operators $C_{1,2 \, LR}$ and the dipole operators was computed in Ref. \cite{Cho:1993zb} for down-type dipoles, 
and it is given by \cite{Cho:1993zb,MisiakPrivate} \footnote{We thank M. Misiak for providing us the expression of $\gamma^{(0)}_{\textrm{dipole,dLR}}$ for general charges $Q_d$ and $Q_u$ \cite{MisiakPrivate}.}
\begin{eqnarray}\label{eq:dipole3}
\gamma^{(0)}_{\textrm{dipole,dLR}} &=& \frac{1}{(4\pi)^2}\left(
\begin{array}{cc}
32 \frac{Q_u}{Q_d} \left( 1 + \frac{2}{3} \frac{Q_d}{Q_u}\right) &  160 \frac{Q_u}{Q_d}\\
\frac{16}{3} & -8  \\
0     &    0 	    
\end{array}
\right)\, .
\end{eqnarray}
$\gamma^{(0)}_{\textrm{dipole, uLR}}$ is obtained by replacing $Q_d$ with $Q_u$ in Eq. \eqref{eq:dipole3}.

The one-loop matching coefficients at the scale $m_W$ are 
\bea
C^{i l}_{\gamma u}(m_W) &=& \frac{4 G_F}{\sqrt{2}}\frac{2}{(4\pi)\sq}\sum_{j \in \{ d,s,b\}}\frac{m_{d_j} }{m_{u_l}Q_u} \xi_{lj}^* V_{ij} \, ,\nn\\
C^{i l}_{g u}(m_W)	&=&  0\, ,\nn\\
C^{i l}_{\gamma d}(m_W)	&=& \frac{4 G_F}{\sqrt{2}}\frac{2}{(4\pi)\sq}\sum_{j \in \{ u,c\} }\frac{m_{u_j}}{m_{d_l}Q_d}  \xi_{jl} V^*_{ji}
+ \frac{4 G_F}{\sqrt{2}}\frac{1}{(4\pi)\sq} \frac{m_{t} Q_u}{m_{d_l}Q_d}  \xi_{tl} V^*_{ti}\left[f_W(x_t)+\frac{1}{Q_u} g_W(x_t)\right]
,\nn\\
C^{i l}_{g d}(m_W)	&=&-\frac{4 G_F}{\sqrt{2}}\frac{1}{(4\pi)\sq} \frac{m_t}{m_{d_l}} \xi_{tl} V^*_{ti} f_W(x_t)\, , \nn \\
C_{\tilde G}(m_W) &=& 0\, ,
\label{Xi1loopMatch}\eea
where $x_t = m_t^2/m_W^2$, and we neglected powers of $x_{u_j}$ and $x_{d_j}$ except for the top quark.
The loop functions are given by 
\bea f_W(x)=
\frac{x^3+3x-4-6x\ln x}{2(x-1)^3}
\,,\qquad
g_W(x) = \frac{4+x(x-11)}{2(x-1)^2}+3\frac{x\sq \ln x}{(x-1)^3}\, .
\eea
Notice that at the scale $m_W$ there is no matching contribution to the Weinberg operator. Two-loop diagrams with internal top and bottom quarks, and a $W$ exchange within the loop,
as the ones computed in Ref.\  \cite{Chang:1990sfa}, are exactly canceled by a diagram in the EFT below $m_W$, with an insertion of a bottom chromo-EDM (CEDM), the $C^{bb}_{gd}$ operator in Eq. \eqref{Xi1loopMatch}, such that $C_{\tilde G}(m_W) = 0$.

\begin{table}
\center
\small$
\begin{array}{c||ccccccccc}
q'\to q\g & \xi_{uq'}V_{uq}^* &\xi_{cq'}V_{cq}^*&\xi_{tq'}V_{tq}^*\\\hline\hline
\frac{m_{q'}}{m_b} v\sq C_{\g d}^{qq'}(m_W) & -3.6\Ex{-5}&-0.019&-3.2
  \\
\frac{m_{q'}}{m_b} v\sq  C_{g d}^{qq'}(m_W) & -&-&-0.48
 \\\hline
\frac{m_{q'}}{m_b}v\sq  C_{\g d}^{qq'}(\mu_b^+) & -4.7\Ex{-5}&-0.025&-2.0
  \\
\frac{m_{q'}}{m_b}v\sq  C_{g d}^{qq'}(\mu_b^+) & -8.0\Ex{-7}&-4.3\Ex{-4}&-0.30
   \\\hline
\frac{m_{q'}}{m_b}v\sq  C_{\g d}^{qq'}(\mu_b^-) &-4.7\Ex{-5}&-0.054&-2.0
   \\
\frac{m_{q'}}{m_b}v\sq  C_{g d}^{qq'}(\mu_b^-) & -8.0\Ex{-7}&-6.2\Ex{-3}&-0.30
\end{array}   $
 \caption{\small Contributions of the right-handed $W$-current to the $q'\to q\g$ dipole operators at $\mu = m_W$ and $\mu=\mu_b^\pm= 2$ GeV. Here $\mu_b^-$ and $\mu_b^+$ differentiate between cases in which the charm quark has been integrated out at $\mu=\mu_b$, or is still present within the EFT, respectively. A $``-"$ denotes the contribution is negligible for our purposes.}\label{BdsCoeffs}  \end{table}

The operators $C^{b q}_{\gamma (g) d}$ and $C^{q b}_{\gamma (g) d}$ contribute to $B \rightarrow X_{q} \gamma$, and their value at the scales $\mu = m_W$ and $\mu = \mu_b = 2$ GeV are given in Table \ref{BdsCoeffs}. The contributions from diagrams with internal top quarks are enhanced by $m_t/m_b$ with respect to the SM, which, as we will see, leads to stringent bounds on the $\xi_{t q}$ elements of the right-handed matrix. 
In order  to match onto the relevant operators for EDMs and $K_L \rightarrow \pi^0 e^+ e^-$ one has to consider the contributions arising at the bottom and charm thresholds.
At the scale $m_b$, four-quark operators such as $C^{ub\, ub}_{1\,LR}$ generate threshold corrections to the up and charm EDM and CEDMs, while the bottom CEDM contributes to the Weinberg
operator. Thus, at the $b$ threshold
\begin{eqnarray} 
C^{i l}_{\gamma u}(m_b^-)  &=& C^{i l}_{\gamma u}(m_b^+) +  \frac{2}{(4\pi)\sq} \frac{m_{b} Q_d}{m_{u_l}Q_u} \left( C^{{lb\, ib}\, *}_{1 LR}(m_b)  + N_c\, C^{{lb\,ib}\, *}_{2\, LR}(m_b)\right)\, ,\nn\\
C^{i l}_{g u}(m_b^-)  &=& -  \frac{2}{(4\pi)\sq} \frac{m_{b} }{m_{u_l}} \, C^{{lb\,ib}\, *}_{1 LR}(m_b) \, , \nn \\
C_{\tilde G}(m_b^-)  &=&  - \frac{\alpha_s}{8 \pi} \textrm{Im}\, C^{bb}_{gd} (m_b)\, ,  
\end{eqnarray}
where we took into account that, with the initial condition of Eq. \eqref{Xi1loopMatch}, the running from $m_W$ to $m_b$ does
not induce an up-type CEDM, or a Weinberg operator. There is no $b$-threshold contribution to $C^{i l}_{g d}$ and $C^{i l}_{\gamma d}$.

At a scale $\mu_c \sim m_c$, one similarly integrates out the charm quark, and obtains additional threshold corrections to the $d$ and $s$ dipoles, and to the Weinberg operator 
\begin{eqnarray}\label{matchcharm}
C^{i l}_{\gamma d}(\mu_c^-)  &=& C^{i l}_{\gamma d}(\mu_c^+) + \frac{2}{(4\pi)\sq} \frac{m_{c} Q_u}{m_{d_l}Q_d} \left( C^{c l\, ci}_{1 LR}(\mu_c)  + N_c \, C^{c l\, ci\, }_{2\, LR}(\mu_c)\right)\, ,\nn\\
C^{i l}_{g d}(\mu_c^-)  &=& C^{i l}_{g d}(\mu_c^+) - \frac{2}{(4\pi)\sq} \frac{m_{c} }{m_{d_l}}  C^{c l\, ci}_{1 LR}(\mu_c)\,  , \nn \\
C_{\tilde G}(\mu_c^-)  &=& C_{\tilde G}(\mu_c^+) - \frac{\alpha_s}{8 \pi}\textrm{Im}\,  C^{cc}_{gu} (\mu_c)\, . 
\end{eqnarray}
In order to avoid large perturbative corrections, we run to the scale $\mu_c = 2$ GeV.

The numerical solution of the RGEs is shown in Table \ref{MQCDXi}, where, for convenience, we introduced the short-hand notation $\tilde c_i =v\sq\, \textrm{Im}\, C_i$. As the RHCC operator does not undergo QCD renormalization, the results in Table \ref{MQCDXi}, as those in Table \ref{BdsCoeffs}, are independent of $\Lambda$ to good approximation.
Some of these results, especially the contributions of $\xi_{ub,cb,tb}$ to the light quark EDMs, are rather sensitive to the choice of renormalization scale. This effect is due to a partial cancellation between matching contributions and contributions from the CEDMs. In these cases, however, the largest contributions to hadronic EDMs come from other operators ($\tilde c_{gu}^{\,uu}$ for $\xi_{ub}$ and $C_{\tilde G}$ for $\xi_{cb,tb}$) that are less sensitive to the details of the running such that the impact of perturbative uncertainties is still minor. We expect the large hadronic uncertainties related to these operators,  which we discuss in  Section \ref{sec:EDMs},
to dominate the  theoretical uncertainties on hadronic EDMs.

{\renewcommand{\arraystretch}{1.4} \begin{table}
\small$
\begin{array}{c||ccccccccc}
& V_{ud}^* \xi _{ud} & V_{us}^* \xi
   _{us} &V_{ub}^* \xi _{ub} & V_{cd}^* \xi
   _{cd}& V_{cs}^* \xi _{cs} & V_{cb}^* \xi
   _{cb} & V_{td}^* \xi _{td}& V_{ts}^* \xi
   _{ts} & V_{tb}^* \xi _{tb} \\\hline\hline
\tilde c_{\g l}^{\,ee} & - & - & - & - & - & -& - & - & 6.8\times 10^{-6}
   \\
\tilde c_{\g u}^{\,uu} & -0.033 & -0.65 & 7.1 &- & - & -1.7\Ex{-8} & - &
- & -1.5\Ex{-5} \\
\tilde c_{g u}^{\,uu} & 3.6\Ex{-3}& 0.073 & 47 & - & - & -2.3\Ex{-7} & - &
 -& -2.0\Ex{-4} \\
\tilde c_{\g d}^{\,dd}& -0.047 & - & - & -54 & - & -1.7\Ex{-8} & -2029 &
- & -1.5\Ex{-5} \\
\tilde c_{g d}^{\,dd} & -8.0\Ex{-4} & - & - & -6.2 & - & -2.3\Ex{-7} &
   -298 & - & -2.0\Ex{-4} \\
\tilde c_{\g d}^{\,ss} & - & -2.3\Ex{-3} & - & -& -2.7 & -1.7\Ex{-8} & - &
   -102 & -1.5\Ex{-5} \\
\tilde c_{g d}^{\,ss} & - & -4.0\Ex{-5} & - & - & -0.31 & -2.3\Ex{-7} &
   - & -15 &-2.0\Ex{-4} \\
v\sq C_{\tilde G} & - & - & - & - & -& -1.2\Ex{-3} & - & - &
   2.2\Ex{-3} \\
\tilde C_{1 \, LR}^{ud\,ud} & 1.8 & - & - & - & - & - & - & - & - \\
\tilde C_{2 \, LR}^{ud\,ud} & 0.91 & - & - & - & - & - & - & - & - \\
\tilde C_{1  \, LR}^{us\,us} &-& 1.8 & - & - & - & - & - & - & - \\
\tilde C_{2 \, LR}^{us\,us}  & - & 0.91 & - & - & - & - & - & - & - \\
\end{array}
$\caption{\small Contributions of the CP-odd combinations, Im$\, (V_{ij}^*\xi_{ij})$, to the operators at $\mu = 2$ GeV. Here $\tilde c_i \equiv v\sq\,{\rm Im}\, C_i$ and a $``-"$ denotes that the contribution is negligible for our purposes.
}
\label{MQCDXi}
\end{table}

\section{Leptonic and semileptonic charged-current decays}\label{sec:tree-level}

The right-handed current matrix $\xi_{ij}$ is strongly constrained by leptonic and semileptonic meson decays, and 
semileptonic decays of baryons. Leptonic decays of pseudoscalar mesons, such as $\pi^+ \rightarrow \mu^+ \nu_\mu$ or $D^+ \rightarrow \mu^+ \nu_\mu$, are sensitive to the axial component of the weak current, 
while semileptonic decays of pseudoscalar mesons into pseudoscalar mesons and leptons, such as $K \rightarrow \pi l  \nu_l$, probe the vector component. 
For the $B$ system, one can in addition study semileptonic decays  of pseudoscalar to vector mesons, such as $B \rightarrow D^* l \nu_l$, and  further orthogonal information is provided 
by inclusive $B$ decays, $B \rightarrow X_{u,c} l \nu_l$. $\beta$ decays of heavy and light baryons, such as $n \rightarrow p e^- \bar{\nu}$ or $\Lambda_b \rightarrow \Lambda_c \mu^- \bar{\nu}$, 
give, in principle, a direct handle on the chiral structure of the interactions, and allow one to construct observables that are sensitive to CP violation in the matrix $\xi$. An example is the triple correlation, $D_n$, measured in neutron $\beta$ decay.

\begin{table}[t]
\center
\begin{tabular}{||c|c|| c |c ||}
\hline
               &Decay constant &   & Form Factor \\
\hline
$f_\pi$  		       &  $130.2 \pm 1.4$ MeV & $f^{K\pi}_+(0)$ 	  	& $0.9677 \pm 0.0027$\\ 
$f_{K}/f_{\pi} $     &  $1.192 \pm 0.005$   &  		  	& \\
$f_D  $ 		       &  $209.2 \pm 3.3$ MeV & $f_{+}^{D\pi}(0)$ 	& $0.666 \pm 0.029$ \\
$f_{D_s}$ 		       &  $249.8 \pm 2.3$ MeV & $f_+^{D K}(0)$    	& $0.747 \pm 0.019$ \\ 
$f_B $ 	 		       &  $192.0 \pm 4.3$ MeV &	$\mathcal F_{D}(1)$ 	& $1.035 \pm 0.040$			\\
$f_{B_s} $ 	 	       &  $228.4 \pm 3.7$ MeV & $\mathcal F_{D^*}(1)$ 	& $0.906 \pm 0.004 \pm 0.012$ 			\\
\hline
\end{tabular}
\caption{Lattice input on pseudoscalar meson decay constants and form factors. We use the FLAG lattice averages with $n_f = 2 + 1$.\cite{Aoki:2016frl} }
\label{LQCDinput}
\end{table}

From the theoretical point of view, leptonic and semileptonic decays are very clean observables.
Leptonic decays are characterized by a single nonperturbative parameter, the meson decay constant, whose values are nowadays precisely computed with lattice QCD (LQCD) \cite{Aoki:2016frl}.
Semileptonic transitions have also been the subject of intense lattice study, and the required form factors are known to high accuracy. In Table \ref{LQCDinput} we list the values of the pseudoscalar
meson decay constants and form factors that we use in our analysis. The LQCD input has been taken from the FLAG review \cite{Aoki:2016frl}.

We now list the experimental information we use to constrain the elements of the matrix $\xi$.

\paragraph{$u\rightarrow d$ and $u \rightarrow s$ transitions}\label{ud&us}
$V_{ud}$ is extracted from superallowed nuclear $\beta$ decay, which is only sensitive  to the vector component of the weak current,  
and from leptonic decays of the pion, which probe the axial component of the current. 
We use the following experimental input \cite{Olive:2016xmw,Hardy:2014qxa}
\begin{eqnarray}\label{Vud_exp}
|V_{ud}(0^+ \rightarrow 0^+)|_{\textrm{exp}}
 &=& 0.97425 \pm 0.00022\, , \nn \\
|V_{ud}(\pi \rightarrow \mu \nu) f_{\pi} |_{\textrm{exp}} 
&=& (127.13 \pm  0.02 \pm 0.13)\, \textrm{MeV}\, , 
\end{eqnarray}
where $f_{\pi} \sim 130$ MeV is the  pion decay constant.
The first uncertainty in the second line of Eq.\ \eqref{Vud_exp} is experimental, while the second is due to radiative corrections.

For the determination of $V_{us}$, we use two quantities that are experimentally very well determined  \cite{Aoki:2013ldr,Antonelli:2010yf}.
From semileptonic kaon decays, one can extract 
\begin{eqnarray}\label{Vus_vec}
(|V_{us} | f^{K\pi}_+(0))_{\textrm{exp}} = 0.2163 \pm 0.0005\, ,
\end{eqnarray}
where $f^{K \pi}_+(0)$ is the form factor entering the $K^0 \rightarrow \pi^- l \nu_l$ decay at zero momentum transfer.
The ratio of the pion and kaon leptonic decays gives  
\begin{eqnarray}\label{Vus_ax}
\left(\left|\frac{V_{us}}{V_{ud}} \right| \frac{f_{K}}{f_{\pi}} \right)_{\textrm{exp}}
= 0.2758 \pm 0.0005\, .
\end{eqnarray}

Because leptonic and semileptonic decays are sensitive to either the axial or the vector components of the current, we can easily modify the relevant expressions in the presence of a RHCC to obtain
\begin{eqnarray}
\left| V_{ud} + \xi_{ud} \right| &=& 0.97425 \pm 0.00022\, ,  \quad
|V_{ud} - \xi_{ud} |f_{\pi}  = (127.13 \pm  0.02 \pm 0.13)\, \textrm{MeV}\, ,  \nn \\
|V_{us} + \xi_{us} | f^{K\pi}_+(0) &=& 0.2163 \pm 0.0005\, , \qquad 
\frac{|V_{us} - \xi_{us} | f_{K}}{|V_{ud} -  \xi_{ud} | f_{\pi}} =  0.2758 \pm 0.0005\, . \label{Vudus} 
\end{eqnarray}
Using the LQCD input in Tab.~\eqref{LQCDinput}, Eq.\ \eqref{Vudus} provides four constraints on $V_{ud}$, $V_{us}$, $\xi_{ud}$, and $\xi_{us}$. 
Two additional constraints on the imaginary part of $\xi_{ud}$ and $\xi_{us}$ can be obtained from neutron and hyperon $\beta$ decays.
Time-reversal violation can be measured in neutron $\beta$ decay by reconstructing the triple correlation $\langle \vec J\, \rangle\cdot ( \vec p_e \times \vec p_\nu)$,
where $\vec J$ is the neutron polarization. Current measurements give $D_n = (-1\pm 2.1)\cdot 10^{-4}$ \cite{Mumm:2011nd}. This observable is contaminated by fake $T$-odd signals from final-state interactions, which, with current experimental accuracy,  can still be neglected (see Ref.~\cite{Vos:2015eba} for a more detailed discussion). 
The same correlation was measured in the decay of the $\Sigma$ baryon, $\Sigma^- \rightarrow n e^- \bar{\nu}$, with a much weaker bound, $D_{\Sigma} = 0.11 \pm 0.10$.
Following Ref.~\cite{Jackson:1957zz}, the $D_n$ and $D_\Sigma$ coefficients can be calculated as 
\begin{eqnarray}
D_n      = \frac{4 g_A}{1+3g_A\sq}\, \mathrm{Im}\,\frac{\xi_{ud}}{V_{ud}}\simeq 0.87\, \mathrm{Im}\,\frac{\xi_{ud}}{V_{ud}}\,,  \quad
D_\Sigma = \frac{4 g_{A\, \Sigma n}}{1+3g_{A\,  \Sigma n}\sq}\, \mathrm{Im}\,\frac{\xi_{us}}{V_{us}}\simeq 1.01 \, \mathrm{Im}\,\frac{\xi_{us}}{V_{us}}\,,  
\end{eqnarray}
where $g_A$ is the nucleon axial coupling, $g_A = 1.27$, and $g_{A\, \Sigma n}$ is the axial coupling of a $\Sigma$ to a neutron, measured to be $0.340 \pm 0.017$ \cite{Olive:2016xmw}. 
$D_n$ gives a strong bound on $\textrm{Im}\, \xi_{ud}/V_{ud}$, at the $10^{-4}$ level. The constraint on $\textrm{Im}\, \xi_{us}$ is at the few-percent level.
As we will see, both bounds are significantly weaker than bounds from EDMs and direct CPV in kaon decays.

\paragraph{$c \rightarrow d$ and $c\rightarrow s$ transitions}\label{sec:cd&cs}
Analogously to the $ud$ and $us$ case, we can use the leptonic and semileptonic decays of the $D$ and $D_s$ mesons,
$D^+ \rightarrow \mu^+ \nu_\mu$, $D^+_s \rightarrow \mu^+ \nu_\mu$,  $D \rightarrow \pi l \nu_l$, and $D \rightarrow K l \nu_l$, to constrain the vector and axial couplings of a charm quark to $s$ and $d$ quarks. The leptonic decays of the pseudoscalar $D$ and $D_s$ mesons probe the axial current, while the semileptonic decays probe the vector current.
The experimental input is \cite{Olive:2016xmw}
\begin{eqnarray}
f_D |V_{cd} - \xi_{cd}| &=& 45.91 \pm 1.05 \, \textrm{MeV}\, , \qquad  f_{D_s} |V_{cs} - \xi_{cs}| = 250.9 \pm 4.0 \, \textrm{MeV}\, , \\
f_+^{D\pi}(0) |V_{cd} + \xi_{cd}| &=& 0.1425 \pm 0.0019 \, ,   \qquad  f_+^{DK}(0) |V_{cs} + \xi_{cs}| = 0.728 \pm 0.005\, ,
\end{eqnarray}
and the LQCD input for the $D$ and $D_s$ decay constants and form factors is given in Table \ref{LQCDinput}.

\paragraph{$b \rightarrow u$ and $b \rightarrow c$ transitions}\label{sec:ub&cb}

In the case of $b\rightarrow c$ transitions, the vector component of the charged current is constrained by the semileptonic decay $B \rightarrow D l \nu_l$. 
For the axial component, the purely leptonic decay of the $B_c$ meson has not yet been observed. The decay $B \rightarrow D^* l \nu_l$ depends on both  the vector and axial current. In the zero-recoil limit,   
when $w = v \cdot v^\prime = 1$, where $v$ and $v^\prime$ are the $B$ and $D$ mesons four-velocities, only the axial contribution survives \cite{Manohar:2000dt}.  
Using the HFAG averages \cite{Amhis:2016xyh}, we can write
\begin{eqnarray}
\eta_{EW}\mathcal F_{D}(1) |V_{cb} + \xi_{cb}| &=&   \left( 42.65 \pm 0.72 \pm 1.35 \right) \cdot 10^{-3} \, , \nonumber \\
\eta_{EW}'\mathcal F_{D^*}(1) |V_{cb} - \xi_{cb}| &=&  \left( 35.81 \pm 0.11 \pm 0.44 \right) \cdot 10^{-3}\, ,
\end{eqnarray}
where $\eta_{EW}=1.012\pm0.005$ and $\eta_{EW}^{\prime}=1.015\pm0.005$ \cite{Olive:2016xmw} are electroweak corrections.
$\mathcal F_{D}(1)$ and $\mathcal F_{D^*}(1)$ denote the form factors, evaluated at $w=1$, for which we used the FLAG averages in Table~\ref{LQCDinput}.
Angular distributions in $B \rightarrow D^* l \nu_l$ could provide additional information on the Lorentz structure of the $Wbc$ vertex \cite{Becirevic:2016hea}.

The inclusive decays $\bar{B} \rightarrow X_c l \bar\nu_l$ also constrain $\xi_{cb}$. Neglecting power corrections of order $\mathcal O(\Lambda_{\textrm{QCD}}/m_b)$, the inclusive semileptonic width into charmed final states is given by
\begin{eqnarray}
\Gamma(B \rightarrow X_c l \nu) &=& \frac{G_F^2 m_b^5 |V_{cb}|^2}{192 \pi^3} \left[ \left( 1 + \left|\frac{\xi_{cb}}{V_{cb}}\right|^2\right) \left( 1 - 8 \rho + 8 \rho^3 - \rho^4 - 12 \rho^2 \log\rho \right)  \right. \nn \\
& & \left. - 4 \frac{m_c}{m_b}  \textrm{Re} \left( \frac{\xi_{cb}}{V_{cb}} \right)\,  \left(  1 + 9 \rho - 9 \rho^2 - \rho^3  + 6 \rho (1+\rho) \log\rho \right)
\right]\, ,\label{VcbWidth}
\end{eqnarray}
where $\rho = m_c^2/m_b^2$. We then set constraints by using the PDG average \cite{Olive:2016xmw},
\bea
|V_{cb}^{\rm eff}| = (42.2 \pm 0.8)\Ex{-3}\qquad (B\to X_c l\nu) \, , \label{VcbIncl}
\eea
where $|V_{cb}^{\rm eff}|\sq = |V_{cb}|^2\, \Gamma(B \rightarrow X_c l \nu) /\Gamma^{{\rm SM}}(B \rightarrow X_c l \nu) $.

The constraints from the inclusive decays we obtain in this way should only be viewed as order-of-magnitude constraints for a number of reasons. 
First of all, we should  take into account power corrections, which are not included in Eq.\ \eqref{VcbWidth}, in order to obtain $V_{cb}$ from inclusive decays \cite{Bauer:2002sh,Gambino:2011cq,Gambino:2013rza,Alberti:2014yda}.  Furthermore, Eq.\ \eqref{VcbIncl} relies on fits to the leptonic and hadronic moments of the decay distribution. As the dependence on the lepton-energy is not the same for $\xi_{cb}$ and $V_{cb}$ after applying cuts, the SM fit will be altered in the presence of right-handed currents.
We should therefore refit the leptonic moments, while taking into account contributions from right-handed currents. 
Such an analysis is beyond the scope of the current work. We will use Eq.\ \eqref{VcbIncl} to estimate the limits from the inclusive measurements, and
refer to Refs. \cite{Dassinger:2008as,Feger:2010qc,Crivellin:2014zpa} for a more detailed discussion.

In the case of $b\to u$ transitions, the leptonic channel $B^+ \rightarrow \tau^+ \nu_\tau$ allows us to determine the axial current $|V_{ub} - \xi_{ub}|$, while the vector current is probed by $B \rightarrow \pi l \nu_l$.
Additional exclusive decays, such as $B \rightarrow \rho l \nu_l$, can be used to further improve the sensitivity to RHCC \cite{Bernlochner:2014ova,Crivellin:2014zpa}.
For the leptonic decays, we use the HFAG average of the BaBar and Belle results, $\textrm{Br} (B^+ \rightarrow \tau \nu) = (1.06 \pm 0.19 ) \cdot 10^{-4}$, and we employ the FLAG extraction for the semileptonic case,
\bea
|V_{ub} -\xi_{ub}| f_B &=&  0.77 \pm 0.07\, ,\nn \\
|V_{ub} + \xi_{ub}|  &=& (3.62 \pm 0.14 ) \cdot 10^{-3}\, ,
\eea
where the decay constant, $f_B$, is given in Table \ref{LQCDinput}.

The inclusive determination from $B\to X_u l\nu_l$ decays suffers from the same problems as the charm-bottom transition. In principle, power corrections should be included \cite{Bauer:2001rc,Lange:2005yw} and the leptonic spectrum should be refitted taking into account a right-handed current. Since such an analysis is beyond the scope of our work, we take a similar approach as in the case of $V_{cb}$. We thus estimate constraints from inclusive decays by \cite{Olive:2016xmw},
\bea
\sqrt{|V_{ub}|\sq +|\xi_{ub}|\sq}= (4.49 \pm 0.18^{+0.16}_{-0.18})\Ex{-3}\qquad (B\to X_u l\nu)\, .
\eea

Another exclusive determination of $V_{cb}$ and $V_{ub}$ is provided by measurements of $\textrm{Br}(\Lambda^0_b \rightarrow p \mu^- \bar{\nu})_{q\sq > 15\, {\rm GeV}}/\allowbreak\textrm{Br}(\Lambda^0_b \rightarrow \Lambda_c^+ \mu^- \bar{\nu})_{q\sq > 7\, {\rm GeV}}$. This ratio of branching fractions can be calculated using lattice determinations of the relevant form factors \cite{Detmold:2015aaa}.
Following the procedure outlined in Ref.\ \cite{Detmold:2015aaa}  we obtain for the partially integrated decay widths,
\bea \label{eq:LambdaB}
\Gamma(\Lambda_b^0\to p\mu^- \bar \nu)_{q\sq > 15\, {\rm GeV}} &=&4.17\, {\rm ps}^{-1} \,|V_{ub}+\xi_{ub}|\sq +8.17\, {\rm ps}^{-1} \,|V_{ub}-\xi_{ub}|\sq\pm \sigma_{\rm stat}^{(p)}\pm \sigma_{\rm syst}^{(p)}\,,\nn\\\Gamma(\Lambda_b^0\to \Lambda_c^+\mu^- \bar \nu)_{q\sq > 7\, {\rm GeV}} &=&1.41\, {\rm ps}^{-1} \,|V_{cb}+\xi_{cb}|\sq +6.99\, {\rm ps}^{-1} \,|V_{cb}-\xi_{cb}|\sq\pm \sigma_{\rm stat}^{(\Lambda_c^+)}\pm \sigma_{\rm syst}^{(\Lambda_c^+)}\,,\nn\\
\eea
where the lattice uncertainties are given by
\bea
(\sigma_{\rm stat}^{(p)}) \sq &=& \left( 0.10\, |V_{ub}+\xi_{ub}|^4+0.33\, |V_{ub}-\xi_{ub}|^4 +0.16\,  |V_{ub}\sq-\xi_{ub}\sq| \sq\right)\, {\rm ps}^{-2}\,,\nn\\
(\sigma_{\rm syst}^{(p)}) \sq &=& \left( 0.10\, |V_{ub}+\xi_{ub}|^4+0.44\, |V_{ub}-\xi_{ub}|^4 +0.050\,  |V_{ub}\sq-\xi_{ub}\sq|\sq \right)\, {\rm ps}^{-2}\,,\nn\\
(\sigma_{\rm stat}^{(\Lambda_c^+)}) \sq &=& \left(0.0023\, |V_{cb}+\xi_{cb}|^4+0.017\, |V_{cb}-\xi_{cb}|^4 +0.0052\,  |V_{cb}\sq-\xi_{cb}\sq|\sq \right)\, {\rm ps}^{-2}\,,\nn\\
(\sigma_{\rm syst}^{(\Lambda_c^+)}) \sq &=& \left(0.0053\, |V_{cb}+\xi_{cb}|^4 +0.11\, |V_{cb}-\xi_{cb}|^4+0.0027\,  |V_{cb}\sq-\xi_{cb}\sq| \sq\right)\, {\rm ps}^{-2}\,.
\eea
The ratio of these decay widths is experimentally determined to be \cite{Aaij:2015bfa}
\bea
\frac{\textrm{Br}(\Lambda^0_b \rightarrow p \mu^- \bar{\nu})_{q\sq > 15\, {\rm GeV}}}{\textrm{Br}(\Lambda^0_b \rightarrow \Lambda_c^+ \mu^- \bar{\nu})_{q\sq > 7\, {\rm GeV}} }= \left( 1.00\pm 0.04\pm0.08\right)\Ex{-2}\,.
\eea
In principle, measurements of angular distributions and correlations in semileptonic $\Lambda_b$ decays could  provide more detailed probes of the Lorentz structure of the $Wbc$
vertex \cite{Manohar:1993qn}.


\section{$\Delta F = 0$ processes: Electric Dipole Moments}\label{sec:EDM}
\label{sec:EDMs}
\begin{table}[t]
\begin{center}\small
\begin{tabular}{||c||cccccc||}
\hline
&$d_e$   & $d_n$& $d_{\mathrm{Hg}}$ & $d_{\mathrm{Xe}}$ & $d_{\mathrm{Ra}}$ & $d_{p,D}$  \\
\hline
\rule{0pt}{3ex}
current limit  &$8.7 \cdot 10^{-16} $  &$ 3.0 \cdot 10^{-13}$  & $6.2 \cdot 10^{-17}$    & $5.5 \cdot 10^{-14}$   & $1.2\cdot 10^{-10}$ & x \\
expected limit &$5.0 \cdot 10^{-17} $ & $1.0 \cdot 10^{-15} $ & $6.2 \cdot 10^{-17}$   & $5.0 \cdot 10^{-16}$ & $1.0 \cdot 10^{-14}$ &$ 1.0 \cdot 10^{-16}$ \\
\hline
\end{tabular}
\end{center}
\caption{\small Current limits on the electron\cite{Baron:2013eja}, neutron \cite{Baker:2006ts,Afach:2015sja}, mercury \cite{Griffith:2009zz,Graner:2016ses}, xenon \cite{PhysRevLett.86.22} and radium \cite{Bishof:2016uqx,Parker:2015yka}
EDMs in units of $e$ fm ($90\%$ confidence level). We also show future sensitivities~\cite{Kumar:2013qya,Chupp:2014gka}.}
\label{tab:EDMexps}  
\end{table}

Permanent EDMs of leptons, nucleons, nuclei, atoms, and molecules provide stringent bounds on flavor-diagonal CPV interactions. 
The right-handed charged-current couplings $\xi_{ij}$ contribute mostly to hadronic and nuclear EDMs. Right-handed couplings of the light quarks, such as $\xi_{ud}$ and $\xi_{us}$, induce EDMs through the tree-level operators  $C^{ud\, ud}_{1,2\, LR}$ and $C^{us\, us}_{1,2\, LR}$, while the couplings to heavier quarks, such as  $\xi_{ub}$ or $\xi_{tb}$,
induce loop corrections to the light quark EDMs, CEDMs, and the Weinberg operator.

In Tab. \ref{tab:EDMexps} we summarize the current limits on the EDMs of the electron, nucleons, $^{199}$Hg,  $^{225}$Ra, and $^{129}$Xe that we used in our analysis, as well  
as projected sensitivities for these systems, and for the EDMs of the proton and deuteron that are targets for storage-ring experiments.

The calculation of nucleon and nuclear EDMs in terms of the operators in  Eqs. \eqref{LagTree} and \eqref{eq:dipole1}  involves two steps. The operators are   first matched to an extension of chiral perturbation theory (ChPT) that contains CPV hadronic interactions \cite{deVries:2012ab,Bsaisou:2014oka}.
The most important interactions are short-range contributions to the nucleon EDM and CPV pion-nucleon couplings. The latter give rise to long-range contributions to the nucleon EDM (from pion loops) and, for chiral-symmetry-breaking operators like the quark CEDMs and four-quark operators, dominate the CPV nucleon-nucleon potential. This CPV potential provides the dominant contribution to the EDMs of nuclei and diamagnetic atoms.

The chiral  power counting predicts that the four-quark operators $C^{ud\, ud}_{1,2\, LR}$ and $C^{us\, us}_{1,2\, LR}$  contribute mainly to the isospin-breaking pion-nucleon coupling $\bar g_1$ \cite{deVries:2012ab,Cirigliano:2016yhc}.  $C^{ud\, ud}_{1,2\, LR}$ do not induce the isoscalar coupling $\bar g_0$, while 
$C^{us\, us}_{1,2\, LR}$ give corrections that, while formally LO, are small with respect to $\bar g_1$. 
As discussed in Ref.\ \cite{Cirigliano:2016yhc}, it is possible to calculate the sizes of $\bar g_{0,1}$ by noticing that these couplings receive large contributions from tadpole diagrams, which involve the coupling of the neutral pion to the vacuum. 
The tadpole coupling can be related, at leading order in ChPT,  
to the $K \rightarrow \pi \pi$ matrix element of the SM electroweak penguin operators $\mathcal Q_{7,8}$.
Using  recent LQCD of these matrix elements \cite{Blum:2012uk,Blum:2015ywa}, it then becomes possible to give a solid estimate of the sizes of $\bar g_{0,1}$ \cite{Cirigliano:2016yhc}. The error on $\bar g_{0,1}$
is at the moment dominated by ChPT uncertainties, which we conservatively estimate at the 50\% level. 

$\bar g_{0,1}$ also receive contributions from the dipole operators $C^{\,uu}_{g u}$, $C^{\, d d}_{g d}$ and $C^{\,s s}_{g d}$.
These contributions can be in principle computed in LQCD  \cite{deVries:2016jox}, but, at the moment, the best estimate comes from QCD sum rules. The sum rules estimates have roughly (50-100)\% uncertainties \cite{Pospelov_qCEDM, Pospelov_deuteron, Pospelov_review, Hisano1}.
We thus find that the pion-nucleon couplings induced by the operators in Eqs.\ \eqref{LagTree} and \eqref{eq:dipole1} are
\begin{eqnarray}\label{couplings0}
\frac{\bar{g}_1}{2 F_\pi} &=& - \left(  (4.5 \pm 2.2 ) \,  \left( \tilde{C}^{us\, us}_{1\, LR} + 2 \tilde{C}^{ud\, ud}_{1\, LR} \right)   + ( 22.0 \pm 11.0 )  \, \left( \tilde{C}^{us\, us}_{2\, LR}
+ 2 \tilde{C}^{ud\, ud}_{2\, LR}\right) \nn \right. \\ & & \left. 
  + (0.2^{+0.4}_{-0.1})  ( 0.7\,  \tilde c^{\,u u}_{g u} -  1.5\,  \tilde c^{\, d d}_{g d})  \right) \times 10^{-6}\, , \\
  \frac{\bar{g}_0}{2 F_\pi} &=&  - \left(  (0.3 \pm 0.1 )\, \tilde C^{us\, us}_{1\, LR}  + ( 1.3 \pm 0.6 ) \, \tilde C^{us\, us}_{2\, LR}   
+ (0.05\pm 0.10 ) ( 0.7\,  \tilde c^{\, u u}_{g u} + 1.5\,  \tilde  c^{\, dd }_{g d}) \right)  \, \times 10^{-6} 
\,,  \nn \\
\label{couplings1}
\end{eqnarray}
where $F_\pi=f_\pi/\sqrt{2}$, all the couplings are evaluated at $\mu = 2$ GeV, and their values in terms of $\xi_{ij}$ can be read off from Table \ref{MQCDXi}. 
As in Table \ref{MQCDXi}, we have defined $\tilde c_i = v^2 \, \textrm{Im}\, C_i$.
Eqs.\ \eqref{couplings0} and \eqref{couplings1}
assume that the strong CP problem is solved by the Peccei-Quinn mechanism \cite{Peccei:1977hh} which somewhat affects the values of the matrix elements.

A variety of techniques are available for the calculation of the nucleon EDM induced by four-quark and dipole operators in Eqs. \eqref{LagTree} and \eqref{eq:dipole1}.
In Refs. \cite{Cirigliano:2016yhc,Seng:2014pba}, we estimated the nucleon EDM induced by the four-quark operators  in Eq. \eqref{LagTree}
by considering long-range contributions induced by the pion-nucleon couplings $\bar g_{0,1}$ (see also Ref.~\cite{Maiezza:2014ala}). This estimate has intrinsically large uncertainties mainly due to uncertainties on $\bar g_{0,1}$ and our ignorance of the size of short-range contributions that appear at the same chiral order. Here we use the uncertainty estimate of Ref.\ \cite{Cirigliano:2016yhc}.

The contributions of $\tilde c_{\gamma u} ^{\, u u}$ and $\tilde c_{\g d}^{\, d d}$, the up- and down-quark EDMs,  are known with $\Or(15\%)$ uncertainties \cite{Bhattacharya:2015esa, Bhattacharya:2015wna,Bhattacharya:2016zcn},
while the strange contribution is still highly uncertain.  While considerable effort is underway for the calculation of the qCEDM contribution to the nucleon EDM
\cite{Bhattacharya:2016oqm,Bhattacharya:2016rrc,Abramczyk:2017oxr}, the best estimate at the moment comes from QCD sum rules, and has an estimated $50\%$ uncertainty  \cite{Pospelov_qCEDM, Pospelov_deuteron, Pospelov_review, Hisano1}.
Finally, the Weinberg operator appears with the largest uncertainty, $\Or(100\%)$, based on a combination of QCD sum-rules \cite{Pospelov_Weinberg} and naive dimensional analysis estimates \cite{Weinberg:1989dx}. 
Combining these results, we find
\bea\label{eq:nucleonEDM}
d_n&=& 
\Big( ( 43  \pm 27 )  \tilde C^{us\, us}_{1\, LR}  + (210   \pm 130) \, \tilde C^{us\, us}_{2\, LR}  + 
      ( 22   \pm 14)   \, \tilde C^{ud\, ud}_{1\, LR}   + (110 \pm 70) \, \tilde C^{ud\, ud}_{2\, LR} \nn \\ & &   
-(0.93\pm0.05)\,  \tilde c_{\g u}^{\, u u}- (4.0\pm0.2 ) \,  \tilde c_{\g d}^{\, d d} - (0.8\pm0.9) \,  \tilde c_{\g d}^{\,s s} \nn \\ & & 
-(3.9\pm2.0)\,  \tilde c_{g u}^{\,u u}-(16.8\pm8.4)\, \tilde c_{g d}^{\, d d} \pm (320 \pm 260) v^2 C_{\tilde G} \Big)\times 10^{-9} \, e\,  \textrm{fm}\, ,\nn\\
d_p&=& \Big
( 
 -( 56   \pm 30)  \, \tilde C^{us\, us}_{1\, LR}  - (280 \pm 150) \, \tilde C^{us\, us}_{2\, LR}  
 -( 
 42  \pm 26)  \, \tilde C^{ud\, ud}_{1\, LR}   - (
 210 \pm 130) \, \tilde C^{ud\, ud}_{2\, LR}  \nn \\
& &  + 
(3.8\pm0.2) \,  \tilde c_{\g u}^{\, u u} + (1.0\pm0.1 )\,  \tilde c_{\g d}^{\,d d}-(0.8\pm0.9) \,  \tilde c_{\g d}^{\,s s)}\nn\\
&&+(9.3\pm4.6)\, \tilde c_{g u}^{\, u u}+(9.2\pm4.2)\,  \tilde c_{g d}^{\, dd} \mp (320 \pm 260) v^2 C_{\tilde G} \Big)\times 10^{-9} \, e\,  \textrm{fm} .
\eea

Finally, using the nuclear calculations of Refs. \cite{deJesus:2005nb,Dobaczewski:2005hz,Ban:2010ea,Dzuba:2009kn,Engel:2013lsa,deVries2011b,Bsaisou:2014oka, Singh:2014jca,Singh:2015aba,Yamanaka:2015qfa,Yamanaka:2017mef,XeEDM} we can predict nuclear EDMs in terms of $\bar g_{0,1}$ and $d_{n,p}$
\bea\label{eq:NuclEDM}
d_D &=&(0.94\pm 0.01)(d_n + d_p)  - (0.18 \pm 0.02) \frac{\bar g_1}{2F_\pi}\, e \, {\rm fm}\ ,\nn\\ 
d_{\mathrm{Hg}} &=& - (2.8\pm 0.6)\Ex{-4} \cdot \bigg[  (1.9\pm0.1)d_n + (0.20\pm 0.06)d_p  \nn \\ & & - \left(  0.13^{+0.5}_{-0.07}\, \frac{\bar g_0}{2F_\pi} + 0.25^{+0.89}_{-0.63}\,\frac{\bar g_1}{2F_\pi}  \right) e\, {\rm fm} \bigg]\ ,  \nn \\
d_{\mathrm{Xe}} &=& (0.33\pm 0.05)\Ex{-4}\cdot\bigg[ (-0.32 \pm0.02)d_n + (0.0061\pm 0.001)d_p  \nn \\ & & +\left( 0.10_{-0.037}^{+0.53}\, \frac{\bar g_0}{2F_\pi} + 0.076_{-0.038}^{+0.55}\, \frac{\bar g_1}{2F_\pi}\right)\bigg]e\, {\rm fm}\ , \nn\\
d_{\mathrm{Ra}} &=& (7.7\pm 0.8)\Ex{-4}\cdot\left(-19^{+6.4}_{-57}\,\frac{\bar g_0}{2F_\pi} + 76^{+227}_{-25}\,\frac{\bar g_1}{2F_\pi}\right)e\, {\rm fm}\ ~. 
\eea
The nucleon EDM contributions to $d_{\mathrm{Ra}}$ have, as far as we know, not been calculated but are expected to be small compared to the large pion-exchange contributions.

The estimates of the nucleon and nuclear EDMs in Eqs.\ \eqref{eq:nucleonEDM} and \eqref{eq:NuclEDM} are affected by large hadronic and nuclear uncertainties.
Several matrix elements are consistent with zero and the large uncertainties allow for cancellations between different contributions, which  can significantly affect the constraints on  the $\xi_{ij}$ couplings. Therefore, when setting constraints, we vary the hadronic and nuclear matrix elements within their allowed ranges in order to minimize the total $\chi\sq$. This corresponds to the Rfit approach for treating theoretical errors as defined in Ref.~\cite{Charles:2004jd}.

\section{$\Delta S = 1$ processes }\label{sec:DeltaS1}

\begin{table}
\center
\begin{tabular}{||c|c||c|c||}
\hline
Re $A_0$ & $33.201\Ex{-8} \,{\rm GeV}$  & ${\rm Re}\, A_2 $&  $1.479\Ex{-8} \,{\rm GeV}$ \\
$|\ep |$ & $(2.228\pm0.011)\Ex{-3}$ & ${\rm Arg}\, \ep$ & $0.75957\, {\rm rad}$\\
${\rm Re}\, \left({\ep'}/{\ep}\right)$ & $(16.6 \pm 2.3) \Ex{-4}$ &  $\textrm{Br} (K^+ \rightarrow \pi^0 e^+ \nu)$ & $ (5.07 \pm 0.04) \cdot 10^{-2}$  \\
$\tau(K_L)$ & $(5.116  \pm 0.021) \Ex{-8}$ s & $\tau(K^+)$ & $(1.2380 \pm 0.0020) \Ex{-8}$ s \\
\hline
\end{tabular}
\caption{Experimental input for the $\Delta S =1$ processes $\ep^\prime/\ep$ and $K_L \rightarrow \pi^0 e^+ e^-$ \cite{Olive:2016xmw}.}
\label{DS1exp}
\end{table}

In this section we discuss the contribution of RHCC to direct CP violation in kaon decays and to the FCNC decay  $K_L \rightarrow \pi^0 e^+ e^-$.
While the real parts of the $\xi_{ij}$ elements are well constrained by the leptonic and semileptonic charged-current decays discussed in Section \ref{sec:tree-level}, 
$\ep^\prime/\ep$ and $K_L \rightarrow \pi^0 e^+ e^-$ provide additional information on the imaginary parts of the $\xi_{i s}$ and $\xi_{i d}$ elements. 
$\ep^\prime/\ep$ is dominated by tree-level contributions from the four-quark operators $C^{ud\, us}_{1,2\, LR}$ and $C^{us\, ud}_{1,2\, LR}$, while $K_L \rightarrow \pi^0 e^+ e^-$ receives
correction at one loop. The latter arise from matching the RHCC  to the flavor-changing dipole operators $C^{ds}_{\gamma d}$ and $C^{sd}_{\gamma d}$, which are particularly important for internal charm and top quarks.
Other $\Delta S=1$ FCNC decays, such as $K_L \rightarrow \pi^0 \nu \bar \nu$,  receive contributions from RHCC that are only quadratic in $\xi$ and therefore play a less important role. We discuss them briefly in Appendix \ref{sec:Bmumu}. In Table \ref{DS1exp} we list the experimental input needed in Sections \ref{eppovep} and \ref{KLpi0}.

\subsection{$\ep' / \ep$}\label{eppovep}

$\xi_{ud}$ and $\xi_{us}$ give large contributions to direct CP violation in $K_L\to \pi\pi$ decays, while indirect CP violation in kaon mixing is not significantly affected  \cite{Cirigliano:2016yhc}.
Direct CP violation is quantified by $\ep'$, which can be expressed as 
\bea
{\rm Re}\, \bigg(\frac{\ep'}{\ep}\bigg) = {\rm Re}\,\bigg(\frac{i \omega e^{i(\delta_2-\delta_0)}}{\sqrt{2}\ep}\bigg)\bigg[\frac{{\rm Im}\,A_2}{{\rm Re}\,A_2}-\frac{{\rm Im}\,A_0}{{\rm Re}\,A_0}\bigg]\label{epsPrime}\ .
\eea
Here $A_{0,2} e^{i\delta_{0,2}}=\frac{1}{\sqrt{2}} \langle (\pi\pi)_{I=0,2}| \mathcal H|K\rangle$ are the amplitudes for final-state pions with total isospin $I=0,2$, the corresponding strong phases are denoted by $\delta_{0,2}$, $ \mathcal H$ is the weak Hamiltonian, and $\omega \equiv {{\rm Re}\,A_2}/{{\rm Re}\,A_0} = 0.04454$.

In the SM, $A_{0}$ and $A_{2}$ are sensitive to contributions from charged-current operators, $\mathcal Q_{1-2}$, strong penguin operators, $\mathcal Q_{3-6}$, and electroweak penguin operators, $\mathcal Q_{7-10}$. The values of their NLO Wilson coefficients have been calculated in Refs.\ \cite{Ciuchini:1992tj,Buras:1993dy,Buchalla:1995vs,Ciuchini:1995cd}, while lattice determinations of the necessary matrix elements are given in Refs.\ \cite{Blum:2015ywa,Bai:2015nea,Blum:2012uk}. Combining these results with the  experimental values in Table \ref{DS1exp} 
and lattice determinations of the strong phases, $\delta_0 = (23.8\pm4.9\pm 1.2)^\circ$, $\delta_2 = -(11.6\pm2.5\pm 1.2)^\circ$, leads to the SM prediction \cite{Bai:2015nea}
\bea
{\rm Re}\, \bigg(\frac{\ep'}{\ep}\bigg)_{\rm SM}=(1.38\pm 5.15\pm 4.59)\Ex{-4} \simeq (1.4\pm 6.9)\Ex{-4}\, ,
\eea
where we combined the statistical and systematical errors in quadrature. 

As noticed in Ref. \cite{Cirigliano:2016yhc}, chiral symmetry  relates the contributions to $\ep'/\ep$
of the four-quark tree-level operators induced by $\xi_{ud}$ and $\xi_{us}$, given  in Eq. \eqref{LagTree}, 
to those of the electroweak penguin operators $\mathcal Q_7$ and $\mathcal Q_8$.
Such a determination  in principle still  suffers from  higher-order, $\Or(m_K\sq)$, corrections. Fortunately,  the $I=3/2$ parts of the LR operators, $O_{1\, LR}^{ud\, us}$ and $O_{2\, LR}^{ud\, us}$,  coincide after an isospin decomposition with those of $\mathcal Q_{7}$ and $\mathcal Q_8$, respectively. Isospin symmetry therefore implies a stronger relation between the contributions of the left-right operators to the $I=2$ amplitude and the matrix elements of  $\mathcal Q_{7,8}$ \cite{Chen:2008kt,Bertolini:2013noa}.
As this relation depends on isospin arguments, it is only subject to  $\Or((m_d-m_u)/\Lambda_\chi)$  and  
$\Or(\alpha/\pi)$  corrections, expected at the few-percent  level. The resulting expression for the $I=2$ amplitude is \cite{Cirigliano:2016yhc}
\bea
{\rm Im}\, A_2(\xi) &=&\frac{1}{6\sqrt{2}}{\rm Im}\,\bigg[\big(C_{1LR}^{ud\, us}-C_{1LR}^{us\, ud^*}\big)\mathcal \, \langle (\pi\pi)_{I=2}|\mathcal  Q_7|K^0\rangle
\nn\\ &&
+\big(C_{2LR}^{ud\, us}-C_{2LR}^{us\, ud^*}\big)\mathcal \, \langle (\pi\pi)_{I=2}|\mathcal  Q_8|K^0\rangle\bigg]\ ,\label{eq:A2expr}
\eea
where \cite{Blum:2015ywa,Blum:2012uk}
\begin{eqnarray}
\langle (\pi\pi)_{I=2}|\mathcal  Q_7|K^0\rangle
= (0.36\pm0.02) \, {\rm GeV}^2\ ,\qquad
\langle (\pi\pi)_{I=2}|\mathcal  Q_8|K^0\rangle
= (1.6\pm0.094) \, {\rm GeV}^2\ . 
\label{LEC}
\end{eqnarray}
Such a relation does not exist for the $I=0$ amplitude, however, at leading order in ChPT we obtain
$A_0(\xi) = -2\sqrt{2}A_2(\xi)$. We thus find 
\bea\label{finaleps'}
{\rm Re}\, \bigg(\frac{\ep'}{\ep}\bigg)={\rm Re}\, \bigg(\frac{\ep'}{\ep}\bigg)_{\rm SM}+{\rm Re}\,\bigg(\frac{i \omega e^{i(\delta_2-\delta_0)}}{\sqrt{2}\ep}\bigg)\bigg[\frac{{\rm Im}\,A_2(\xi)}{{\rm Re}\,A_2}-\frac{{\rm Im}\,A_0(\xi)}{{\rm Re}\,A_0}\bigg]\ ,
\eea
where we use the experimental values for ${\rm Re}\,A_{0,2}$. The expression for $A_0(\xi)$ might suffer from relatively large SU(3) corrections. However, it is the $A_2(\xi)$ term that constitutes the dominant $\xi$ contribution to $\ep'$, while the $A_0(\xi)$ term is suppressed by $2\sqrt{2}\omega\simeq 0.1$. We therefore expect Eq.~\eqref{finaleps'} to be accurate up to the lattice uncertainties in Eq.~\eqref{LEC}.

\subsection{$K_L \rightarrow \pi^0 e^+ e^-$}\label{KLpi0}
In the SM, the decay $K_L \rightarrow \pi^0 e^+ e^-$ has a large direct CPV component dominated by the penguin operators $C_{7V} \bar s \gamma^\mu d \, \bar e \gamma_\mu e $ and $C_{7A} \bar s \gamma^\mu d \, \bar e \gamma_\mu \gamma_5 e $ \cite{Buchalla:1995vs}.
In addition, there is a CP-even long-distance component dominated by two-photon exchange and an indirect CPV contribution proportional to the mixing parameter $\ep_K$.
Finally, in the presence of right-handed currents, this decay gets contributions from the dipole operators $C^{ds}_{\gamma d}$ and $C^{sd}_{\gamma d}$. Due to the large factors of $m_t/m_{s,d}$ and $m_c/m_{s,d}$ that appear in the matching coefficients
\eqref{Xi1loopMatch}, this $K_L$ decay is particularly sensitive to the imaginary part of the couplings $\xi_{td}$, $\xi_{ts}$ and, to a lesser extent, $\xi_{cd}$ and $\xi_{cs}$.

The decay rate can be expressed in terms of the vector and tensor form factors
\begin{eqnarray}
\langle \pi^0 | \bar s \gamma^\mu d | K_L \rangle &=&  \frac{1}{\sqrt{2}} f^{K^0\pi^+}_+(q^2) (p^\mu_K + p^\mu_\pi)\, , \nn \\
\langle \pi^0 | \bar s \sigma^{\mu\nu} d | K_L \rangle &=& i f^{K\pi}_{T}(q^2) \frac{\sqrt{2}}{m_K + m_\pi} (p^\mu_\pi p^\nu_K - p_K^\mu p_\pi^\nu)\, ,
 \end{eqnarray}
where $f^{K\pi}_+$ (see Table~\ref{LQCDinput}) is related to the vector form factor in $K^+ \rightarrow \pi^0 e^+ \nu$, while $f^{K\pi}_T$ has been computed on the lattice. 
We will use the evaluation of Ref. \cite{Baum:2011rm}, $f_{T}^{K\pi}= 0.417 \pm 0.015$, at a renormalization scale $\mu = 2$ GeV.

The RHCC contribution to the branching ratio is determined by the coupling $C_T$
\begin{equation}
C_T(\mu) = -\frac{Q_d}{4} \left( m_s C^{ds *}_{\gamma d}(\mu) + m_d C^{sd}_{\gamma d}(\mu) \right)\, ,
\end{equation}
where the values of the coefficients at $\mu = 2$ GeV are given in Table \ref{BdsCoeffs}.
The SM contribution is expressed by the functions $\tilde y_{7V}$ and $\tilde y_{7A}$ \cite{Buchalla:1995vs} given in Appendix \ref{sec:Bmumu}. 
The $\xi$ operators also contribute to the penguin operators $C_{7V}$ and $C_{7A}$, as discussed in Appendix \ref{sec:Bmumu}, but these
 contributions are quadratic in $\xi$ and not enhanced by $m_{t,c}/m_s$. We therefore do not include them in our analysis.

In terms of $\tilde y_{7V}$, $\tilde y_{7A}$, and $C_T$, the branching ratio becomes
\begin{eqnarray}\label{eq:KLpiee}
\textrm{Br}(K_L \rightarrow \pi^0 e^+ e^-) = \kappa_e \left[ \left( \textrm{Im} \lambda_t\, \tilde y_{7V} +  \frac{2}{m_K + m_\pi}  \frac{f_T^{K\pi}(0)}{f^{K\pi}_+(0)} 
16 \pi^2 \textrm{Im} (v^2 C_T)  \right)^2 + \textrm{Im} \lambda_t^2\, \tilde{y}^2_{7A}\right] \, ,
\end{eqnarray}
where $\lambda_t = V^*_{ts} V_{td}$.
The factor $\kappa_e$ is introduced to cancel the SM dependence on the vector form factor $f^{K\pi}_+$ by normalizing to the $K^+ \rightarrow \pi^0 e^+ \nu$ decay rate. $\kappa_e$ is defined as
\begin{eqnarray}
\kappa_e = \frac{1}{|V_{us} + \xi_{us}|^2} \frac{\tau(K_L)}{\tau(K^+)} \left( \frac{\alpha_{\textrm{em}}}{2\pi} \right)^2 \textrm{Br} (K^+ \rightarrow \pi^0 e^+ \nu) 
\sim \left(\frac{0.225}{|V_{us} + \xi_{us}|} \right)^2\, 6 \cdot 10^{-6}\, ,
\end{eqnarray}
where we used the experimental values in Table \ref{DS1exp}. The expression in Eq.\ \eqref{eq:KLpiee} involves only the direct CPV contributions from the SM. However, since the experimental limit is currently only sensitive to branching ratios that are roughly two orders of magnitude larger than the SM prediction \cite{Olive:2016xmw},
\bea
{\rm BR}(K_L\to \pi^0e^+e^-) < 2.8 \Ex{-10} \quad (90\% \, {\rm C.L.)}\,\, ,
\eea 
we simply use Eq.\ \eqref{eq:KLpiee} to estimate the branching ratio.

\section{$\Delta B =1$ and $\Delta B =2$ processes}\label{sec:DeltaB1}

\begin{table}
\center
\begin{tabular}{||c|c||c|c||}
\hline
$\text{BR}\,(B\to X_d\g)$ & $(14.1\pm 5.7)\cdot 10^{-6}$  & $\text{BR}\,(B\to X_s\g)$& $(3.32\pm 0.15)\times 10^{-4}$ \\
$A_{CP}(B\to X_{d+s}\g)$ & $0.032 \pm 0.034$ & $A_{CP}(B\to s\g)$ &  $0.015\pm 0.02$  \\
& & $S_{K^*\g}$ & $ -0.16\pm0.22$      \\
$\Delta m_d$ & $\left(0.5064 \pm 0.0019\right)  \, \textrm{ps}^{-1}$& $\Delta m_s$ &$\left(17.757 \pm 0.021\right)  \, \textrm{ps}^{-1}$ \\
$\Delta \Gamma^{(d)}$ & $ (-1.3\pm 6.7)\Ex{-3} \,{\rm ps}^{-1}$&  $\Delta \Gamma^{(s)} $ &  $(0.086\pm 0.006)\,{\rm ps}^{-1}$\\
$a_{\rm fs}^d $ & $ -0.0020 \pm 0.0016$& $ a_{\rm fs}^s$ & $ -0.0006 \pm 0.0028$ \\
\hline
\end{tabular}
\caption{Experimental input for the processes discussed in Section \ref{sec:DeltaB1}  \cite{Olive:2016xmw,Amhis:2016xyh}. 
The branching ratios $\text{BR}\,(B\to X_{d,s}\g)$ have a cut on the photon energy,  $E_\gamma > 1.6$ GeV.}
\label{DB1exp}
\end{table}

$\Delta B = 1$ FCNC processes such as $B \rightarrow X_{s,d}\, \gamma$ 
lead to very strong constraints on RHCC in the top sector. 
These processes are described by the effective Hamiltonian 
\begin{equation}\label{HamC7}
\mathcal H_{\textrm{eff}} = - \frac{4 G_F}{\sqrt{2}} V_{tb} V^*_{tq} \left[ C_7 \mathcal O_7 + C^\prime_7 \mathcal O^\prime_7 
+ C_8 \mathcal O_8 + C^\prime_8 \mathcal O^\prime_8 
\right]\, ,
\end{equation}
with 
\begin{eqnarray}
\mathcal O_7 &=& \frac{e}{(4\pi)^2} m_b \bar q_L \sigma^{\mu \nu} b_R\, F_{\mu \nu}\, , \qquad 
\mathcal O_8 = - \frac{g_s}{(4\pi)^2} m_b \bar q_L \sigma^{\mu \nu} G_{\mu \nu}^at^a b_R\, . \label{C7C8def}
\end{eqnarray}
$\mathcal O_{7,8}^\prime$ have analogous definitions with $L \leftrightarrow R$.
Relations between the coefficients in Eq. \eqref{HamC7} and the coefficients of the dipole operators in Eq. \eqref{eq:dipole1} are given by
\begin{eqnarray}\label{eq:C7}
C_7(m_W) &=&-\frac{4\pi\sq Q_d }{V_{tb}V_{tq}^*}\, v\sq C_{\g d}^{qb} \qquad C^\prime_7(m_W) = -\frac{4\pi\sq Q_d }{V_{tb}V_{tq}^*}\frac{m_q}{m_b}\, \big(v\sq C_{\g d}^{bq}\big)^* \, ,  \nn \\
C_8(m_W)& =&  \frac{4\pi\sq  }{V_{tb}V_{tq}^*}\, v\sq C_{g d}^{qb}   \qquad  C^\prime_8(m_W) = \frac{4\pi\sq  }{V_{tb}V_{tq}^*}\frac{m_q}{m_b}\, \big(v\sq C_{g d}^{bq}\big)^*\, .
\end{eqnarray}
The coefficients at the scales $\mu = \mu_b =2$ GeV are given in Table \ref{BdsCoeffs}.
From Eq. \eqref{Xi1loopMatch} we see that the contribution of $\xi_{tb}$ to $C_{7,8}$
and of $\xi_{ts}$ and $\xi_{td}$ to $C^\prime_{7,8}$ are enhanced by $m_t/m_b$ with respect to the SM, and therefore give rise to large effects in 
the $B \rightarrow X_{s,d} \gamma$ branching ratios. Information on the phases of the $\xi_{tb}$ and $\xi_{ts}$ elements can be gained by studying 
the CP asymmetries in inclusive  $B\to X_{d,s}\g$ decays, and in the exclusive channel $B\to K^{*0}\g$. We discuss the $B\to X_{d,s}\g$ branching ratios in Section \ref{BRbsg},
and the inclusive and exclusive CP asymmetries in Sections \ref{ACPinc} and \ref{ACPex}, respectively.

$B\to X_{d,s}\g$ is not very sensitive to RHCC in the $Wbc$ vertex. In Section \ref{BBar}, we therefore study the corrections 
from RHCC to $B_q - \bar B_q$ mixing with $q = d,s$. While the contributions to the mass differences $\Delta m_d$ and $\Delta m_s$ 
are either quadratic in $\xi$, or suppressed by $m_b/m_t$ with respect to the SM, corrections to the real and imaginary part of the width are more important and lead 
 to constraints on $\textrm{Im}\, \xi_{cb}$ that are comparable to those obtained from the tree-level processes discussed in Section \ref{sec:tree-level}.

The experimental input used in this Section is taken from Refs. \cite{Olive:2016xmw,Amhis:2016xyh} and is summarized in Table \ref{DB1exp}.

\subsection{The $B\to X_{d,s}\g$  branching ratio}\label{BRbsg}

For the $B \rightarrow X_{d,s}\gamma$ branching ratios, we employ the expressions derived in Ref.\ \cite{Hurth:2003dk} rescaled by the SM predictions of Ref.\ \cite{Misiak:2006zs,Misiak:2015xwa,Czakon:2015exa},
\bea
 \text{BR}\,(B\to X_q\g) &=& r_q\frac{\mathcal N}{100} \frac{\left|V_{tq}^* V_{tb}\right|^2}{|V_{cb}|\sq+{|\xi_{cb}|\sq}}\bigg[ a+a_{77}(|R_7|\sq+|R_7'|\sq)+a_7^r \,{\rm Re}\, R_7+a_7^i \,{\rm Im}\, R_7\nn\\&&+
a_{88}(|R_8|\sq+|R_8'|\sq)+a_8^r \,{\rm Re}\, R_8+a_8^i \,{\rm Im}\, R_8 +a_{\epsilon\epsilon}|\epsilon_q|\sq +a_\epsilon^r \,{\rm Re}\, \epsilon_q\nn\\
&&+a_\epsilon^i \,{\rm Im}\, \epsilon_q + a_{87}^r\,{\rm Re}\,(R_8 R_7^*+R_8' R_7^{\prime\,*})+a_{87}^i\,{\rm Im}\,(R_8 R_7^*+R_8' R_7^{\prime\,*})\nn\\
&&+a_{7\epsilon}^r\,{\rm Re}\,(R_7 \epsilon_q^*)+a_{7\epsilon}^i\,{\rm Im}\,(R_7 \epsilon_q^*)+a_{8\epsilon}^r\,{\rm Re}\,(R_8 \epsilon_q^*)+a_{8\epsilon}^i\,{\rm Im}\,(R_8 \epsilon_q^*)\bigg]\, ,
\label{BR}\eea
where $R_{7,8} = \frac{C_{7,8}(m_t)}{C_{7,8}^{\rm SM}(m_t)}$, $R_{7,8}' = \frac{C'_{7,8}(m_t)}{C_{7,8}^{\rm SM}(m_t)}$, $C_7^{\rm SM}(m_t) = -0.189$, and $C_8^{\rm SM}(m_t) = -0.095$. Furthermore, $\mathcal N =  2.567(1\pm 0.064)\cdot 10^{-3}$. $r_q$ is a factor that rescales the above expression to the SM predictions of Refs.\ \cite{Misiak:2006zs,Misiak:2015xwa,Czakon:2015exa}. It is given by $r_s = \frac{3.36}{3.61}$ and $r_d = \frac{1.73}{1.38}$. Finally, $\epsilon_q = \frac{V_{uq}^*V_{ub}}{V_{tq}^*V_{tb}}$ and  the numerical values of $a_{ij}$ can be found in Ref.\ \cite{Hurth:2003dk}. In our analysis, we applied the expressions relevant for the following cut on the photon energy $E_\gamma > 1.6 $ GeV.
For $B \rightarrow X_d\, \gamma$ this requires extrapolating the branching ratio quoted in Ref. \cite{Amhis:2016xyh}, as discussed in Ref. \cite{Misiak:2015xwa},

The branching ratios in Eq. \eqref{BR} should be compared with the current experimental world averages  \cite{Olive:2016xmw,Amhis:2016xyh}, which we give in Table \ref{DB1exp}.
To derive constraints we  follow  Refs.\ \cite{Altmannshofer:2012az,Altmannshofer:2011gn} and apply the relative uncertainties on the SM predictions $\sigma_d = \frac{0.22}{1.73}{\rm BR}(B\to X_d\g) $ $\sigma_s = \frac{0.23}{3.36}{\rm BR}(B\to X_s\g) $. These theoretical uncertainties are then added in quadrature to the experimental ones.

\subsection{The $B\to X_{d,s}\g$  CP asymmetry}\label{ACPinc}
The phase of $\xi_{tb}$ can be probed by the $B\to X_s\g$ CP asymmetry. We employ the expression derived in Ref.~\cite{Benzke:2010tq},
\bea
\frac{A_{CP}(B\to s\g)}{\pi}&\equiv & \frac{1}{\pi}\frac{\Gamma(\bar B\to X_s\g)-\Gamma(B\to X_{\bar s}\g)}{\Gamma(\bar B\to X_s\g)+\Gamma(B\to X_{\bar s}\g)} \nn\\&\approx & \bigg[\bigg(\frac{40}{81}-\frac{40}{9}\frac{\Lambda_c}{m_b}\bigg)\frac{\al_s}{\pi}+\frac{\Lambda_{17}^c}{m_b}\bigg]\text{Im}\,\frac{C_2}{C_7}-\bigg(\frac{4\al_s}{9\pi}+4\pi\al_s\frac{\Lambda_{78}}{3m_b}\bigg)\text{Im}\,\frac{C_8}{C_7}\nn\\
&&-\bigg(\frac{\Lambda_{17}^u-\Lambda_{17}^c}{m_b}+\frac{40}{9}\frac{\Lambda_c}{m_b}\frac{\al_s}{\pi}\bigg)\text{Im}\,\bigg( \ep_s \frac{C_2}{C_7}\bigg)\ ,
\eea 
where $C_2$ is the coefficient of the charged-current operator $\mathcal O^{cb\, cs}_{1\, LL}$, $C_2 = C^{cb\, cs}_{1\, LL}/(V_{cb} V^*_{cs})$, which, along with $C_{7,8}$, should be evaluated at the factorization scale $\mu_b\simeq 2$ GeV. We employ the following SM values for these coefficients  \cite{Benzke:2010tq},
\bea
C_2^{\rm SM}(2\, {\rm GeV}) = 1.204\, ,\qquad C_7^{\rm SM}(2\, {\rm GeV}) = -0.381\, ,\qquad C_8^{\rm SM}(2\, {\rm GeV}) = -0.175\, .\label{BcoeffSM}
\eea
In addition, the CP asymmetry depends on the scale, $\Lambda_c\simeq 0.38\, \text{GeV}$, and on three hadronic parameters that are estimated to lie in the following ranges \cite{Benzke:2010tq},
\bea
\Lambda_{17}^u\in [-0.33,\, 0.525]\, \text{GeV},\qquad \Lambda_{17}^c\in [-0.009,\, 0.011]\, \text{GeV},\qquad \Lambda_{78}\in [-0.017,\, 0.19]\, \text{GeV}.\label{lambdas}
\eea
We use the Rfit procedure to deal with these uncertainties \cite{Charles:2004jd}.

In the case of $B\to X_d\g$ decays, instead of the CP asymmetry $A_{CP}(B\to X_d\g)$, the combined asymmetry $A_{CP}(B\to X_{d+s}\g)$ is measured. This combination can be expressed as \cite{Hurth:2003dk}, 
\bea
A_{CP}(B\to X_{d+s}\g) = \frac{A_{CP}(B\to X_{s}\g)+R_{ds} A_{CP}(B\to X_{d}\g)}{1+R_{ds}}\, ,
\eea
with $R_{ds} = (\Gamma(B\to X_{d}\g)+\Gamma(\bar B\to X_{d}\g))/(\Gamma(B\to X_{s}\g)+\Gamma(\bar B\to X_{s}\g))$. Since the branching ratio of $B\to X_s\g$ is significantly larger than that of $B\to X_d\g$, $R_{ds}$ is expected to be at the percent level and $A_{CP}(B\to X_{d+s}\g)$ is therefore mainly sensitive to $A_{CP}(B\to s \g)$. In addition, the experimental precision on the determination of $A_{CP}(B\to X_{d+s}\g)$ is of the same order as $A_{CP}(B\to X_s\g)$, such that the latter does not provide any additional constraints.

\subsection{The $B\to K^{*0}\g$  CP asymmetry}\label{ACPex}
The time-dependent CP asymmetry in the exclusive decay 
 $B\to K^{*0}\g$ can be described by
\bea
\frac{\Gamma(\bar B\to \bar K^{*0}\g)-\Gamma(B\to K^{*0}\g)}{\Gamma(\bar B\to \bar K^{*0}\g)+\Gamma(B\to K^{*0}\g)}
=S_{K^*\g}  \cos(\Delta m_d t)+C_{K^*\g}  \sin(\Delta m_d t)\, .\eea
Here we are interested in the parameter $S_{K^*\g}$, which is given by  
\bea
S_{K^*\g}  = 2\frac{{\rm Im}\, \lambda_{K^{*}\g}}{1+|\lambda_{K^{*}\g}|\sq}, \qquad \lambda_{K^{*}\g} = \frac{q}{p} \frac{A( \bar B\to \bar K^{*0}\g)}{A(B\to  K^{*0}\g)}\, ,
\eea
where the ratio  $\frac{q}{p} =\frac{V_{tb}V_{td}^*}{V_{tb}^*V_{td}}$ arises from $B_d-\bar B_d$ mixing. 
At leading order the coefficient $S_{K^*\g}$ is generated by the electromagnetic dipole operators, $C_7$ and $C_7^\prime$. The dependence on $C_7'$ is particularly interesting as this Wilson coefficient is induced by right-handed currents, while being suppressed by $m_s/m_b$ in the SM.  In fact, the leading-order expression is \cite{Altmannshofer:2011gn,Paul:2016urs},
\bea
S_{K^*\g} = \frac{2\, {\rm Im}\bigg(\frac{V_{tb}V_{td}^*}{V_{tb}^*V_{td}} \frac{V_{tb}V_{ts}^*}{V_{tb}^*V_{ts} }\,C_7 C_7'\bigg)}{|C_7|\sq+|C_7'|\sq}\, ,
\eea 
such that $S_{K^*\g}$ vanishes unless $C'_7$ is nonzero. As mentioned above, the SM prediction is rather small \cite{Ball:2006cva,Ball:2006eu}
\bea
S_{K^*\g}^{\rm SM} = (-2.3\pm1.6)\Ex{-2}\, .
\eea
The  experimental value for $S_{K^*\g}$ is given in Table \ref{DB1exp}.

\subsection{$B_q-\bar B_q$ mixing}\label{BBar}
Right-handed currents can affect $B_q-\bar B_q$ oscillations through insertions of $\xi$ in $\Delta B=2$ box diagrams that govern this mixing. The contributions to the dispersive part of these amplitudes, $M_{12}$, are either quadratic in $\xi_{ij}$ or suppressed with respect to the SM by a factor of the external quark mass, which is at most $m_b\sq/m_t\sq$. We will therefore neglect the contributions to $M_{12}$ which are linear in $\xi$, as well as the dimension-eight effects discussed in Appendix \ref{sec:DeltaF2}. 
In contrast, the $\xi_{ij}$ contributions to the absorptive part of the mixing amplitude are not suppressed with respect to the SM, as both are proportional to $m_b\sq$. 
The largest contributions come from $\xi_{cb}$ and $\xi_{cq}$, for which we find
\bea
\Gamma_{12}^{(q)}(\xi) &=& -\frac{G_F\sq m_b\sq  m_{B_q}f_{B_q}\sq}{\pi}\sqrt{z} \left(\lambda^{(q)\,2}_c\big(\sqrt{1-4z}-(1-z)\sq\big)-\lambda^{(q)}_c\lambda^{(q)}_t (1-z)\sq\right)\times\nn\\
&& \bigg[\left(\left[\frac{2}{3}B_1-\frac{5}{6}B_2R \right]\frac{\xi_{cb}}{V_{cb}}+\frac{1}{3}B_5 R\frac{\xi_{cq}^*}{V_{cq}^*}\right) \eta_{11LL}\eta_{11LR} \nn\\
&&+\left(\left[\frac{2}{3}B_1+\frac{1}{6}B_3 R\right]\frac{\xi_{cb}}{V_{cb}}+B_4 R\frac{\xi_{cq}^*}{V_{cq}^*}\right)\big(\eta_{11LL}\eta_{21LR}+\eta_{21LL}\eta_{11LR} + 3 \,\eta_{21LL}\eta_{21LR}\big)\bigg]\, , \nn\\
\eea
where $z\equiv m_c\sq/m_b\sq$, $\lambda_i^{(q)} = V_{ib}V_{iq}^*$, and $R = m_{B_q}\sq/(m_b+m_q)\sq$. The $B_i$ are given in Appendix \ref{sec:DeltaF2} and represent the bag factors of the $\Delta B=2$ operators in Eq.\ \eqref{DeltaF2basis}. Finally, the $\eta$ factors originate from the RG evolution, between $m_W$ and $m_b$, of the four-fermion operators in Eq.\ \eqref{LagTree}. These factors  relate the four-fermion operators at different scales through $C_{i\, LL(LR)}(m_b)=\eta_{ij LL(LR)}C_{j \,LL(LR)}(m_W)$, and are determined by Eq.\ \eqref{eq:4q3}. Explicitly we have
\bea
\eta_{11\,LL} &=& \frac{1}{2}\big(\eta^{6/23}+\eta^{-12/23}\big),\qquad \eta_{11\,LR} = \eta^{3/23},\nn\\
\eta_{21\,LL} &=& \frac{1}{2}\big(\eta^{6/23}-\eta^{-12/23}\big),\qquad \eta_{21\,LR} = \frac{1}{3}\big(\eta^{-24/23}-\eta^{3/23}\big),
\eea
where $\eta = \al_s(m_W)/\al_s(m_b)$.

The real part of the right-handed contribution to $\Gamma_{12}$ can be constrained by the width difference between the mass eigenstates, whereas the imaginary parts are probed by the measure of CP violation, $a_{\rm fs}^q$ \cite{Buras:1997fb},
\bea
\Delta \Gamma^{(q)}=4\frac{{\rm Re}\, \big(\Gamma_{12}^{(q)*} M^{(q)}_{12}\big)}{\Delta m_{B_q}}\, ,\qquad a_{\rm fs}^q=1-\bigg|\frac{q}{p}\bigg|\sq = {\rm Im}\bigg(\frac{\Gamma_{12}^{(q)} }{M^{(q)}_{12}}\bigg)\, .
\eea
As mentioned above, the right-handed corrections to $M_{12}$ are small and we neglect them here, while the SM expression for $M_{12}$ can be found in Appendix \ref{sec:DeltaF2}.
The SM values for these quantities are given by \cite{Artuso:2015swg},
\bea
& \Delta \Gamma^{(d)}_{\rm SM} = (2.61\pm0.59)\Ex{-3} \,{\rm ps}^{-1}\, ,\qquad  \Delta \Gamma^{(d)}_{\rm SM} = (0.085\pm0.015) \,{\rm ps}^{-1} \, ,&\nn\\
&a_{\rm fs}^d\big|_{\rm SM} = (-4.7\pm 0.6)\Ex{-4}\, ,\qquad a_{\rm fs}^s\big|_{\rm SM} = (2.22\pm 0.27)\Ex{-5}\, ,&
\eea
while the current experimental determinations are given in Table \ref{DB1exp}.

\section{Single-coupling constraints}\label{Single}
In this section we discuss the constraints on the various right-handed couplings in the case that a single $\xi_{ij}$ element dominates at the scale of new physics. 
To obtain bounds we construct a $\chi\sq$ involving the observables described in Sections \ref{sec:tree-level} - \ref{sec:DeltaB1}. Furthermore, we assume that the CKM matrix is SM-like and apply the Wolfenstein parametrization to write the CKM matrix in terms of $A,\,\lambda,\,\bar\rho$, and $\bar\eta$, up to $\Or(\lambda^6)$ corrections \cite{Buras:1998raa} \footnote{The higher-order terms are mainly important for $\epsilon_K$, which we employ to constrain the CKM parameters. The imaginary part of the $V_{cs}^*V_{cd}$ term only appears after expanding $V$ to $\Or(\lambda^5)$. }. 
Since the standard extraction of the CKM elements can be modified by the inclusion of right-handed currents, we determine the SM CKM parameters along with the $\xi_{ij}$ from the $\chi\sq$. Thus, for each $\xi_{ij}$ we simultaneously fit for $A,\,\lambda,\,\bar\rho$, and $\bar\eta$ as well as the real and imaginary parts of $\xi_{ij}$.

Apart from the observables discussed in the sections above, we include $B \rightarrow J/\psi K$,  $B^0_q \rightarrow \mu^+ \mu^-$ and the $\Delta F = 2$ processes $\ep_K$, $\Delta m_d$, and $\Delta m_s$. As discussed in Appendix \ref{sec:Beta}, \ref{sec:Bmumu} and  \ref{sec:DeltaF2}, these processes do not get large corrections from the RHCC operators. However, we include these observables in our analysis as they provide an important role in determining the SM CKM parameters.

We do not include other non-leptonic $B$ decays, such as $B \rightarrow \pi \pi$. These processes are affected by right-handed currents at tree level, and 
a reliable estimate of the corrections requires non-perturbative information on the matrix elements of the four-quark operators $C_{1\, LR}$ and $C_{2\, LR}$, which, at the moment, is not available. As a result, even without taking into account $\xi$ contributions, we find wider ranges for $\bar\rho$ and $\bar\eta$  compared to Ref. \cite{Olive:2016xmw}, but we expect these differences to have small impact on the bounds on $\xi$.   

Finally, most of the observables in Sections \ref{sec:tree-level} - \ref{sec:DeltaB1} involve theory uncertainties, which we treat by adding them in quadrature to the experimental errors. However, there are several cases in which these uncertainties are large and allow for cancellations, notably in $\ep^\prime/\ep$, $\epsilon_K$,  $d_n$, $d_{\rm Hg}$, and $A_{CP}(b\to s\g)$. In these specific cases, where cancellations between different contributions can significantly affect the constraints, we treat the theoretical errors using the Rfit approach as defined in \cite{Charles:2004jd}. We vary the matrix elements within their allowed ranges and apply those values of the matrix elements that minimize the $\chi^2$. This procedure leads to conservative constraints as it allows for cancellations between different contributions.

Using the approach described above, we find the following  90\% C.L. constraints on the real and imaginary part of $\xi_{ij}$ 
\begin{eqnarray}
\textrm{Re}\, \xi_{ij} &\in& \left(
\begin{array}{ccc}
[-7.0,\,  1.6] \Ex{-4} & [-2.1,\, 0.05 ]  \Ex{-3} & [-1.4,\, 1.3 ] \Ex{-3} \\ 
\lbrack-1.0 ,\, 0.8] \Ex{-2} & [-4.2,\, 0.55] \Ex{-2} &   [0.1, \, 3.5]   \Ex{-3}\\
\lbrack-1.0,\, 1.0]  \Ex{-4} & [-2.1,\, 2.5  ]  \Ex{-4} &  [-1.4,\, 1.2] \Ex{-3} \\
\end{array}
\right)\, ,   \label{boundsreal} \\
\textrm{Im}\, \xi_{ij} &\in& \left(
\begin{array}{ccc}
[0.15,\, 3.4 ]\Ex{-6} & [ 0.5,\, 7.9 ] \Ex{-7} & [ -0.4,\, 0.7 ]\Ex{-3} \\ 
\lbrack -8.5,\, 7.2 ] \Ex{-6} & [ -5.7,\, 7.0 ] \Ex{-3} &   [ -1.5,\, 0.6 ] \Ex{-2}\\
\lbrack -4.2,\, 4.2 ]\Ex{-5} & [ -2.5,\, 1.9] \Ex{-4} &  [ -2.4,\, 2.3 ] \Ex{-3} \\
\end{array}
\right)\, ,  \label{boundsimag}
\end{eqnarray}
where  we stress that the bounds are obtained turning on one complex $\xi_{ij}$ element at a time.

For the CKM parameters we obtain the $90\%$ C.L.\ allowed ranges in case of the $\xi_{ud}$ fit 
\begin{eqnarray}\label{CKMfit}
\lambda \in [0.2232,\, 0.2255], \qquad A \in [0.787,\, 0.827], \qquad \bar\rho\in [0.060,\, 0.20], \, \qquad \bar\eta\in [0.33,\, 0.40].
\end{eqnarray}
These values are in agreement with those found in Ref. \cite{Olive:2016xmw}, although the constraints found here are generally weaker. As mentioned above, this is to be expected as the fit of Ref. \cite{Olive:2016xmw} includes several more observables than we take into account here.  This affects $\bar \rho$ the most, which is reflected by the observation that our allowed ranges are roughly twice as wide. 
In addition, at $68\%$ C.L.\ our upper limit for $\bar \eta$ extends roughly one standard deviation upwards  compared to Ref.\ \cite{Olive:2016xmw}, while our lower limit for $\lambda$ extends one standard deviation downwards. For most of the $\xi_{ij}$ couplings, the ranges in Eq.\ \eqref{CKMfit} are rather stable and do not vary significantly between the different $\xi_{ij}$. The upper and lower ranges of both $\lambda$ and $A$ vary by less than $1\%$ between the different fits, while $\bar \rho$ and $\bar\eta$ exhibit variations of up to a few percent. The exception occurs in the case of $\xi_{cb}$, where the upper ranges of $A$ and $\bar \rho$ widen by about $5\%$, while the allowed lower range for $\bar \eta$ is widened by roughly $5\%$.

While the $\chi^2$ function used to obtain Eqs.\ \eqref{boundsreal} and \eqref{boundsimag} includes all the observables described in Sections  \ref{sec:tree-level} - \ref{sec:DeltaB1}, the bounds on
most entries of the $\xi_{ij}$ elements are dominated by a smaller set of processes. Below we briefly describe which observables drive the constraints for each $\xi$ element.

\paragraph{$\boldsymbol{\xi_{ud}}$ and $\boldsymbol{\xi_{us}}$}\hfill
\\
\\
\indent An overview of the constraints on $\textrm{Re}\, \xi_{ud}$ is shown in blue in the left panel of Fig.\ \ref{fig:ud_us_chart}. These constraints are obtained by setting the CKM parameters to the values of Ref.~\cite{Olive:2016xmw} and assuming that only $\textrm{Re}\, \xi_{ud}$ is turned on. As these limits indicate, in the single coupling analysis, the best constraints on $\textrm{Re}\, \xi_{ud}$ come from semileptonic decays, in particular, superallowed $\beta$ decay.
Indeed, the unitarity of the CKM matrix $V$ and the absence of modifications of the $us$ element allow one to use leptonic and semileptonic kaon  decays to accurately determine $\lambda$, and then extract $\xi_{ud}$ 
from superallowed $\beta$ decays without relying on leptonic pion decays, which suffer $\sim 1\%$ percent theoretical uncertainty from the LQCD determination of the pion decay constant.  
As is shown in red in the left panel of Fig.\ \ref{fig:ud_us_chart}, $\textrm{Im}\, \xi_{ud}$ is constrained by the $D$ coefficient, $\ep^\prime/\ep$ and EDMs. The stronger constraint comes from $\ep^\prime/\ep$, followed closely by the neutron EDM.
The limit from the $D$ coefficient is two orders of magnitude weaker, $\textrm{Im}\, \xi_{ud} \in [-2.9, 5.1] \Ex{-4}$. This translates into a difference of one order of magnitude between the semileptonic and EDM constraints on $\Lambda$ in Fig.\ \ref{fig:ud_us_chart}. 

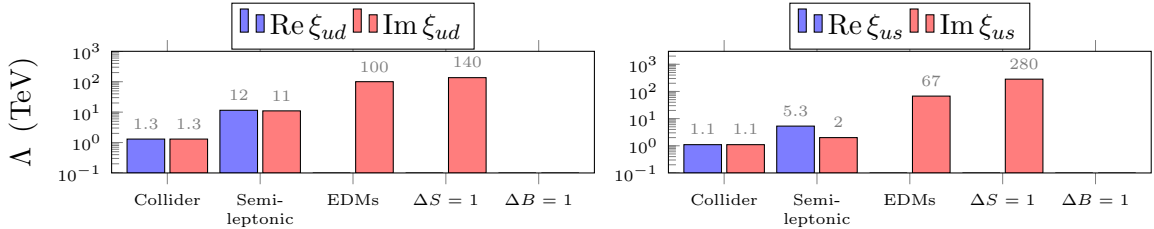
\begin{figure}[h!]
\pgfplotstableread[row sep=\\,col sep=&]{
    interval & carT & carD  \\
    Collider    & 1.3& 1.3 \\
    Semi-leptonic    & 11.5 & 11   \\
    EDMs    &.1 & 100.3  \\
   S & .1 &136.1 \\
    B& .1  & .1 \\
    }\mydata
\begin{tikzpicture}
    \begin{axis}[
            ybar,ymode=log,
           log origin=infty    ,         
            bar width=.5cm, 
            width=.5\textwidth,
            height=.20\textwidth,
            legend style={at={(0.5,1.40)},
                anchor=north,legend columns=-1},
           symbolic  x coords={Collider,Semi-leptonic,EDMs, S,B},
           enlarge x limits=.15,
            xtick=data,
            xticklabel style={align=center,text width=1cm, font=\tiny},
            xticklabels={Collider,Semi-leptonic,EDMs, $\Delta S=1$,$\Delta B=1$},
            tick label style={font=\tiny},
            ytick={0.1,1,10,100,1000},
            ylabel={$\Lambda\,$ (TeV)}, 
       nodes near coords/.append style={font=\tiny},                    
              ymin=.1,ymax=1000,
                          ytick pos=left,
            nodes near coords align={vertical},            
            point meta=rawy  ]			
        \addplot +[black,fill=blue, fill opacity=.51,nodes near coords={\ifnum\coordindex=2\else{\ifnum\coordindex=3\else{\ifnum\coordindex=4\else{\pgfmathprintnumber[fixed,fixed zerofill,fixed relative, precision=2]{\pgfplotspointmeta}}\fi}\fi}\fi}]  table[x=interval,y=carT]{\mydata};
        \addplot   +[black,fill=red, fill opacity=.51,nodes near coords={\ifnum\coordindex=4\else{\pgfmathprintnumber[fixed,fixed zerofill,fixed relative, precision=2]{\pgfplotspointmeta}}\fi}]  table[x=interval,y=carD]{\mydata};
       \legend{Re$\,\xi_{ud}$,Im$\,\xi_{ud}$}
    \end{axis}

\end{tikzpicture}
\pgfplotstableread[row sep=\\,col sep=&]{
    interval & carT & carD  \\
    Collider    & 1.1& 1.1 \\
    Semi-leptonic    & 5.3 & 2.0   \\
    EDMs    &.1 & 67.4  \\
   S & .1 &283.3 \\
    B& .1  & .1 \\
    }\mydata
\begin{tikzpicture}
    \begin{axis}[
            ybar,ymode=log,
             log origin=infty    ,    
            bar width=.5cm, 
            width=.5\textwidth,
            height=.20\textwidth,
            legend style={at={(0.5,1.40)},
                anchor=north,legend columns=-1},
           symbolic  x coords={Collider,Semi-leptonic,EDMs, S,B},
                                 enlarge x limits=.15,
            xtick=data,
            xticklabel style={align=center,text width=1cm, font=\tiny},
            xticklabels={Collider,Semi-leptonic,EDMs, $\Delta S=1$,$\Delta B=1$},
            tick label style={font=\tiny},
            ytick={0.1,1,10,100,1000},
       nodes near coords/.append style={font=\tiny},                    
              ymin=.1,ymax=3000,
                          ytick pos=left,
            nodes near coords align={vertical},            
            point meta=rawy  ]			
        \addplot +[black,fill=blue, fill opacity=.51,nodes near coords={\ifnum\coordindex=2\else{\ifnum\coordindex=2\else{\ifnum\coordindex=3\else{\ifnum\coordindex=4\else{\pgfmathprintnumber[fixed,fixed zerofill,fixed relative, precision=2]{\pgfplotspointmeta}}\fi}\fi}\fi }\fi}]  table[x=interval,y=carT]{\mydata};
        \addplot   +[black,fill=red, fill opacity=.51,nodes near coords={\ifnum\coordindex=4\else{\pgfmathprintnumber[fixed,fixed zerofill,fixed relative, precision=2]{\pgfplotspointmeta}}\fi}]  table[x=interval,y=carD]{\mydata};
       \legend{Re$\,\xi_{us}$,Im$\,\xi_{us}$}
    \end{axis}
\end{tikzpicture}
\caption{\small The figure shows naive constraints on the real (in blue) and imaginary (in red) parts of $\xi_{ud}$ and $\xi_{us}$ converted to an effective scale $\Lambda$ by using $\xi_{ij}\equiv v^2/\Lambda^2$. These limits are obtained by setting the CKM parameters to the values of Ref.~\cite{Olive:2016xmw}, and turning on only one real or imaginary part of $\xi_{ij}$ at a time. In the cases where the bounds are asymmetrical, we show the limit that results in the lowest value of $\Lambda$.
The different bars represent the limits from the collider, semileptonic, EDM ($\Delta F=0$), $\Delta S=1$, and $\Delta B=1$ observables, discussed in Sections \ref{sec:collider}, \ref{sec:tree-level}, \ref{sec:EDMs}, \ref{sec:DeltaS1}, and \ref{sec:DeltaB1}, respectively. }
\label{fig:ud_us_chart}
\end{figure}

The situation is very similar for $\xi_{us}$ as shown the right panel of Fig.\ \ref{fig:ud_us_chart}. Here the real part is constrained by leptonic and semileptonic kaon decays, while the imaginary part is constrained by EDMs and $\ep'/\ep$, 
with the latter giving again the stronger bound. In this case, the semileptonic constraint on the imaginary part is weaker due to the fact that the experimental limit on $D_\Sigma$ is significantly less stringent than that on $D_n$.
As noticed in Ref. \cite{Cirigliano:2016yhc}, an imaginary part of $\xi_{ud}$ or $\xi_{us}$ can solve the $2\sigma$ discrepancy between
the measured value of $\ep^\prime/\ep$ and the SM predictions of Ref.~\cite{Bai:2015nea,Buras:2015yba,Buras:2015xba,Kitahara:2016nld,Buras:2016fys}, without conflict with EDM and other low- and high-energy constraints. This manifests in the non-zero values for $\textrm{Im} \,\xi_{ud}$ and $\textrm{Im} \, \xi_{us}$ in Eq.\ \eqref{boundsimag}.

\paragraph{$\boldsymbol{\xi_{cd}}$ and $\boldsymbol{\xi_{cs}}$}\hfill
\\
\\
\indent As shown in the left (right) panel of Fig.\ \ref{fig:cd_cd_chart}, the real part of $\xi_{cd}$  ($\xi_{cs}$) is mainly constrained by leptonic and semileptonic $D$  ($D_s$) decays, while the collider limits are weaker by a factor of a few.
The larger theoretical and experimental errors  cause the  bounds from semileptonic decays to be less stringent than for $\xi_{ud}$ and $\xi_{us}$. 
The bound on $\textrm{Im} \,\xi_{cd}$ is dominated by the neutron EDM, while the small imaginary part of $V_{cd}$ gives rise to a (much weaker) EDM bound on Re$\, \xi_{cd}$ as well. Im$\,\xi_{cd}$ also contributes to $K_L \rightarrow \pi^0 e^+ e^-$, but the bound is three orders of magnitude weaker,  $|\textrm{Im}\, \xi_{cd} | < 2 \cdot 10^{-3}$. 
$\textrm{Im}\, \xi_{cs}$ mainly contributes to the nucleon EDM  by generating a strange quark EDM. 
As shown in Eq. \eqref{eq:nucleonEDM}, the matrix element linking the neutron EDM to the strange EDM is consistent with zero \cite{Bhattacharya:2015esa,Bhattacharya:2015wna},
which, in the Rfit approach, leads to no constraint on $\textrm{Im}\, \xi_{cs}$. The  bound in Eq. \eqref{boundsimag} therefore comes from $K_L \rightarrow \pi^0 e^+ e^-$.  

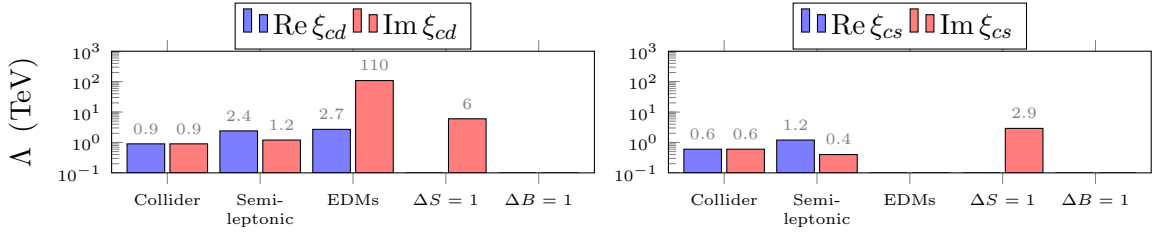
\begin{figure}[h!]
\pgfplotstableread[row sep=\\,col sep=&]{
    interval & carT & carD  \\
    Collider    & 0.9& 0.9 \\
    Semi-leptonic    & 2.4 & 1.2\\
    EDMs    &2.7 & 108.2  \\
   S & .1 &6.0 \\
    B& .1  & .1 \\
    }\mydata
\begin{tikzpicture}
    \begin{axis}[
            ybar,ymode=log,
           log origin=infty    ,         
            bar width=.5cm, 
            width=.5\textwidth,
            height=.20\textwidth,
            legend style={at={(0.5,1.40)},
                anchor=north,legend columns=-1},
           symbolic  x coords={Collider,Semi-leptonic,EDMs, S,B},
                                 enlarge x limits=.15,
            xtick=data,
            xticklabel style={align=center,text width=1cm, font=\tiny},
            xticklabels={Collider,Semi-leptonic,EDMs, $\Delta S=1$,$\Delta B=1$},
            tick label style={font=\tiny},
            ytick={0.1,1,10,100,1000},
            ylabel={$\Lambda\,$ (TeV)}, 
       nodes near coords/.append style={font=\tiny},                    
              ymin=.1,ymax=1000,
                          ytick pos=left,
            nodes near coords align={vertical},            
            point meta=rawy  ]			
        \addplot +[black,fill=blue, fill opacity=.51, nodes near coords={\ifnum\coordindex=3\else{\ifnum\coordindex=3\else{\ifnum\coordindex=3\else{\ifnum\coordindex=4\else{\pgfmathprintnumber[fixed,fixed zerofill,fixed relative, precision=2]{\pgfplotspointmeta}}\fi}\fi}\fi }\fi}]  table[x=interval,y=carT]{\mydata};
        \addplot   +[black,fill=red, fill opacity=.51,nodes near coords={\ifnum\coordindex=4\else{\ifnum\coordindex=4\else{\ifnum\coordindex=4\else{\pgfmathprintnumber[fixed,fixed zerofill,fixed relative, precision=2]{\pgfplotspointmeta}}\fi}\fi}\fi }]  table[x=interval,y=carD]{\mydata};
       \legend{Re$\,\xi_{cd}$,Im$\,\xi_{cd}$}
    \end{axis}

\end{tikzpicture}
\pgfplotstableread[row sep=\\,col sep=&]{
    interval & carT & carD  \\
    Collider    & 0.6& 0.6 \\
    Semi-leptonic    & 1.2 & 0.4   \\
    EDMs    &.1 & .1  \\
   S & .1 &2.9 \\
    B& .1  & .1\\
    }\mydata
\begin{tikzpicture}
    \begin{axis}[
            ybar,ymode=log,
             log origin=infty    ,    
            bar width=.5cm, 
            width=.5\textwidth,
            height=.20\textwidth,
            legend style={at={(0.5,1.40)},
                anchor=north,legend columns=-1},
           symbolic  x coords={Collider,Semi-leptonic,EDMs, S,B},
                                 enlarge x limits=.15,
            xtick=data,
            xticklabel style={align=center,text width=1cm, font=\tiny},
            xticklabels={Collider,Semi-leptonic,EDMs, $\Delta S=1$,$\Delta B=1$},
            tick label style={font=\tiny},
            ytick={0.1,1,10,100,1000},
       nodes near coords/.append style={font=\tiny},                    
              ymin=.1,ymax=1000,
                          ytick pos=left,
            nodes near coords align={vertical},            
            point meta=rawy  ]			
        \addplot +[black,fill=blue, fill opacity=.51,nodes near coords={\ifnum\coordindex=2\else{\ifnum\coordindex=2\else{\ifnum\coordindex=3\else{\ifnum\coordindex=4\else{\pgfmathprintnumber[fixed,fixed zerofill,fixed relative, precision=2]{\pgfplotspointmeta}}\fi}\fi}\fi }\fi}]  table[x=interval,y=carT]{\mydata};
        \addplot   +[black,fill=red, fill opacity=.51,nodes near coords={\ifnum\coordindex=2\else{\ifnum\coordindex=2\else{\ifnum\coordindex=2\else{\ifnum\coordindex=4\else{\pgfmathprintnumber[fixed,fixed zerofill,fixed relative, precision=2]{\pgfplotspointmeta}}\fi}\fi}\fi }\fi}]  table[x=interval,y=carD]{\mydata};
       \legend{Re$\,\xi_{cs}$,Im$\,\xi_{cs}$};
    \end{axis}
\end{tikzpicture}
\caption{\small The figure shows naive constraints on $\xi_{cd}$ and $\xi_{cs}$. Notation is the same as in Fig.\ \ref{fig:ud_us_chart}. }
\label{fig:cd_cd_chart}
\end{figure}

\paragraph{$\boldsymbol{\xi_{cb}}$ and $\boldsymbol{\xi_{ub}}$}\hfill
\\
\\
\indent 
As can be seen in Fig.\ \ref{fig:cb_ub_chart}, 
$\xi_{cb}$ and $\xi_{ub}$ are both constrained by the inclusive and exclusive semileptonic $B$ decays. 
Furthermore, the $B_q-\bar B_q $ oscillation observables, $\Delta \Gamma_q$ and $a_{\rm fs}^q$, constrain the real and imaginary parts of $\xi_{cb}$, while EDMs only constrain the imaginary part. Instead, for $\xi_{ub}$ both the real part and imaginary parts are constrained by $d_n$ (due to the sizable imaginary part of the relevant CKM element, $V_{ub}$), while the contributions to $B_q-\bar B_q $ mixing are negligible. 

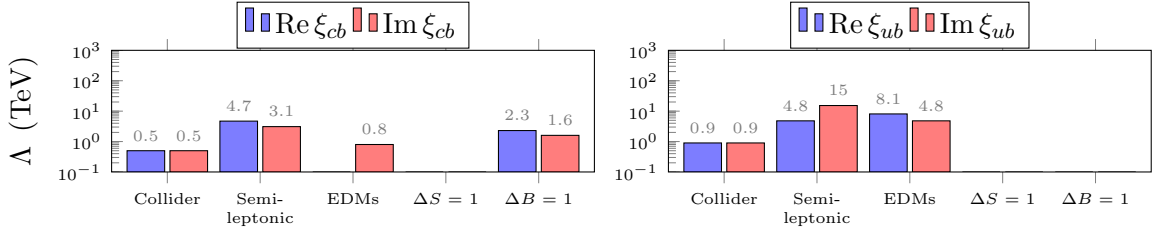
\begin{figure}[h!]
\pgfplotstableread[row sep=\\,col sep=&]{
    interval & carT & carD  \\
    Collider    & .5& .5 \\
    Semi-leptonic    & 4.7 & 3.1  \\
    EDMs    &.1 & 0.8\\
   S & .1 &.1\\
    B& 2.3  & 1.6 \\
    }\mydata
\begin{tikzpicture}
    \begin{axis}[
            ybar,ymode=log,
           log origin=infty    ,         
            bar width=.5cm, 
            width=.5\textwidth,
            height=.20\textwidth,
            legend style={at={(0.5,1.40)},
                anchor=north,legend columns=-1},
           symbolic  x coords={Collider,Semi-leptonic,EDMs, S,B},
                                 enlarge x limits=.15,
            xtick=data,
            xticklabel style={align=center,text width=1cm, font=\tiny},
            xticklabels={Collider,Semi-leptonic,EDMs, $\Delta S=1$,$\Delta B=1$},
            tick label style={font=\tiny},
            ytick={0.1,1,10,100,1000},
            ylabel={$\Lambda\,$ (TeV)}, 
       nodes near coords/.append style={font=\tiny},                    
              ymin=.1,ymax=1000,
                          ytick pos=left,
            nodes near coords align={vertical},            
            point meta=rawy  ]			
        \addplot +[black,fill=blue, fill opacity=.51, nodes near coords={\ifnum\coordindex=3\else{\ifnum\coordindex=2\else{\ifnum\coordindex=3\else{\ifnum\coordindex=3\else{\pgfmathprintnumber[fixed,fixed zerofill,fixed relative, precision=2]{\pgfplotspointmeta}}\fi}\fi}\fi }\fi}]  table[x=interval,y=carT]{\mydata};
        \addplot   +[black,fill=red, fill opacity=.51,nodes near coords={\ifnum\coordindex=3\else{\ifnum\coordindex=3\else{\ifnum\coordindex=3\else{\ifnum\coordindex=3\else{\pgfmathprintnumber[fixed,fixed zerofill,precision=1]{\pgfplotspointmeta}}\fi}\fi}\fi }\fi}]  table[x=interval,y=carD]{\mydata};
       \legend{Re$\,\xi_{cb}$,Im$\,\xi_{cb}$};
    \end{axis}

\end{tikzpicture}
\pgfplotstableread[row sep=\\,col sep=&]{
    interval & carT & carD  \\
    Collider    & 0.9& 0.9 \\
    Semi-leptonic    & 4.8 & 15.3\\
    EDMs    &8.1 & 4.8  \\
   S & .1 &.1 \\
    B& .1  & .1 \\
    }\mydata
\begin{tikzpicture}
    \begin{axis}[
            ybar,ymode=log,
             log origin=infty    ,    
            bar width=.5cm, 
            width=.5\textwidth,
            height=.20\textwidth,
            legend style={at={(0.5,1.40)},
                anchor=north,legend columns=-1},
           symbolic  x coords={Collider,Semi-leptonic,EDMs, S,B},
                                 enlarge x limits=.15,
            xtick=data,
            xticklabel style={align=center,text width=1cm, font=\tiny},
            xticklabels={Collider,Semi-leptonic,EDMs, $\Delta S=1$,$\Delta B=1$},
            tick label style={font=\tiny},
            ytick={0.1,1,10,100,1000},
                        ytick pos=left,
       nodes near coords/.append style={font=\tiny},                    
              ymin=.1,ymax=1000,
            nodes near coords align={vertical},            
            point meta=rawy  ]			
        \addplot +[black,fill=blue, fill opacity=.51,nodes near coords={\ifnum\coordindex=3\else{\ifnum\coordindex=3\else{\ifnum\coordindex=3\else{\ifnum\coordindex=4\else{\pgfmathprintnumber[fixed,fixed zerofill,fixed relative, precision=2]{\pgfplotspointmeta}}\fi}\fi}\fi }\fi}]  table[x=interval,y=carT]{\mydata};
        \addplot   +[black,fill=red, fill opacity=.51,nodes near coords={\ifnum\coordindex=4\else{\ifnum\coordindex=3\else{\ifnum\coordindex=4\else{\pgfmathprintnumber[fixed,fixed zerofill,fixed relative, precision=2]{\pgfplotspointmeta}}\fi}\fi}\fi }]  table[x=interval,y=carD]{\mydata};
       \legend{Re$\,\xi_{ub}$,Im$\,\xi_{ub}$};
    \end{axis}
\end{tikzpicture}
\caption{\small The figure shows naive constraints on $\xi_{cb}$ and $\xi_{ub}$. Notation is the same as in Fig.\ \ref{fig:ud_us_chart}. }
\label{fig:cb_ub_chart}
\end{figure}

For both ${ub}$ and ${cb}$ elements there is some tension between the determination via inclusive and exclusive decays. We find that adding a right-handed current improves the $\chi^2$. As shown in Eq.\ \eqref{boundsreal}, in the case of $cb$, our fit prefers a non-zero value of $\textrm{Re}\, \xi_{cb}$, while for the $ub$ element, both real and imaginary part are compatible with zero. The reason being that the nonzero values of $\xi_{ub}$ that are preferred by the semileptonic decays are disfavored by the neutron EDM.
We caution though that our analysis of the inclusive decays is incomplete, and, in particular, we did not repeat the fits to the lepton spectrum, which receive different contributions from left- and right-handed currents.
We notice that the bounds on $\xi_{ub}$ and $\textrm{Im}\, \xi_{cb}$ are rather weak when compared to the magnitudes of $V_{ub}$ and $V_{cb}$, and sizable right-handed corrections
(up to 50\% for $V_{ub}$ and 30\% for $V_{cb}$) are still allowed.

\paragraph{$\boldsymbol{\xi_{td}}$, $\boldsymbol{\xi_{ts}}$, and $\boldsymbol{\xi_{tb}}$}\hfill
\\
\\
\indent 
We collect the naive constraints on $\xi_{td}$, $\xi_{ts}$, and $\xi_{tb}$ in, respectively, the top-left, top-right, and bottom panels of Fig.\ \ref{fig:top_line_chart}.
The figure shows that all the top-row elements are strongly constrained by $\Delta B=1$ observables. In particular, $\xi_{td}$ is constrained by $B \rightarrow X_d \gamma$, while for $\xi_{ts,tb}$ stringent limits arise from $B \rightarrow X_s \gamma$. \footnote{It should be noted that, apart from the allowed range given in Eq.\ \eqref{boundsreal}, the flavor and low-energy observables allow for larger negative values of Re$\, \xi_{tb}$ namely, Re$\, \xi_{tb}\in [-0.034,\,-0.031]$. However, this possibility is excluded by LHC constraints on $h\to b\bar b$.} A comparable limit on Im$\, \xi_{tb}$ comes from EDMs, while for $\xi_{td}$  the EDM limits   are stronger than the $\Delta B=1$ constraints by an order of magnitude. Due to the imaginary part of $V_{td}$, EDMs constrain the real part of $\xi_{td}$ as well. 
In contrast, due to the poorly known matrix element related to $\tilde c_{\g d}^{(ss)}$, there are no EDM constraints on $\xi_{ts}$.
Finally, $\xi_{td}$ ($\xi_{ts}$) also contributes to $K_L \rightarrow \pi^0 e^+ e^-$, but the bounds are weaker by roughly a factor $10$ ($100$). For all $\xi_{tj}$ elements the indirect bounds are stronger than the direct collider bounds, by at least an order of magnitude. 
\begin{figure}[h!]
\pgfplotstableread[row sep=\\,col sep=&]{
    interval & carT & carD  \\
    Collider    & 0.7& .7 \\
    Semi-leptonic    & .1 & .1   \\
    EDMs    &82.7 & 129.7\\
   S & 1.0 &7.5\\
    B& 24.9  & 24.9 \\
    }\mydata
\begin{tikzpicture}
    \begin{axis}[
            ybar,ymode=log,
           log origin=infty    ,         
            bar width=.5cm, 
            width=.5\textwidth,
            height=.20\textwidth,
            legend style={at={(0.5,1.40)},
                anchor=north,legend columns=-1},
           symbolic  x coords={Collider,Semi-leptonic,EDMs, S,B},
                                 enlarge x limits=.15,
            xtick=data,
            xticklabel style={align=center,text width=1cm, font=\tiny},
            xticklabels={Collider,Semi-leptonic,EDMs, $\Delta S=1$,$\Delta B=1$},
            tick label style={font=\tiny},
            ytick={0.1,1,10,100,1000},
                        ytick pos=left,
            ylabel={$\Lambda\,$ (TeV)}, 
       nodes near coords/.append style={font=\tiny},                    
              ymin=.1,ymax=1000,
            nodes near coords align={vertical},            
            point meta=rawy  ]			
        \addplot +[black, fill=blue, fill opacity=.51,nodes near coords={\ifnum\coordindex=1\else{\ifnum\coordindex=1\else{\pgfmathprintnumber[fixed,fixed zerofill,fixed relative, precision=2]{\pgfplotspointmeta}}\fi}\fi}]  table[x=interval,y=carT]{\mydata};
        \addplot   +[black,fill=red, fill opacity=.51,nodes near coords={\ifnum\coordindex=1\else{\ifnum\coordindex=1\else{\pgfmathprintnumber[fixed,fixed zerofill,fixed relative, precision=2]{\pgfplotspointmeta}}\fi}\fi}]  table[x=interval,y=carD]{\mydata};
       \legend{Re$\,\xi_{td}$,Im$\,\xi_{td}$}
    \end{axis}

\end{tikzpicture}
\pgfplotstableread[row sep=\\,col sep=&]{
    interval & carT & carD  \\
    Collider    & .5& .5 \\
    Semi-leptonic    & .1 & .1   \\
    EDMs    &.1 & .1  \\
   S & 2.1 &3.4\\
    B& 27.9  & 29.2 \\
    }\mydata
\begin{tikzpicture}
    \begin{axis}[
            ybar,ymode=log,
             log origin=infty    ,    
            bar width=.5cm, 
            width=.5\textwidth,
            height=.20\textwidth,
            legend style={at={(0.5,1.40)},
                anchor=north,legend columns=-1},
           symbolic  x coords={Collider,Semi-leptonic,EDMs, S,B},
                                 enlarge x limits=.15,
            xtick=data,
            xticklabel style={align=center,text width=1cm, font=\tiny},
            xticklabels={Collider,Semi-leptonic,EDMs, $\Delta S=1$,$\Delta B=1$},
            tick label style={font=\tiny},
            ytick={0.1,1,10,100,1000},
                        ytick pos=left,
       nodes near coords/.append style={font=\tiny},                    
              ymin=.1,ymax=1000,
            nodes near coords align={vertical},            
            point meta=rawy  ]			
        \addplot +[black,fill=blue, fill opacity=.51,nodes near coords={\ifnum\coordindex=1\else{\ifnum\coordindex=2\else{\ifnum\coordindex=1\else{\pgfmathprintnumber[fixed,fixed zerofill,fixed relative, precision=2]{\pgfplotspointmeta}}\fi}\fi}\fi}]  table[x=interval,y=carT]{\mydata};
        \addplot   +[black,fill=red, fill opacity=.51,nodes near coords={\ifnum\coordindex=2\else{\ifnum\coordindex=1\else{\pgfmathprintnumber[fixed,fixed zerofill,fixed relative, precision=2]{\pgfplotspointmeta}}\fi}\fi}]  table[x=interval,y=carD]{\mydata};
       \legend{Re$\,\xi_{ts}$,Im$\,\xi_{ts}$}
    \end{axis}

\end{tikzpicture}

\pgfplotstableread[row sep=\\,col sep=&]{
    interval & carT & carD  \\
    Collider    & .7& .6 \\
    Semi-leptonic    & .1 & .1   \\
    EDMs    &.1 & 5.0  \\
   S & .1 &.1 \\
    B& 6.3  & 3.2 \\
    }\mydata
\begin{center}
\begin{tikzpicture}
    \begin{axis}[
            ybar,ymode=log,
             log origin=infty    ,    
            bar width=.5cm, 
            width=.5\textwidth,
            height=.20\textwidth,
            legend style={at={(0.5,1.40)},
                anchor=north,legend columns=-1},
           symbolic  x coords={Collider,Semi-leptonic,EDMs, S,B},
                      enlarge x limits=.15,
            xtick=data,
            xticklabel style={align=center,text width=1cm, font=\tiny},
            xticklabels={Collider,Semi-leptonic,EDMs, $\Delta S=1$,$\Delta B=1$},
            tick label style={font=\tiny},
            ytick={0.1,1,10,100,1000},
            ytick pos=left,
            ylabel={$\Lambda\,$ (TeV)}, 
       nodes near coords/.append style={font=\tiny},                    
              ymin=.1,ymax=1000,
            nodes near coords align={vertical},            
            point meta=rawy  ]			
        \addplot +[black,fill=blue, fill opacity=.51,nodes near coords={
        \ifnum\coordindex=1    
        \else{\ifnum\coordindex=2\else{\ifnum\coordindex=3\else{\ifnum\coordindex=3\else{\pgfmathprintnumber[fixed,fixed zerofill,fixed relative, precision=2]{\pgfplotspointmeta}}\fi}\fi}\fi }\fi}]  table[x=interval,y=carT]{\mydata};
        \addplot   +[black,fill=red, fill opacity=.51,nodes near coords={\ifnum\coordindex=1\else{\ifnum\coordindex=3\else{\ifnum\coordindex=3\else{\ifnum\coordindex=3\else{\pgfmathprintnumber[fixed,fixed zerofill,fixed relative, precision=2]{\pgfplotspointmeta}}\fi}\fi}\fi }\fi}]  table[x=interval,y=carD]{\mydata};
       \legend{Re$\,\xi_{tb}$,Im$\,\xi_{tb}$}
    \end{axis}
\end{tikzpicture}
\end{center}
\caption{\small The figure shows naive constraints on $\xi_{td}$, $\xi_{ts}$ and $\xi_{tb}$. Notation is the same as in Fig.\ \ref{fig:ud_us_chart}. }
\label{fig:top_line_chart}
\end{figure}
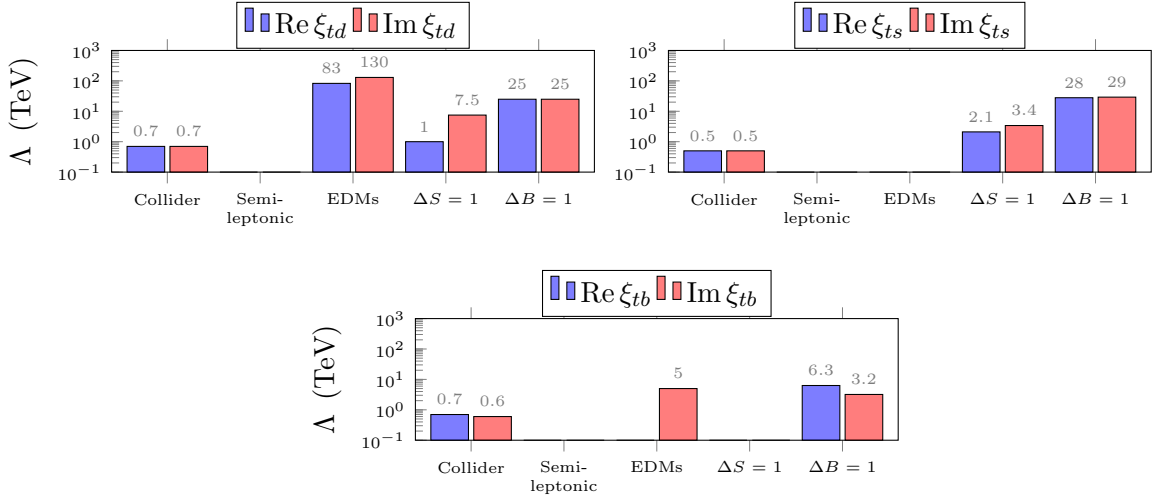

\subsection{Summary}
We summarize the strongest constraints on the real and imaginary parts in the left and right panels of Fig.\ \ref{fig:summary_chart}, respectively. The solid bars depict the constraints derived using the Rfit approach for the EDM uncertainties as outlined at the end of Section \ref{sec:EDMs}. Instead, the dashed bars indicate the `central' case, in which we set the theory errors in $d_n$ and $d_{\rm Hg}$ to zero. The difference in the strengths of the constraints  illustrates the impact of the hadronic and nuclear uncertainties. 

The real parts of $\xi_{td}$ and $\xi_{ts}$ are the most stringently constrained elements (by $b\to q \g$), while the weakest constraints are obtained in the case of $\xi_{cd,cs}$. In the latter case, the  precision of the semileptonic decays and the corresponding lattice input is at the percent level, thus allowing for couplings of order $\Or(10^{-2})$. The remaining real parts are constrained at the sub-percent level. Furthermore, as can be seen from the dashed bars, most of the real parts are  unaffected by the theory uncertainties related to EDMs. The exceptions are $\xi_{ub}$, $\xi_{cd}$, and $\xi_{td}$, which contribute to EDMs as their corresponding CKM elements have sizable imaginary parts. The main effect of neglecting the theoretical errors is that cancellations in the neutron and mercury EDMs are no longer possible.
As a result, the mercury EDM  provides the most stringent limit on the real parts of $\xi_{ub,cd,td}$ in the `central' case. Although the mercury EDM also constrains $\xi_{ts}$ in the `central' scenario it  does not overtake the $b\to s\g$ limits.

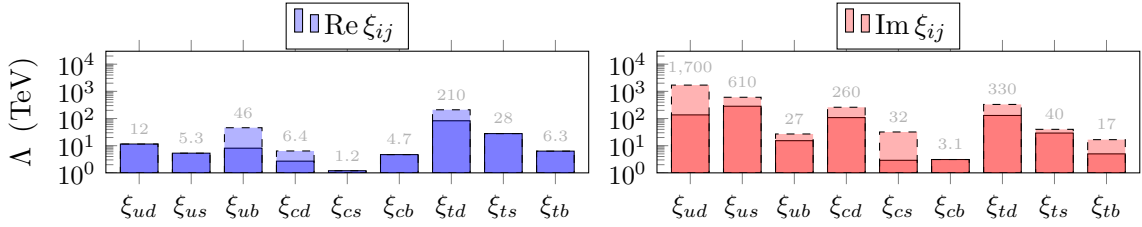
\begin{figure}[h!]
\pgfplotstableread[row sep=\\,col sep=&]{
    interval & carT & carD  \\
    ud    & 11.5& 11.5 \\
    us    & 5.3 & 5.3\\
    ub    &8.1 & 45.9\\
    cd &2.7 & 6.4\\
    cs & 1.2  & 1.2 \\
    cb & 4.7  & 4.7 \\
	td & 82.7 &210\\
    ts & 27.9  & 27.9\\
    tb & 6.3  & 6.3 \\    
    }\mydata
\begin{tikzpicture}
    \begin{axis}[
            ybar,ymode=log,
           log origin=infty    ,         
            bar width=.5cm, 
            bar shift=0pt,
            width=.5\textwidth,
            height=.20\textwidth,
            legend style={at={(0.5,1.40)},
                anchor=north,legend columns=-1},
           symbolic  x coords={ud,us,ub,cd,cs,cb,td,ts,tb},
                                 enlarge x limits=.08,
            xtick=data,
            xticklabels={$\xi_{ud}$,$\xi_{us}$,$\xi_{ub}$,$\xi_{cd}$,$\xi_{cs}$,$\xi_{cb}$,$\xi_{td}$,$\xi_{ts}$,$\xi_{tb}$},
            tick label style={font=\small},
            ytick={0.1,1,10,100,1000,10000},
            ylabel={$\Lambda\,$ (TeV)}, 
       nodes near coords/.append style={font=\tiny},                    
              ymin=1,ymax=30000,
                          ytick pos=left,
            nodes near coords align={vertical},            
            point meta=rawy  ]			
         \addplot[fill=blue, fill opacity=.3]  table[x=interval,y=carT]{\mydata};
        \addplot[nodes near coords={\pgfmathprintnumber[fixed,fixed zerofill,fixed relative, precision=2]{\pgfplotspointmeta}}, dashed, fill=blue, fill opacity=.3]     table[x=interval,y=carD]{\mydata};
       \legend{Re$\,\xi_{ij}$}
    \end{axis}
\end{tikzpicture}
\pgfplotstableread[row sep=\\,col sep=&]{
    interval & carT & carD  \\
    ud    & 136.1 &1700\\
    us    &  283.3 & 606\\
    ub    &15.3  & 27\\
    cd &108.2 & 260\\
    cs & 2.9 & 32\\
    cb &  3.1 & 3.1\\
	td & 129.7 &330 \\
    ts &  29.2 &40\\
    tb &  5.0 & 16.7\\    
    }\mydata
\begin{tikzpicture}
    \begin{axis}[
            ybar,ymode=log,
           log origin=infty    ,         
            bar width=.5cm, 
            bar shift=0pt,
            width=.5\textwidth,
            height=.20\textwidth,
            legend style={at={(0.5,1.40)},
                anchor=north,legend columns=-1},
           symbolic  x coords={ud,us,ub,cd,cs,cb,td,ts,tb},
                                 enlarge x limits=.08,
            xtick=data,
            xticklabels={$\xi_{ud}$,$\xi_{us}$,$\xi_{ub}$,$\xi_{cd}$,$\xi_{cs}$,$\xi_{cb}$,$\xi_{td}$,$\xi_{ts}$,$\xi_{tb}$},
            tick label style={font=\small},
            ytick={0.1,1,10,100,1000,10000}, 
       nodes near coords/.append style={font=\tiny},                    
              ymin=1,ymax=30000,
                          ytick pos=left,
            nodes near coords align={vertical},            
            point meta=rawy  ]			
        \addplot[fill=red, fill opacity=.3]  table[x=interval,y=carT]{\mydata};
        \addplot[nodes near coords={\pgfmathprintnumber[fixed,fixed zerofill,fixed relative, precision=2]{\pgfplotspointmeta}}, dashed, fill=red, fill opacity=.3]     table[x=interval,y=carD]{\mydata};
       \legend{Im$\,\xi_{ij}$}
    \end{axis}
\end{tikzpicture}

\pgfplotstableread[row sep=\\,col sep=&]{
    interval & carT & carD  \\
    ud    & 11.5& 136.1 \\
    us    & 5.3 & 283.3\\
    ub    &8.1 & 15.3  \\
    cd &2.7 & 108.2  \\
    cs & 1.2  & 2.9 \\
    cb & 4.7  & 3.1 \\
	td & 82.7 &129.7 \\
    ts & 27.9  & 29.2 \\
    tb & 5.0  & 5.0 \\    
    }\mydata
\caption{\small The figure summarizes the most stringent naive constraints on $\xi_{ij}$. The dashed bars show the naive constraints in the case that we do not take into account the theory errors that  appear in the EDM expressions Eqs.\  \eqref{eq:nucleonEDM} and \eqref{eq:NuclEDM}.
Notation is the same as in Fig.\ \ref{fig:ud_us_chart}.}
\label{fig:summary_chart}
\end{figure}

Moving on to the imaginary parts, one can compare the left and right panels of Fig.\ \ref{fig:summary_chart} to see that, even when using the Rfit approach, the limits on the imaginary parts are generally better than those on the real parts. All constraints are well below the percent level, apart from those on $ \xi_{cb}$ and $\xi_{cs}$. The weak bounds on the $cb$ and $cs$ elements result partially due to suppressed contributions to $d_n$: the $\xi_{cs}$ contribution depends on the poorly known strange-EDM matrix element, while the $\xi_{cb}$ contributions only arise at the two-loop level. 
Among the stronger constraints are those on $\xi_{td,ts}$ (from  $b\to q\g$ and EDMs), however the most impressive limits are set on $\xi_{ud}$ and $\xi_{us}$ and arise from the neutron EDM and $\ep'/\ep$ which probe effective scales around $\Or(100\, {\rm TeV})$. 

As seen from the dashed bars, most constraints on the imaginary parts are at least somewhat affected when moving from the Rfit approach to the `central' case.  For most couplings this results in an improvement of the EDM limit by a factor of a few to $\Or(10)$.
More drastic changes occur for $\xi_{cs,ts}$, $\xi_{ud}$, and $\xi_{us}$. 
In case of $\xi_{cs}$ this is due to the poorly known strange-EDM matrix element resulting in a vanishing EDM constraint in the Rfit approach, whereas the neutron EDM strongly constraints $\mathrm{Im}\,\xi_{cs}$ in the central case.
The situation is similar for $\xi_{ts}$, although less clear from Fig.\ \ref{fig:summary_chart} as the improved EDM limits do not overtake the $b\to s\g$ constraints. For $\xi_{ud}$ and $\xi_{us}$ the EDM limits improve by factors of $\Or(300)$ and $\Or(100)$, respectively. These large factors arise due to the fact that the four-quark operators $O_{i\, LR}^{udud}$ and $O_{i\, LR}^{usus}$  induce a large pion-nucleon coupling $\bar g_1$, to which the mercury EDM has an increased sensitivity compared to $d_n$. The resulting limits then overtake the $\ep'/\ep$ constraints and naively reach scales up to $\Or(10^3)\, {\rm TeV}$.

As the above discussion shows, in some cases the uncertainties related to the matrix elements can mean the difference between a stringent limit or no bound at all. For example, although the neutron and mercury EDMs are in principle sensitive to the right-handed couplings to strange quarks, the poor knowledge of the strange matrix elements  does not allow us to set EDM bounds on $\xi_{cs}$ and $\xi_{ts}$.
 This also holds true for the nuclear matrix elements related to the pion-nucleon couplings $\bar g_{0,1}$. These matrix elements allow the mercury EDM to vanish for all $\xi_{ij}$ elements, even though the `central' limits on $\xi_{ud}$ and $\xi_{us}$ show that $d_{\rm Hg}$ could  potentially probe scales up to $10^3\, {\rm TeV}$. These observations are similar to those discussed in Ref.\ \cite{Chien:2015xha} in the context of CPV Higgs-quark interactions, and further motivate studies of hadronic and nuclear matrix
elements with lattice QCD and modern nuclear many-body methods. 

\begin{figure}
\center
\includegraphics[width=6cm]{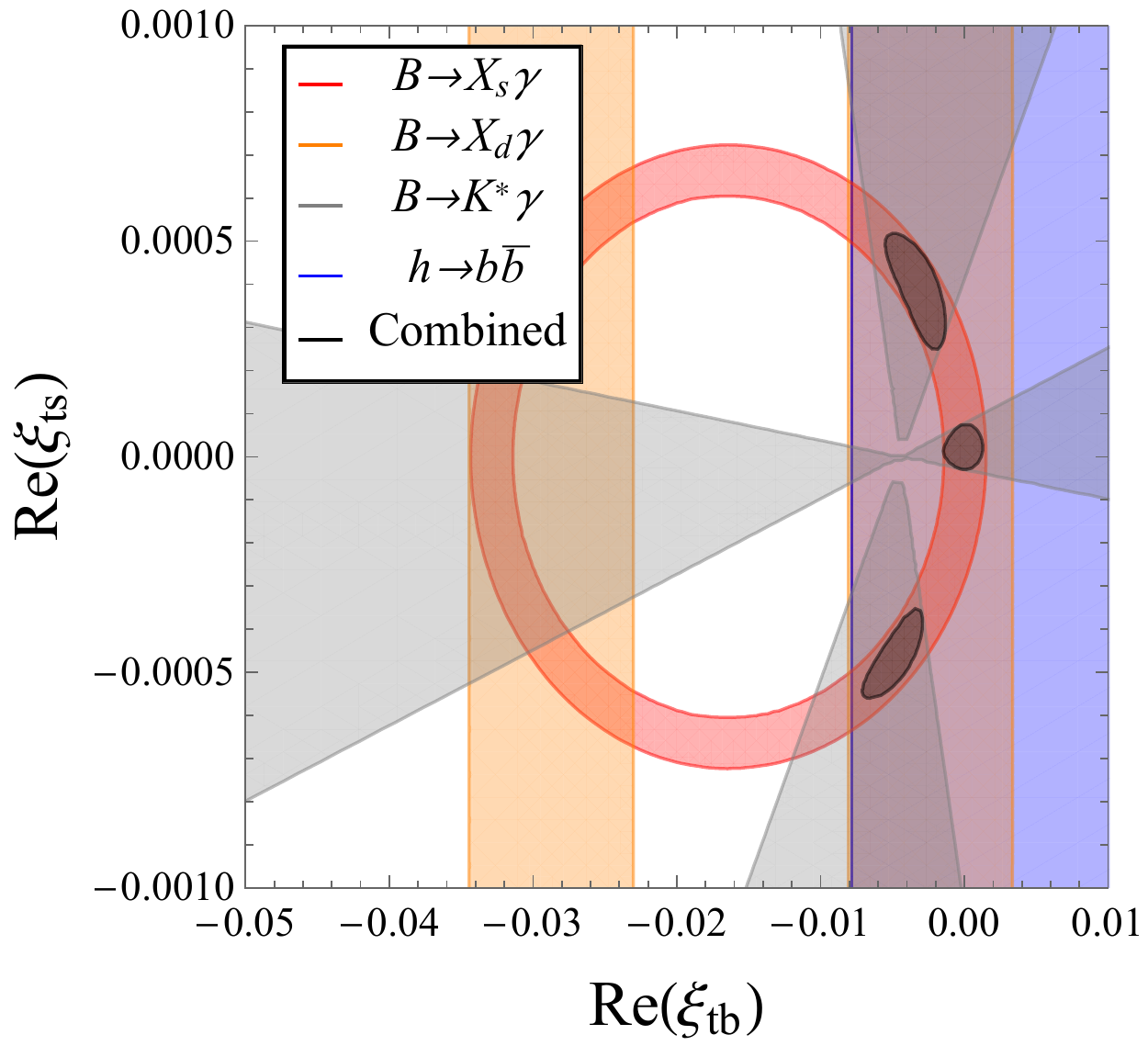}\qquad
\includegraphics[width=6cm]{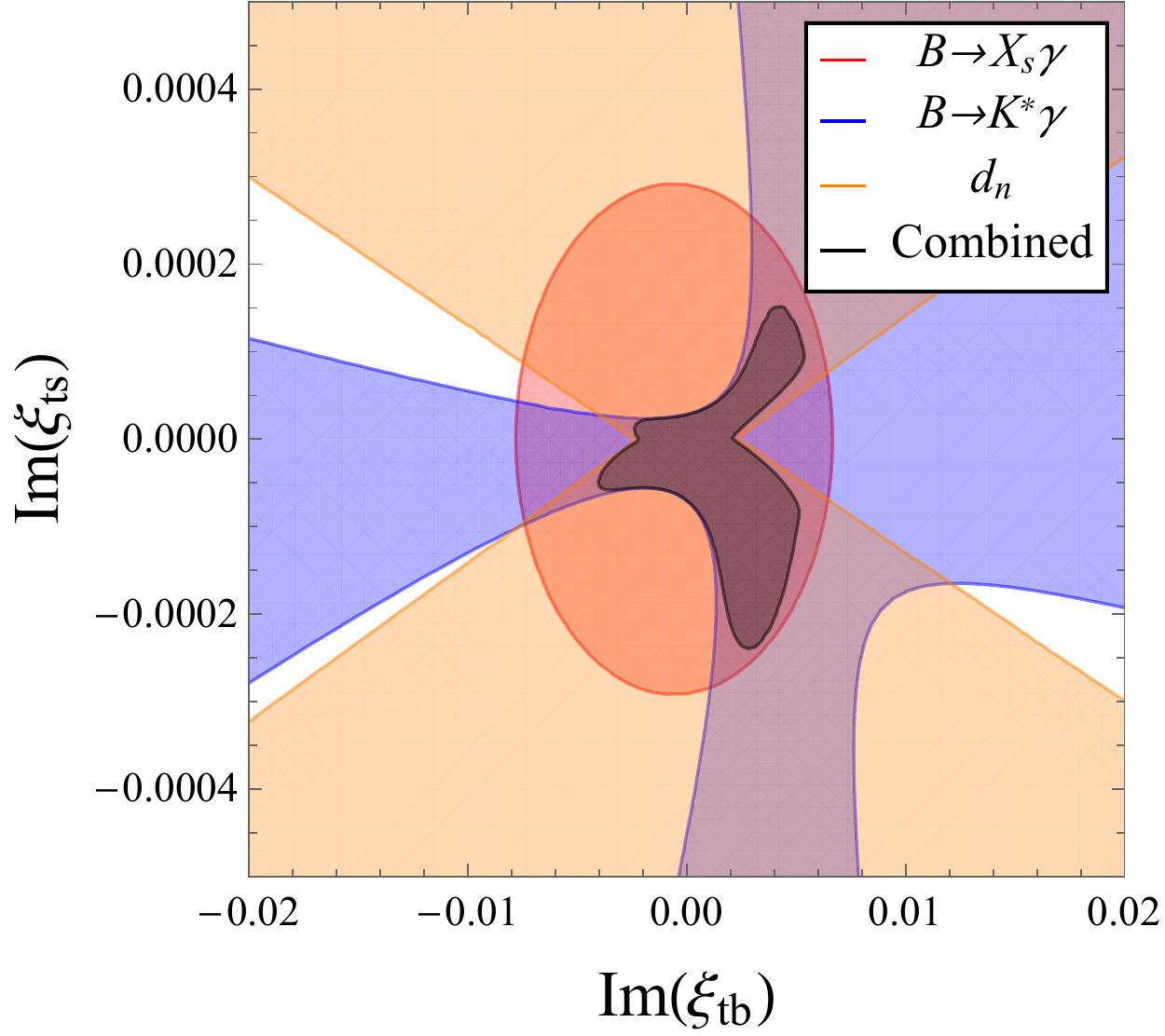}
\caption{\small The left (right) panel shows the constraints on the real (imaginary) parts of $\xi_{tb}$ and $\xi_{ts}$. These limits assume that only the real or imaginary parts of $\xi_{tb}$ and $\xi_{ts}$ are generated at the scale $\Lambda$. }
\label{Fig:tbVSts}\end{figure}

Finally, it is interesting to see whether the limits given in Eq.\ \eqref{boundsreal} and \eqref{boundsimag}  are stable against turning on several $\xi_{ij}$ couplings at the same time. Here we do not commit to a global analysis involving all the $\xi_{ij}$ couplings. Instead, as an example, we briefly discuss the resulting limits when turning on one row of the $\xi_{ij}$ matrix at a time. In this scenario the real parts of the first two rows remain largely unaffected \footnote{The only exception is $\xi_{ub}$ for which nonzero values, $|{\rm Re}\,\xi_{ub}|\approx 2\Ex{-3}$, are preferred due to the discrepancy between inclusive and exclusive semileptonic decays. Given that our determination of the inclusive decay is not entirely consistent, see Section \ref{sec:tree-level}, these nonzero values should not be taken too seriously.
The reason that $\xi_{ub}$ is consistent with zero in the single-coupling analysis, Eq.\ \eqref{boundsreal}, is that nonzero values are disfavored by the neutron EDM. When  several couplings can be nonzero at once, however, the neutron EDM limit can be canceled by Im$\,\xi_{ud}$ and Im$\,\xi_{us}$ contributions.}.

The imaginary parts are more sensitive to the effect of turning on additional couplings, as this allows for cancellations in $d_n$ and $\ep'/\ep$. As a result, the limit on the imaginary part of $\xi_{ud}$  is now determined by the $D_n$ coefficient, giving constraints at the $\Or(10^{-4})$ level. In turn, this leaves room for an imaginary part of $\xi_{us}$ up to $\Or(10^{-3})$. The limit on Im$\,\xi_{cd}$ is weakened by roughly two orders of magnitude due to similar cancellations in $d_n$. Instead, the constraints on the imaginary parts of $\xi_{ub,cb}$ and $\xi_{cs}$ do not change much since they are mainly determined by the semileptonic decays and $K_L\to \pi^0e^+e^-$, respectively. 

For the third row, the limits on both the real and imaginary parts are weakened by factors of a few for $\xi_{td,ts}$, while those on $\xi_{tb}$ deteriorate by an order of magnitude. 
As an example of the interplay between the different $\xi_{tq}$ elements, we show the $\xi_{tb}-\xi_{ts}$ plane for the real (imaginary) parts in the left (right) panel of Fig.\ \ref{Fig:tbVSts}. The constraints shown in the left (right) panel assume that only the real (imaginary) parts $\xi_{tb}$ and $ \xi_{ts}$ are present at the scale $\Lambda$.

\section{Identifying right-handed currents at low and high energy}\label{Disentangling}

In Section \ref{Single} we showed that, under the assumption that the SM is modified dominantly  by a RHCC at high energy, 
low-energy bounds from leptonic and semileptonic charged-current decays, $B \rightarrow X_{s,d} \gamma$, $\ep^\prime/\ep$, and  EDMs  are significantly stronger than 
collider bounds. On the other hand, in explicit models of new physics 
the low-energy observables unavoidably involve some degeneracy \cite{Jiang:2016czg}. In this section we therefore study more general scenarios in which other operators apart from a RHCC
are induced at high energy and how to unambiguously identify RHCCs both at high and low energy.  
In Sections \ref{DisLow} and \ref{DisHigh} we focus on the couplings of the $W$ to $ud$ and $us$ quarks,  
while in Section \ref{sec:topanomalous} we examine the $Wtb$ coupling.

\subsection{Low-energy probes}\label{DisLow}

In Section~\ref{Single} we used superallowed $\beta$ decays, and leptonic and semileptonic pion and kaon decays to put stringent bounds on RHCC involving the $u$ and 
$d$, and the $u$ and $s$ quarks. 
The processes we used to constrain $\textrm{Re}\, \xi_{ud}$ and $\textrm{Re}\, \xi_{us}$ are, however, sensitive not only to RHCC, but can be affected 
by additional  contributions. These can be studied  by considering the most general semileptonic dimension-six Lagrangian at low energy   \cite{Cirigliano:2009wk,Cirigliano:2012ab,Cirigliano:2013xha}
\begin{eqnarray}\label{LagTreeLOW}
\mathcal L &=& -\frac{4 G_F}{\sqrt{2}} V_{ud}  \Bigg [   \left( 1 + \delta V_{ud} + (\varepsilon_L)_{ud} \right) \,  \bar u \gamma^\mu P_L d \, \bar l \gamma_\mu P_L \nu  + 
\frac{\xi_{ud}}{V_{ud}}\,  \bar u \gamma^\mu P_R d \, \bar l \gamma_\mu P_L \nu  \nn \\ 
& & + \frac{1}{2}(\varepsilon_S)_{ud}\,  \bar u d \, \bar l  P_L \nu  -\frac{1}{2}(\varepsilon_P)_{ud}\,  \bar u \gamma_5 d \, \bar l  P_L \nu  +
(\varepsilon_T)_{ud}\,  \bar u \sigma^{\mu \nu} P_L d \, \bar l \sigma_{\mu \nu} P_L \nu  + \textrm{h.c.}\Bigg ]\, ,
\end{eqnarray}
and analogous contributions for the $us$ couplings. In Eq.~\eqref{LagTreeLOW}, we separated the contribution to left-handed currents coming from  corrections to the $W$ couplings to left-handed quark 
or leptons, $\delta V_{ud}$, from a semileptonic four-fermion operator, $\varepsilon_L$. While these operators are degenerate at low energy, they have different manifestations at collider experiments \cite{Cirigliano:2009wk,Cirigliano:2012ab,Cirigliano:2013xha}. The operators in Eq. \eqref{LagTreeLOW} are in direct correspondence with gauge-invariant operators in the basis of Ref. \cite{Grzadkowski:2010es},
and the mapping is discussed  in Refs. \cite{Cirigliano:2009wk,Cirigliano:2012ab}. 
Semileptonic operators arising from vertex corrections, $\delta V_{ud}$ and $\xi_{ud}$, are automatically lepton-flavor universal. For the four-fermion operators, 
$\varepsilon_{L,P,S,T}$, we assumed the couplings to be diagonal in lepton flavor.

The operators in Eq. \eqref{LagTreeLOW} affect all the observables introduced in Section \ref{ud&us}, which we used to bound $\xi_{ud}$ and $\xi_{us}$ (see Ref. \cite{Gonzalez-Alonso:2016etj} for a comprehensive analysis).
For example, superallowed $\beta$ decays receive corrections from the scalar coupling $\varepsilon_S$, which shifts $V_{ud}$ 
into \cite{Bhattacharya:2011qm,Cirigliano:2012ab,Cirigliano:2013xha}
\begin{equation}
|V_{ud}(0^+ \rightarrow 0^+)  |_{\textrm{exp}}
 = | V_{ud} |  \left| 1 + \delta V_{ud} + (\varepsilon_L)_{ud}  + \frac{\xi_{ud}}{V_{ud}}  + \frac{g_S}{2}c^S_{0^+}(Z)    (\varepsilon_S)_{ud} \right|\, ,
\end{equation}
where $g_S$ is the nucleon matrix element of the scalar current, and $c^S_{0^+}(Z)$ is a function which depends on the individual nuclear transition.
The expression for $\pi^\pm \rightarrow \mu^\pm \nu_\mu$ is modified into
\begin{equation}
|V_{ud}(\pi \rightarrow \mu \nu) f_{\pi} |_{\textrm{exp}}  = 
\left| V_{ud} \right| \left| 1 + \delta V_{ud} + (\varepsilon_L)_{ud} - \frac{\xi_{ud}}{V_{ud}}  -  (\varepsilon_P)_{ud} \frac{m^2_\pi}{m_\mu (m_d + m_u)}  \right| f_\pi\, ,
\end{equation}
and, similarly, the ratio of pion and kaon decays
\begin{equation}
\left(\left|\frac{V_{us}}{V_{ud}} \right| \frac{f_{K}}{f_{\pi}} \right)_{\textrm{exp}}= 
\frac{  \left| 1 + \delta V_{us} + (\varepsilon_L)_{us} - \frac{\xi_{us}}{V_{us}}  -  (\varepsilon_P)_{us} \frac{m^2_K}{m_\mu (m_s + m_u)}\right| \left| V_{us} \right| f_K}{\left| 1 + \delta V_{ud} + (\varepsilon_L)_{ud} - \frac{\xi_{ud}}{V_{ud}}  -  (\varepsilon_P)_{ud} \frac{m^2_\pi}{m_\mu (m_d + m_u)}  \right| \left| V_{ud} \right| f_\pi}\, .
\end{equation}
Analogously, the semileptonic decay $K^0 \rightarrow \pi^+ l \nu_l$ receives contributions from $\delta V_{us}$, $(\varepsilon_S)_{us}$, and $(\varepsilon_T)_{us}$. 

The difficulty in identifying a right-handed current at low energies can be illustrated by looking at the degeneracy with anomalous left-handed currents. 
By setting the four-fermion couplings to zero but allowing for nonzero values of  $\delta V_{ud}$ and $\delta V_{us}$, we obtain significantly weaker constraints  on, for example, $\xi_{ud}$
\begin{equation}
\textrm{Re} \, \xi_{ud} \in \left[-1.0, 0.7 \right] \cdot 10^{-2}\, , 
\end{equation}
which is in reach of future collider searches. 
The bound is determined by the theoretical uncertainty of $f_{\pi}$, which is at the percent level \cite{Aoki:2016frl}.
If we only allow BSM effects in the left- and right-handed currents, $\xi_{ud}$ and $\delta V_{ud}$ are completely anticorrelated, since the vector combination $|V_{ud} + \xi_{ud}|$
has to satisfy the stringent constraints from superallowed $\beta$ decays. Introducing additional operators further weakens the bounds on $\xi_{ud}$ and $\xi_{us}$ \cite{Gonzalez-Alonso:2016etj}. Still,  
 in order to not disrupt the agreement between the SM and the data for leptonic and semileptonic decays, strong correlations between the operators in Eq. \eqref{LagTreeLOW} must exist, posing non-trivial constraints on models of new physics.

\begin{figure}
\center
\includegraphics[width=6cm]{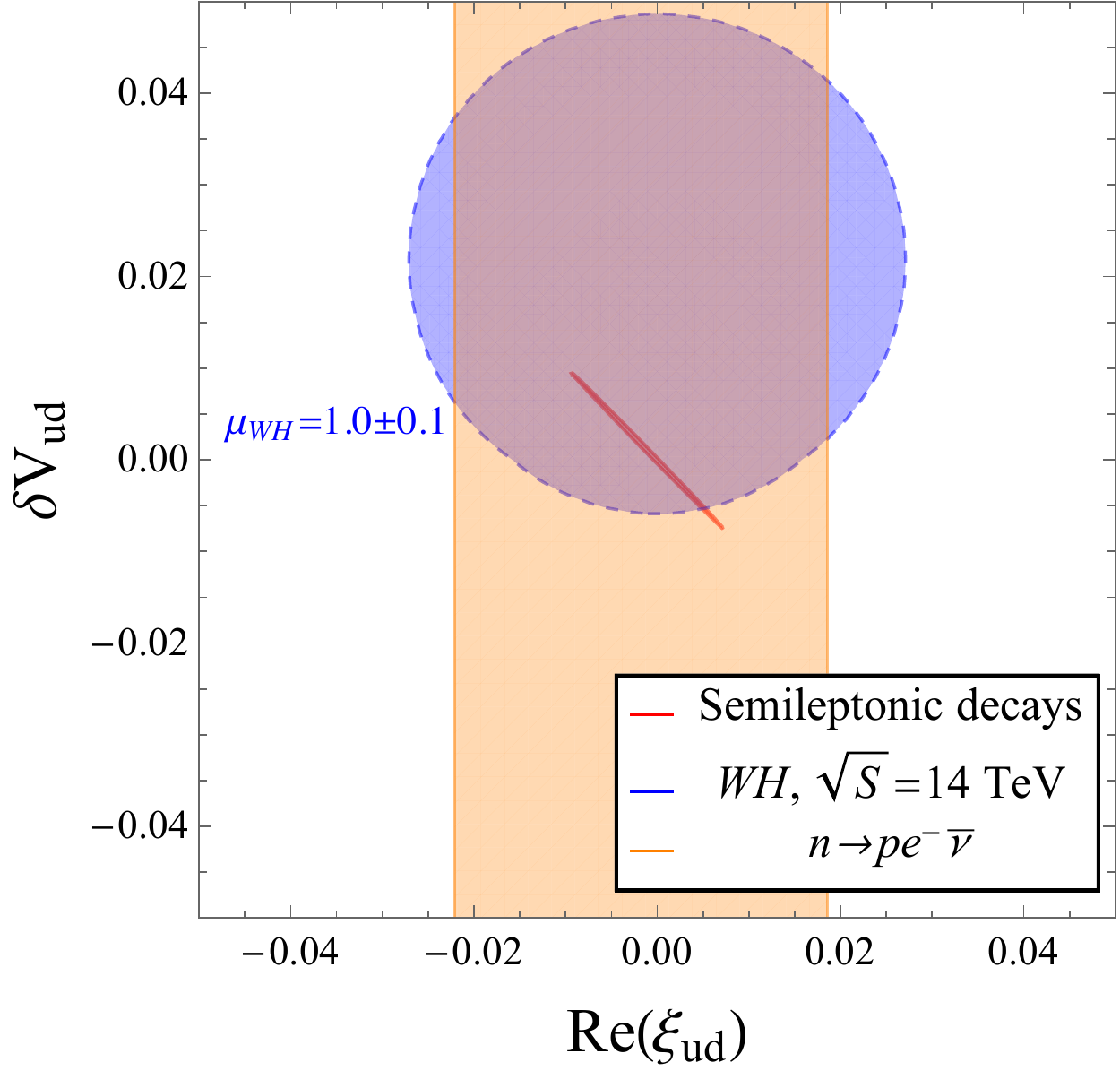}
\caption{\small The figure shows the constraints in the $\xi_{ud}-\delta V_{ud}$ plane, after marginalizing over $\xi_{us}$ and $\delta V_{us}$. The blue line depicts the constraint from $WH$ production, while the red line indicates the limits from superallowed $\bt$ decay and leptonic pion decay. The vertical orange band results from the experimental determination of $\lambda$, from neutron decay correlations, in combination with an assumed lattice determination of $g_A=1.27\pm 0.05$. }
\label{Fig:HWvsLE}\end{figure}

In light of the intrinsic degeneracy of the observables used in Section \ref{ud&us}, one might ask if there is a more direct way to access RHCCs at low energy.
Decay correlations in the neutron and hyperon $\beta$ decays  are particularly sensitive to the Lorentz structure of the quark and lepton coupling \cite{Bhattacharya:2011qm,Cirigliano:2012ab,Cirigliano:2013xha,Gonzalez-Alonso:2013uqa,Gonzalez-Alonso:2016etj}. For example, the $\bt$  and neutrino asymmetry in neutron  $\beta$ decay can be expressed as \cite{Cirigliano:2012ab,Cirigliano:2013xha}
\begin{equation}
A(E_e) =  \frac{2 \lambda (1 - \lambda)}{1 + 3 \lambda^2}\, , \qquad  B(E_e) =  \frac{2 \lambda (1 + \lambda)}{1 + 3 \lambda^2}\, .
\end{equation}
In the presence of the most general modification of the semi-leptonic dimension-six Lagrangian~\cite{Bhattacharya:2011qm,Bernard:2007cf},  we have
\begin{equation}\label{lambda}
\lambda = \frac{g_A}{g_V}  \left| \frac{1 + \delta V_{ud} + (\varepsilon_L)_{ud} - \frac{\xi_{ud}}{V_{ud}}}{1 + \delta V_{ud} + (\varepsilon_L)_{ud} + \frac{\xi_{ud}}{V_{ud}} } \right| 
=  
\frac{g_A}{g_V}  \left( 1 - 2 \textrm{Re} \left(\frac{\xi_{ud}}{V_{ud}} \right)  \right)  + \mathcal O\left(\frac{v^4}{\Lambda^4}\right)\, ,
\end{equation}
where $g_V$ and $g_A$ are the nucleon matrix elements of the vector and axial-vector currents.
Experimentally, the ratio $g_A/g_V$ is determined with per mil uncertainties,  $\lambda = 1.2723 \pm 0.0023$. 
In order to constrain $\xi_{ud}$  one needs   precise information on $g_A$. While this is the subject of intense research in LQCD, current determinations of $g_A$ have about a 4-5\% uncertainty \cite{Bhattacharya:2016zcn,Berkowitz:2017gql}, 
which allows for percent-level right-handed contributions, as first discussed in Ref. \cite{Gonzalez-Alonso:2016etj} .
Setting the central value of $g_A$ to $1.27$, and assigning a $4\%$ theoretical error, $g_A = 1.27 \pm 0.05$, would result in  $\textrm{Re}\,\xi_{ud} \in [-2.1,2.0] \cdot 10^{-2}$,
in the same range as the values probed by $WH$ production. 
To illustrate the interplay between the different low-energy and collider constraints, we show in Fig.\ \ref{Fig:HWvsLE} the limits  in the $\xi_{ud}-\delta V_{ud}$ plane, after marginalizing over $\xi_{us}$ and $\delta V_{us}$. As can be seen from the figure, superallowed $\bt$ decay and $\pi\to\mu\nu$ currently provide the strongest limits. However, as mentioned, these observables  get additional contributions from scalar and pseudo-scalar interactions ($\varepsilon_S$ and $\varepsilon_P$), and do not uniquely probe RHCCs.
The experimental determination of $\lambda$ combined with lattice calculation of $g_A$ provides a direct low-energy probe of right-handed currents in the $ud$ sector. Currently this leads to a constraint that is comparable to future collider limits.

\subsection{Collider probes}\label{DisHigh}

\begin{figure}
\center
\includegraphics[width=\textwidth]{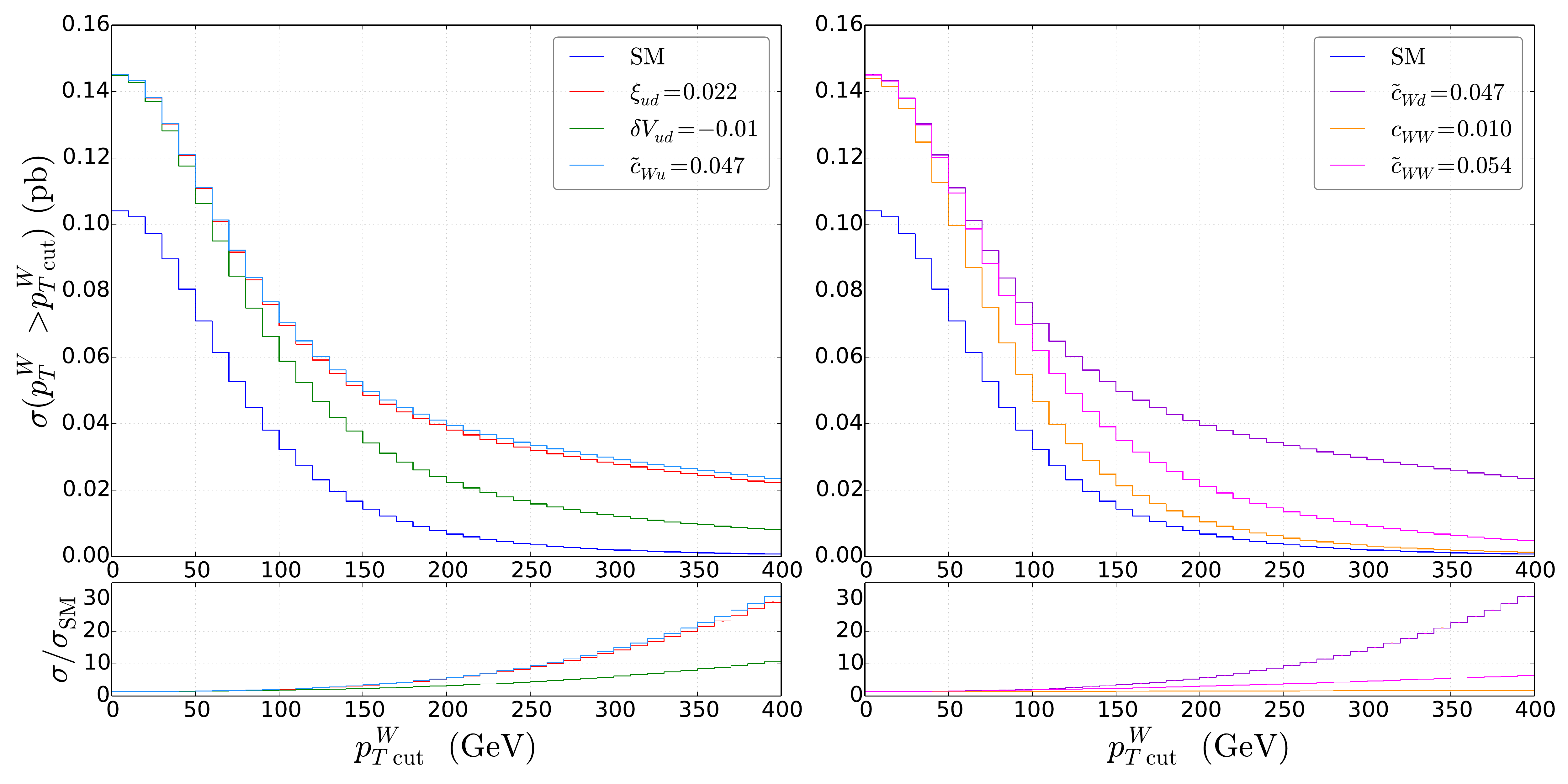}
\caption{$W^+ H$ cross section  for $p_{T}^{W} > p^{W}_{T\, \textrm{cut}}$, at $\sqrt{S}= 14$ TeV. The blue line denotes the SM cross section, the remaining lines include the contributions of the operators in Eq. \eqref{extLag2}. }\label{Fig:HWptcut}
\end{figure}

In similar fashion we can ask whether a discrepancy in a collider setting could be unambiguously attributed to a RHCC. In this section we focus on observables related to $WH$ productions as this process, as discussed in Sect.~\ref{sec:HW}, is particular sensitive to right-handed interactions.
To address the issue of identifying the $\xi$ operator, we explore observables that could disentangle RHCCs from other BSM contributions.
We consider the full set of dimension-six operators that modifies $WH$ production using the basis of Ref. \cite{Grzadkowski:2010es} 
\begin{eqnarray}\label{eq:extLag}
\mathcal L_6  &=&  C_{\varphi W} \varphi^{\dagger} \varphi W_{\mu \nu} W^{\mu \nu}  + C_{\varphi \tilde{W}} \varphi^{\dagger} \varphi \tilde{W}_{\mu \nu} W^{\mu \nu}
+   \varphi^{\dagger} \tau^{I} i \overleftrightarrow{D}_{\mu} \varphi\,  \bar q_L \tau^I  \gamma^{\mu} c^{(3)}_{Q\varphi} q_L 
+   \varphi^{\dagger}  i \overleftrightarrow{D}_{\mu} \varphi\,  \bar q_L   \gamma^{\mu} c^{(1)}_{Q\varphi} q_L\nonumber\\
& &
- \frac{g}{\sqrt{2}} \bar q_L \sigma^{\mu\nu} \Gamma^{u}_W \tau^I  W^I_{\mu\nu}  \tilde \varphi \, u_R  - \frac{g}{\sqrt{2}} \bar q_L \sigma^{\mu\nu}  \Gamma^{d}_W  \tau^I  W^I_{\mu\nu} \varphi \, d_R
+ \frac{2}{v^2} i \tilde{\vp}^{\dagger} D_{\mu} \vp \, \bar{u}_R \gamma^\mu \,\xi d_R+ \mathrm{h.c.}\, ,
\end{eqnarray}
where $\vp$ is the Higgs doublet, $\tau^I$ are Pauli matrices,  $\tilde \vp = i\tau_2 \vp^*$,  $W^I_{\mu\nu}$ denotes the  $SU(2)$ field strengths,
and $\tilde W^{\mu \nu} = \varepsilon^{\mu \nu \alpha \beta} W_{\alpha \beta}/2$. The Higgs covariant derivatives are given by
\begin{equation}
\varphi^{\dagger} i \overleftrightarrow{D}_{\mu} \varphi = i \varphi^{\dagger} ( D_{\mu} - \overleftarrow{D}_{\mu})\varphi\, , \qquad
\varphi^{\dagger} \tau^{I} i \overleftrightarrow{D}_{\mu} \varphi =  i \varphi^{\dagger} (\tau^I  D_{\mu} - \overleftarrow{D}_{\mu} \tau^I)\varphi\, .
\end{equation}

\begin{figure}
\center
\includegraphics[width=16cm]{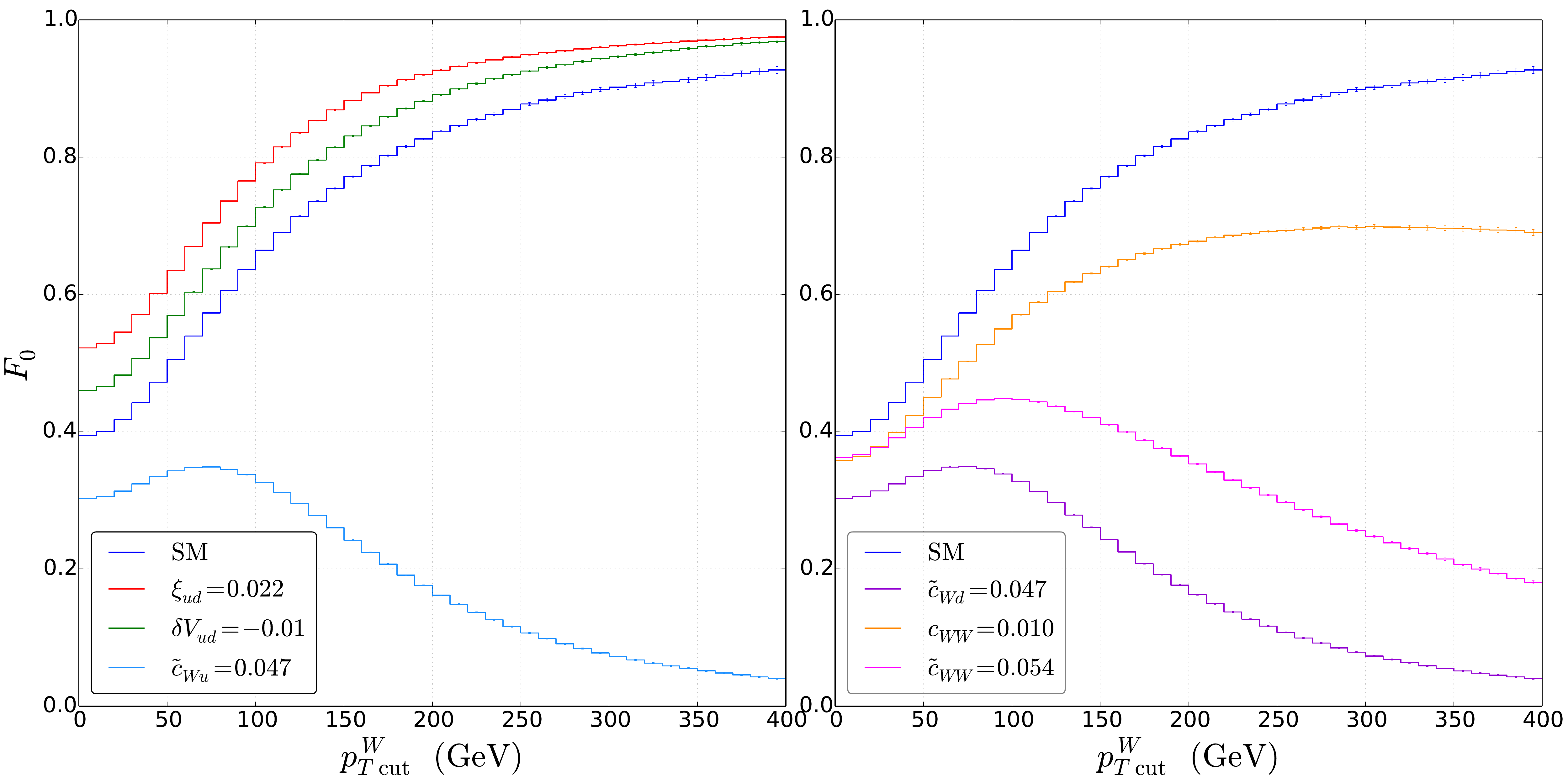} 
\caption{Longitudinal polarization of the $W^+$ boson in $W^+H$ production as a function of $p^{W}_{T\,\textrm{cut}}$, at $\sqrt{S} = 14$ TeV.}\label{Fig:F0}
\end{figure}
\begin{figure}
\center
\includegraphics[width=16cm]{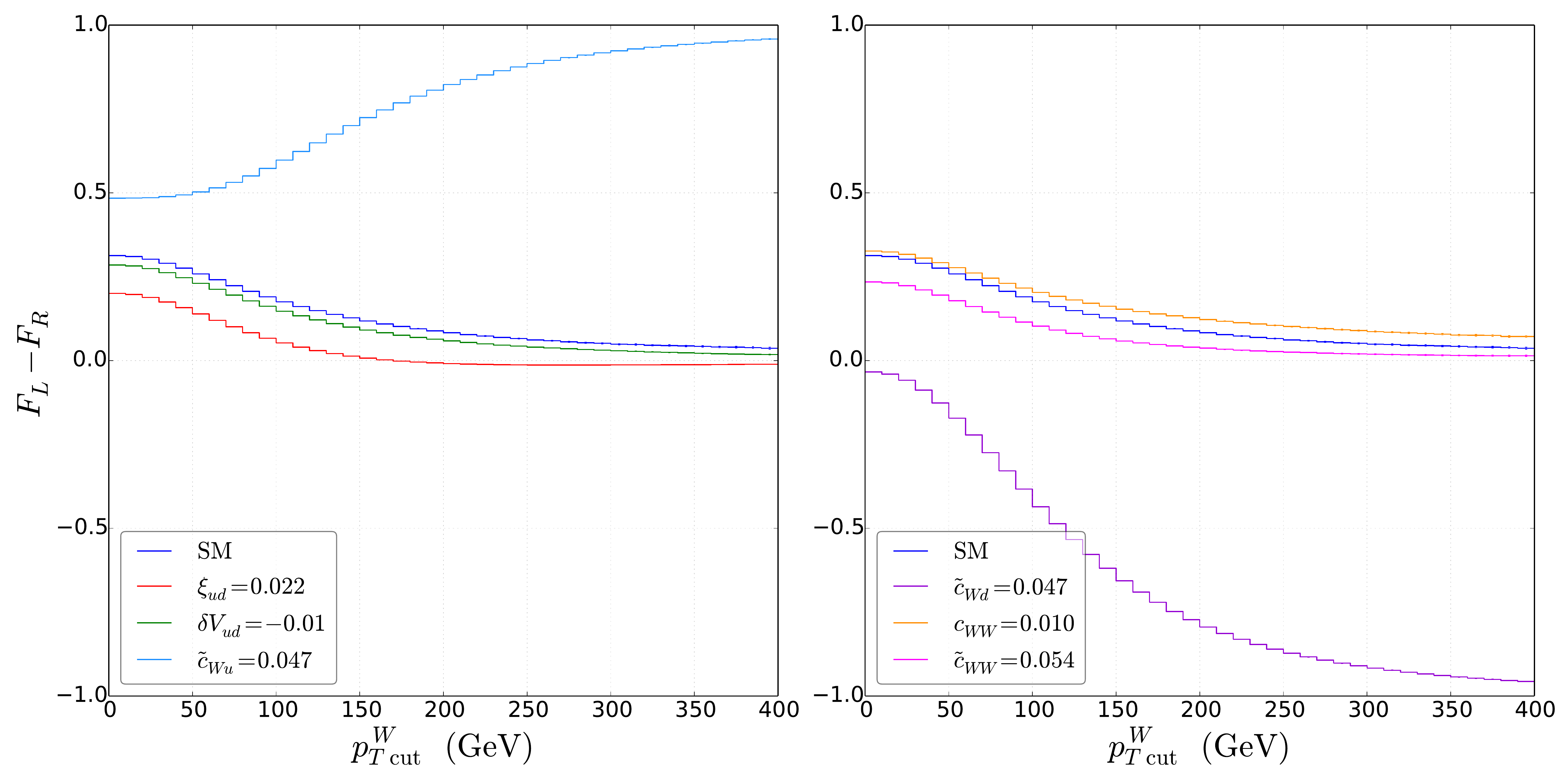} 
\caption{Difference of left and right-handed polarizations of the $W^+$ boson in $W^+H$ production as a function of $p^{W}_{T\,\textrm{cut}}$, at $\sqrt{S} = 14$ TeV.}\label{Fig:FLR}
\end{figure}

For fermionic operators we consider couplings to the $u$ and $d$ quarks (the $us$ couplings can be studied in analogous fashion), which, in the mass basis, can be written as 
\begin{eqnarray}\label{extLag2}
\mathcal L &=&  \frac{h}{v} \left( c_{WW} W^{\mu \nu} W_{\mu \nu} +  \tilde{c}_{WW} \tilde{W}^{\mu \nu} W_{\mu \nu} \right)  + \frac{g}{\sqrt{2}}   V_{ud} \left( 1 + \delta V_{ud} \left( 1 +  \frac{h}{v} \right)^2 \right)  \bar u_L \gamma^\mu d_L  W^+_{\mu}
\nonumber \\  &  & +  \frac{g}{\sqrt{2}}  \xi_{ud} \left( 1 +  \frac{h}{v} \right)^2 \bar u_R \gamma^\mu  d_R     W^+_{\mu}  
  - \frac{g V_{ud}}{\sqrt{2} v} \left(1 + \frac{h}{v}\right)  \left(  c_{Wd} \bar u_L \sigma^{\mu \nu}  d_R +  c_{Wu} \bar u_R \sigma^{\mu \nu}  d_L \right) W^{+}_{\mu \nu} + \mathrm{h.c.}\, \nn \\ 
\end{eqnarray}
The CKM factors in the previous expressions arise from rotating to the mass basis. In order to avoid flavor-changing neutral currents at  tree level, we impose that the matrices  $\Gamma^{u}_{W}$, $\Gamma^{d}_W$, 
$c^{(1)}_{Q\varphi}$, and $c^{(3)}_{Q\varphi}$ are diagonal in the mass basis. 
In addition to $\xi_{ud}$ we  then need to consider the dimensionless couplings 
\begin{eqnarray}
&c_{WW} = v^2 C_{\varphi W}\, , \quad  \tilde{c}_{WW} = v^2 C_{\varphi \tilde{W}}\, ,  \nn \\ 
&\delta V_{ud} = \left( v^2 c^{(3)}_{Q\varphi} \right)_{11}\, , \quad \tilde{c}_{W d}  = \left( v^2 \Gamma^{d}_W \right)_{11},\,  \quad \tilde{c}_{W u}  = \left( v^2 \Gamma^{u}_W \right)_{11}\, .
\end{eqnarray}
These couplings scale as $v^2/\Lambda^2$. Hermiticity implies that $c_{WW}$, $\tilde{c}_{WW}$, and $\delta V_{ud}$ are real, whereas $\xi_{ud}$, $\tilde{c}_{Wu}$, and $\tilde{c}_{Wd}$ in general have real and imaginary parts. Since we are neglecting interference terms proportional to the light quark masses, the cross section only depends on the absolute values of these couplings.

To illustrate the diagnostic power of collider measurements, 
we set $\xi_{ud}=0.022$. This value is still allowed by the 8 and 13 TeV data and produces a 40\% modification of the signal strength at 14 TeV. 
We then proceed by turning on one of the couplings in Eq.~\eqref{extLag2} at a time, and tune
them such that they give the same signal-strength modification as $\xi_{ud}$. In Fig. \ref{Fig:HWptcut} we show the effect of the different couplings on the cumulative $W^+ H$ cross section for
$p_{T}^{W} > p^{W}_{T\,\textrm{cut}}$, where $p_{T}^{W}$ is the transverse momentum of the $W$ boson. In general this observable receives  different corrections from different operators. However, as can be seen from the figure, this observable is not sufficient to lift the degeneracy. In particular, $\xi_{ud}$ and the dipole operators, $\tilde c_{Wu}$ and $\tilde c_{Wd}$, induce very similar corrections.

A more suitable observable to disentangle the effects of the various BSM contributions is the angular distributions of the charged lepton coming from  the decay of the $W$ boson. We work in the $W$-boson rest frame,  with the direction of the $z$-axis along the momentum of the $W$ boson in the lab frame. $\theta^*$ is the polar angle of the charged lepton in this frame.
The  $x$-axis is in the direction orthogonal to the Higgs and $W$ momenta $\hat{x} \sim (\vec{p}_W \times \vec{p}_H)$. In this frame, we define the azimuthal angle $\phi^*$ as the angle between the plane containing the $W$ and the Higgs bosons, and the plane containing the $W$ and its charged decay product. That is
\begin{equation}
\cos\phi^* = \frac{(\vec p_W \times \vec p_H) \cdot (\vec p_W \times \vec p_e)}{|\vec p_W \times \vec p_H|\,  |\vec p_W \times \vec p_e|} \, ,
\end{equation}
 and we note that $\phi^*$ is invariant under boosts along the $W$ momentum $\vec p_W$.

\begin{figure}
\center
\includegraphics[width=10cm]{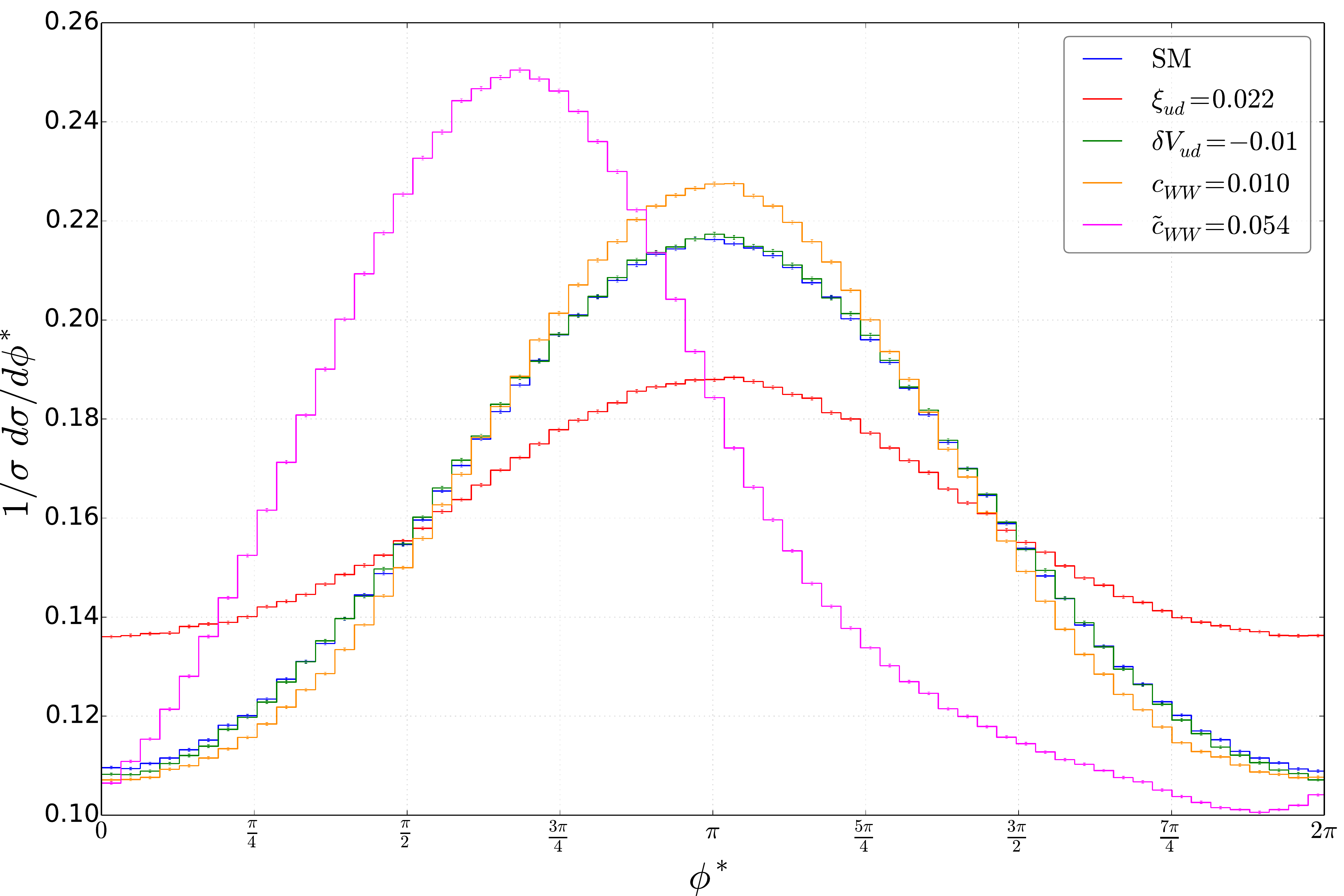}
\caption{$W^+H$ differential cross section with respect to $\phi^*$ at $\sqrt{S} = 14$ TeV, in the SM (blue line), and in the presence of dimension-six operators.}\label{Fig:phi}
\end{figure}

The angular distribution of the $W$ boson in this frame is parameterized by 8 coefficients, which completely characterize the $W$-boson spin-density matrix
\begin{eqnarray}
& & \frac{1}{\sigma} \frac{d\sigma}{d\cos\theta^*\, d\phi^*} = \frac{3}{16 \pi} \Big[  1 + \cos^2\theta^*  + \frac{A_0}{2} (1- 3 \cos^2 \theta^*) + A_1 \sin 2\theta^* \cos\phi^*  
+ \frac{A_2}{2} \sin^2\theta^* \cos 2\phi^*   \nn \\ &&  + A_3 \sin\theta^* \cos\phi^* + A_4 \cos\theta^* + A_5 \sin\theta^* \sin \phi^*  + A_6 \sin2\theta^* \sin \phi^*
+ A_7 \sin^2\theta^* \sin 2\phi^* \label{diffthetaphi}
\Big]\, .
\end{eqnarray}
The differential distributions with respect to  $\theta^*$ and $\phi^*$ are obtained by integrating Eq. \eqref{diffthetaphi}, and are given by 
\begin{eqnarray}
\frac{1}{\sigma} \frac{d\sigma}{d\cos\theta^*} &=& \frac{3}{8} \left[  1 + \cos^2\theta^*  + \frac{A_0}{2} (1- 3 \cos^2 \theta^*)  + A_4 \cos\theta^* \right], \\
\frac{1}{\sigma} \frac{d\sigma}{d\phi^*} &=& \frac{1}{2 \pi} \left[  1 + \frac{3 \pi}{16} (A_3 \cos\phi^* + A_5 \sin\phi^*)  + \frac{1}{4} (A_2\cos 2\phi^* + A_7 \sin 2\phi^*)  \right]\, .
\end{eqnarray}
The coefficients  $A_0$ and $A_4$ are related to the $W$-boson helicity fractions \cite{Bern:2011ie},
\begin{equation}
F_0 = \frac{A_0}{2}\, , \qquad F_L = \frac{1}{4} (2 -A_0 \mp A_4)\, , \qquad F_R = \frac{1}{4} (2 -A_0 \pm A_4)\, ,
\end{equation}
for $W^\pm$, respectively.

In Figs. \ref{Fig:F0} and \ref{Fig:FLR} we plot the longitudinal, and the difference of the left- and right-handed polarization fractions of the $W^+$ boson, for $p^W_T > p^W_{T\textrm{cut}}$. We show these quantities for the pure SM and for the SM modified by one of the dimension-six operators in Eq.~\eqref{extLag2}. For the Wilson coefficients we use the same values as used in Fig.~\ref{Fig:HWptcut} such that the signal strength at 14 TeV is modified by 40\%. As can be glimpsed from the figures, within the SM the W-boson becomes increasingly polarized in the longitudinal direction as the cut on the $W$ transverse momentum increases \cite{Stirling:2012zt}. This behavior is not significantly affected by a right-handed current, $\xi_{ud}$, or a gauge-invariant correction to the left-handed current, $\delta V_{ud}$. The operator $c_{WW}$ would also preferentially induce a longitudinally polarized $W$ at large $p^{W}_{T\,\textrm{cut}}$, but with a smaller fraction.

On the other hand,  dipole couplings of the $W$ boson to the up and down quarks would greatly reduce  the longitudinal fraction at large $p_T$. In Fig. \ref{Fig:FLR} we show that $\tilde{c}_{Wu}$ and $\tilde{c}_{Wd}$
induce, respectively, a left- and right-handed polarized $W$. Finally, a nonzero value of $\tilde{c}_{WW}$ would also reduce the longitudinal fraction and produce equal amount of left- and right-polarized $W$ bosons at large $p_T$.
The $W$-boson helicity fractions would therefore make it possible to identify the effects of the dipole interactions, or perhaps of $\tilde c_{WW}$, but they would not clearly identify a right-handed current.   

\begin{figure}
\center
\includegraphics[width=16cm]{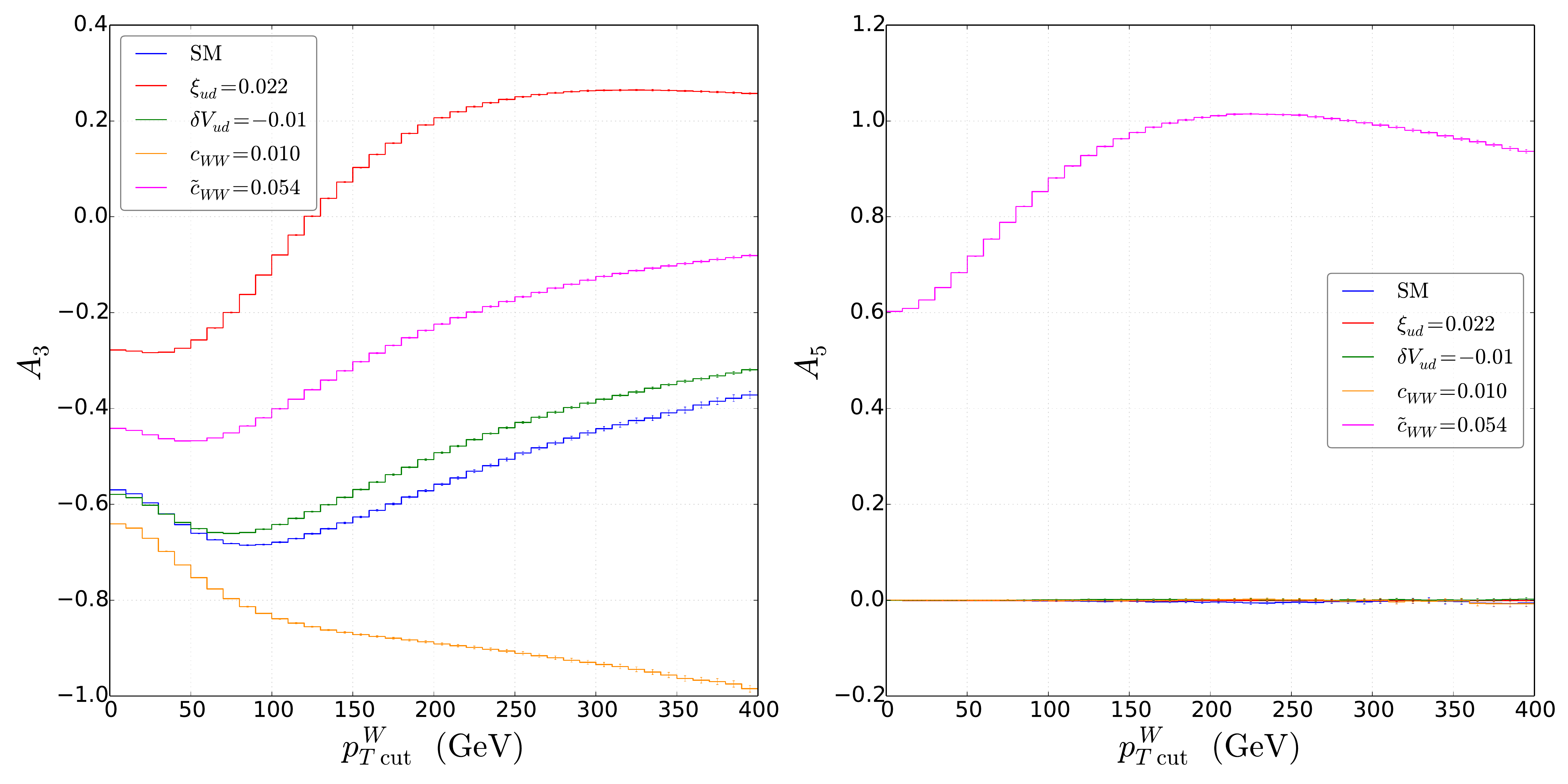}
\caption{$A_3$ and $A_5$ coefficients in $W^+H$ production as a function of $p^W_{T\,\textrm{cut}}$, at $\sqrt{S} = 14$ TeV.}\label{Fig:A3A5}
\end{figure}

We now turn to the azimuthal-angle distribution which does turn out to be sensitive to RHCC.
In Fig. \ref{Fig:phi} we show the normalized differential cross section with respect to $\phi^*$, with no cut on $p^W_{T}$, for the SM, and for the operators 
$\xi_{ud}$, $\delta V_{ud}$, $c_{WW}$, and $\tilde c_{WW}$. In the SM, the cross section is well described by the $\cos\phi^*$ term, with a smaller component proportional to $\cos 2\phi^*$. 
We observe that the left-handed current, $\delta V_{ud}$, does not significantly affect the shape of the $\phi^*$ distribution. 
 $c_{WW}$ induces a slightly larger $\cos 2\phi^*$ component which only mildly modifies the distribution.
$\xi_{ud}$ does not modify the functional form of the distribution, which is proportional to $\cos\phi^*$, but significantly affects the amplitude. This is captured by the coefficient $A_3$, which we show in the left-panel of Fig.~\ref{Fig:A3A5} as a function of $p^W_{T\,\textrm{cut}}$.
$A_3$ vanishes if parity is conserved in the production of the $W$ boson, and, for the chosen value of $\xi_{ud}$, this effectively occurs at $p^W_{T\,\textrm{cut}}= $ 120 GeV. For larger cuts, $\xi_{ud}$ overtakes the SM contribution and the sign of $A_3$ flips.

Fig. \ref{Fig:phi} shows another interesting feature, namely that $\tilde{c}_{WW}$ induces a $\sin\phi^*$ dependence of the cross section. 
Differently from all other operators in Eq. \eqref{extLag2}, for $\tilde{c}_{WW}$ a CPV interference term with the SM survives.
This term does not induce any corrections to the total cross section, but it does induce a large $A_5$ coefficient as shown in the right-panel of Fig.~\ref{Fig:A3A5}. Such a signature is a distinctive feature of a $\varphi^{\dagger} \varphi \tilde{W} W$ operator at the LHC. 

This discussion shows that, in the presence of a deviation of the total $WH$ cross section from the SM prediction, 
a study of the $p^W_{T}$ spectrum and of angular distributions of the charged lepton produced by $W$ decay, would provide important information in identifying the origin of the discrepancy. We should point out that our preliminary study did not include detector effects, background estimates, nor reconstruction efficiencies and a more careful investigation is warranted. Such an investigation would be better performed  within the framework of an experimental collaboration.

\subsection{Anomalous $Wtb$ couplings}\label{sec:topanomalous}
\begin{table}
\center\begin{tabular}{||c |c| c| |c|c| c ||}
\hline \hline
Real &\textbf{Individual} &\textbf{Marginalized}  & Imaginary &\textbf{Individual} &\textbf{Marginalized}  \\
\hline
$\xi_{tb}$ & $[-1.4,1.5]\Ex{-3}$ &$[-0.01,0.12]$ & $\xi_{tb}$& $[-2.4,2.4]\cdot 10^{-3}$ &$[-0.16,0.13]$  \\
$\delta_{LL}$ 	  & $[-0.03,0.04]$  &$[-0.03,0.04]$  & $-$& &  \\
$v^2\,C_{W t}$  & $[-0.09,0.05]$ &$[-0.10,0.04]$& $v^2\,C_{W t}$&  $[-6.0,6.0]\cdot 10^{-4}$ & $[-1.0,0.9]\cdot 10^{-3}$  \\
$v^2\,C_{W b}$  & $[-0.04,0.05]$  & $[-3.5,0.4]$ & $v^2\,C_{Wb}$  & $[-3.5,3.5]\cdot 10^{-2}$ & $[-1.9,2.2]$       \\
\hline \hline
\end{tabular}\\
\caption{\small Allowed regions ($90\%$ C.L.) for the $Wtb$ couplings at the scale $\Lambda = 1$ TeV. The second and fifth columns show the constraints under the assumption that  only a single coupling is generated at the high scale. In the third and sixth columns we assume that all $Wtb$ couplings in Eq.~\eqref{eq:TBoperators} are present at the scale of new physics and we marginalize over all couplings.}
\label{Tab:WtbMargin}
\end{table}

Analogously to the $ud$-$us$ sector it is interesting to see to what extent the constraints  in the $tb$ sector are affected by turning on additional operators. To explore this, we  extend our Lagrangian, involving right-handed ($tb$) currents, with the following set of operators,
\begin{subequations}
\label{eq:TBoperators}
\bea
O_{LL} &=&\frac{g v\sq}{2}\bigg[\frac{1}{\sqrt{2}}\bar t_L\g^\mu b'_L W_\mu^++\frac{1}{\sqrt{2}}\bar b_L'\g^\mu t_L W_\mu^- + \frac{1}{c_W} Z_\mu \bar t_L\g^\mu t_L\bigg] \bigg(1+\frac{h}{v}\bigg)\sq\, ,\\
O_{Wt} \!\!   &=& -g m_t \bigg[  \frac{1}{\sqrt{2}}  \bar{b}_L'   \simu  t_R W_{\mu\nu}^- 
+  \bar t_L\simu t_R \bigg(\frac{1}{2c_W} Z_{\mu\nu}+i g W_\mu^-W_\nu^+\bigg)\bigg]\bigg(1+\frac{h}{v}\bigg)\, , \qquad
\\
O_{Wb} \!\! &=& - g m_b   \bigg[\frac{1}{\sqrt{2}} \bar t_L'  \simu b_R  W_{\mu\nu}^+ 
- \bar b_L\simu b_R \bigg(\frac{1}{2c_W} Z_{\mu\nu}+i g W_\mu^-W_\nu^+\bigg)\bigg]\bigg(1+\frac{h}{v}\bigg) ~,
\label{eq:NSY}
\eea
\end{subequations}
which appear in the Lagrangian with couplings $C_{LL,\,Wt,\, Wb}$, respectively. Here, $b' = V_{tb} b + V_{ts} s + V_{td} d$, $t' = V_{tb}^* t + V_{cb}^* c + V_{ub}^* u$, and $c_W = \cos \theta_W$,  with $\theta_W$ the Weinberg angle. In total, the effective $Wtb$ vertex can then be written as
\bea \label{eq:btw}
\vL_{tb} &=& \frac{g}{\sqrt{2}}\bar t \bigg[ \g^\mu \big(V_{tb}(1+\delta_{LL})\, P_L + \xi_{tb}\, P_R\big) W_\mu^+ - \simu W_{\mu\nu}^+\big(m_tC_{Wt}^*P_L+m_bC_{Wb}P_R\big)\bigg]b
+{\rm h.c.}\quad ,\nn\\
\eea
where\footnote{Within the framework of the SMEFT, $O_{LL}$ arises from the operators $Q_{Hq}^{(1)}$ and $Q_{Hq}^{(3)}$ in the notation of \cite{Jenkins:2013zja,Jenkins:2013wua,Alonso:2013hga}.  We follow Ref.\ \cite{Grzadkowski:2008mf} and assume no flavor-changing neutral currents at tree level, the $Wtb$ coupling is then hermitian and $C_{LL}$ is forced to be real.} $\delta_{LL} = \frac{v\sq}{2} C_{LL}$. 
Eq.~\eqref{eq:btw} provides a general parametrization of the $Wtb$ vertex \cite{AguilarSaavedra:2008zc} and these couplings have been studied in many previous works \cite{Grzadkowski:2008mf,Drobnak:2010ej,AguilarSaavedra:2008zc,GonzalezSprinberg:2011kx,Drobnak:2011aa,Cao:2015doa,Hioki:2015env,Schulze:2016qas,Kamenik:2011dk,Zhang:2012cd,deBlas:2015aea,Buckley:2015nca,Buckley:2015lku,Bylund:2016phk, Castro:2016jjv}. Most of these studies constrain the $Wtb$ vertex by looking at collider processes or flavor constraints (in particular $\Delta B=1$ processes). Here we study  the effect of turning on $\xi_{tb}$ and $C_{LL,\,Wt,\, Wb}$ simultaneously, while taking into account both collider and low-energy constraints including those from EDM experiments which are usually not considered.

To derive the resulting constraints we require several additions to the expressions discussed in previous Sections. In particular, for the $C_{LL,Wb,Wt}$ contributions to the helicity fractions discussed in Section \ref{sec:top}, we employ the expressions given in Ref.\ \cite{Drobnak:2010ej}. We take into account  $\delta_{LL}$ contributions to single-top production by replacing $V_{tq}\to V_{tq}(1+\delta_{LL})$, where $q\in (d,s,b)$.
For the running and matching of $C_{LL,Wt,Wb}$ onto the $C_{7,8}$ operators relevant for $b\to q\g$, we use the expressions in Refs.\ \cite{Grzadkowski:2008mf,Aebischer:2015fzz}. Finally, for the contributions of $C_{Wt}$ and $C_{Wb}$ to EDMs we follow the analysis of Refs.\ \cite{Cirigliano:2016nyn,Cirigliano:2016njn}. 
With this combined input we turn on $\delta_{LL}$, $\xi_{tb}$, and $C_{Wt,Wb}$ simultaneously, while setting $V_{tb}= 1$. The resulting constraints are shown in Table \ref{Tab:WtbMargin}, together with the bounds that result from a single-coupling analysis.

\begin{figure}
\center
\includegraphics[width=6cm]{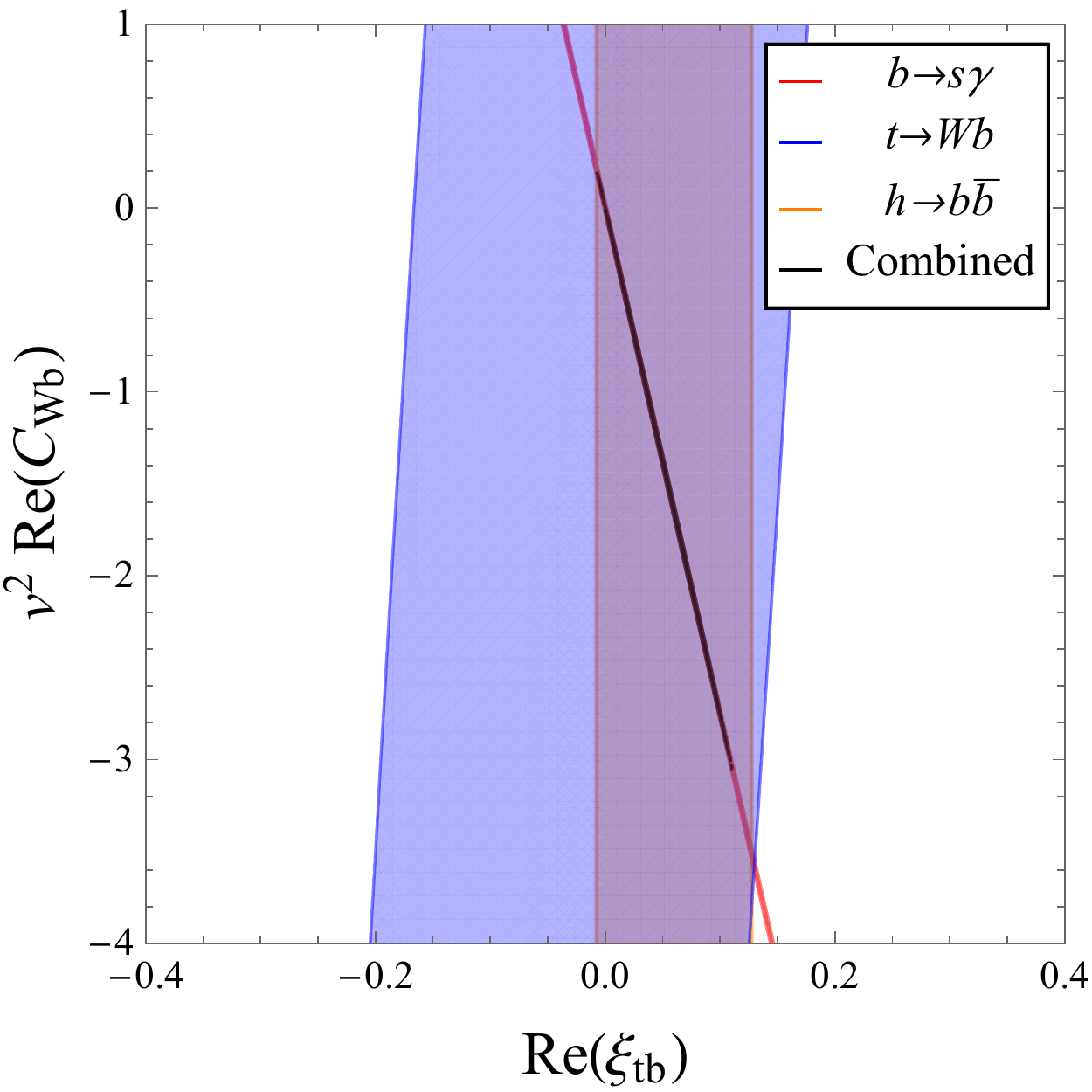}
\includegraphics[width=6cm]{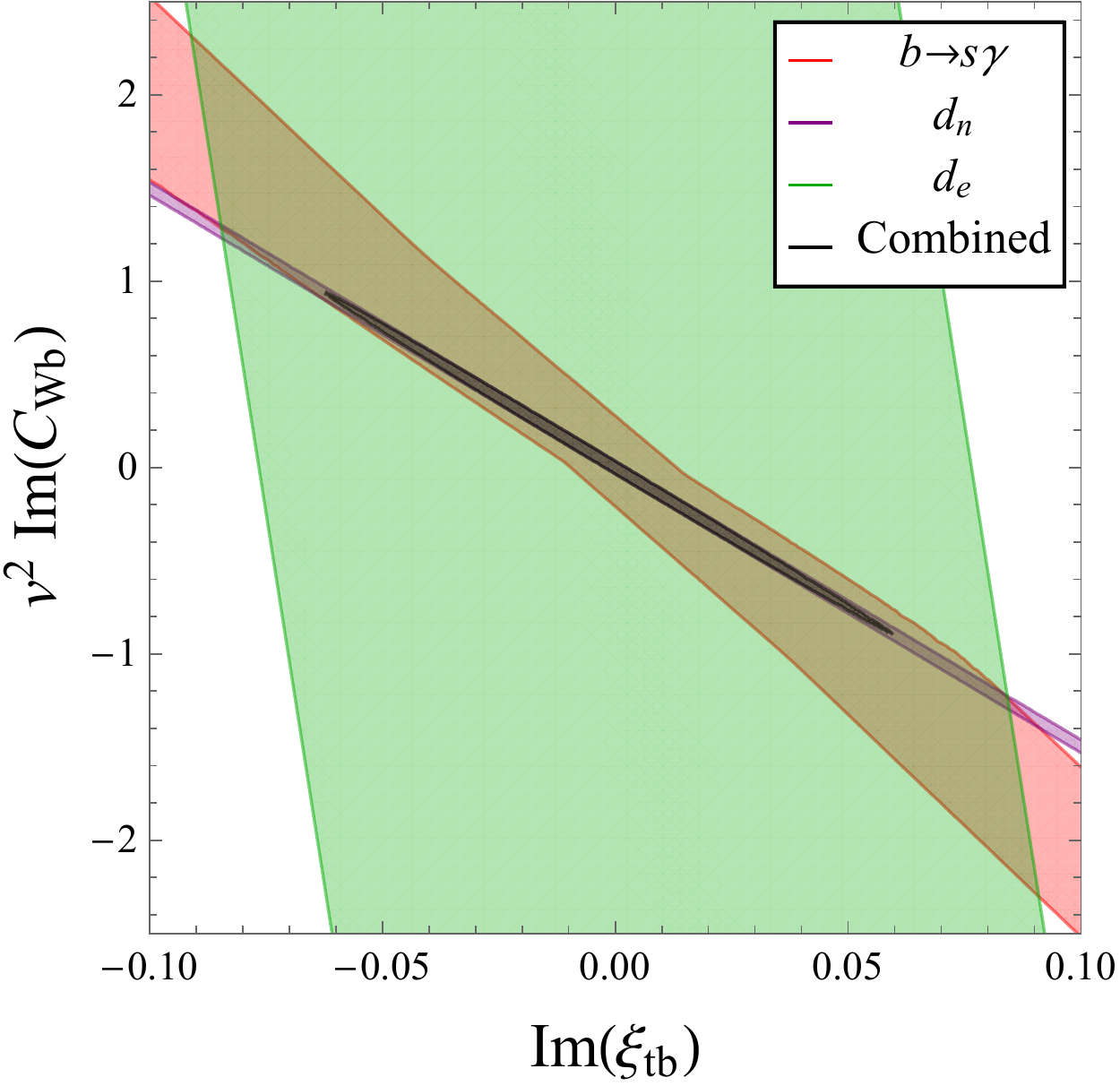}
\caption{\small The figure shows the constraints in the $\xi_{tb}-v\sq\,C_{Wb}$ plane, illustrating the significant cancellations that can occur between the two couplings. The real (imaginary) couplings are shown in the left (right) panel. The exclusion bands (at $90\%$ C.L.) assume that only $C_{Wb}$ and $\xi_{tb}$ are generated at the scale $\Lambda$.}
\label{Fig:Cwb}\end{figure}

The constraints on $\delta_{LL}$ and $C_{Wt}$ are the least affected by the presence of the other operators, and the marginalized bounds are fairly close to the single-coupling analysis. This can be understood by noticing that $\delta_{LL}$ is stringently constrained by single-top production, which does not allow for cancellations against the other couplings. Similarly, Im$\, C_{Wt}$ provides the dominant contribution to the electron EDM such that its constraint survives to large extent the global analysis as well. Re$\, C_{Wt}$ is constrained by several observables with similar strength (electroweak precision tests, W helicity fractions, and $b\to s\g$, see Ref.~\cite{Cirigliano:2016njn}) such that the marginalized constraint is not too different from the individual one.

The situation is significantly different for $C_{Wb}$ and $\xi_{tb}$. In the single-coupling analysis the real parts of these couplings are mainly constrained by $b\to s\g$, while their imaginary parts are constrained by the neutron EDM. When both operators are present the constraints on the real and imaginary parts can be weakened significantly by mutual cancellations in BR$(b\to s\g)$ and $d_n$, respectively. In fact, comparing the second (fifth) and third (sixth) columns of Table\ref{Tab:WtbMargin} we see that the limits on the real (imaginary) part of $\xi_{tb}$ deteriorate by roughly two orders of magnitude. The bounds are similarly weakened for $C_{Wb}$. 

This effect is illustrated in Fig.\ \ref{Fig:Cwb}, where we show the constraints in the $\xi_{tb}-C_{Wb}$ plane for the real and imaginary parts. The results in this figure assumes only $\xi_{tb}$ and $C_{Wb}$ to be present at the scale $\Lambda$, but this is sufficient to see that significant cancellations can occur between these two couplings. The left-panel shows that the CP-even $b\to s\g$ observables allow for a free direction, and the much weaker limits from the helicity fractions and $h\to b\bar b$ are needed to obtain a constraint. In the case of the imaginary parts, shown in the right panel, the neutron EDM allows for a free direction and the electron EDM is needed to obtain a bound. As can be seen from Table \ref{Tab:WtbMargin}, including $C_{Wt}$ and $\delta_{LL}$ hardly affects the bounds on the real parts of $\xi_{tb}$ and $C_{Wb}$ compared to Fig.\ \ref{Fig:Cwb}. The limits on the imaginary parts are weakened by a minor factor compared to Fig.\ \ref{Fig:Cwb}, confirming that the major deterioration between the single-coupling and global constraints are indeed due to cancellations between $\xi_{tb}$ and $C_{Wb}$. 
A comparison of the single-coupling and global constraints for the Re$\,\xi_{tb}-$Im$\,\xi_{tb}$ plane is shown in Fig.\ \ref{Fig:xitb}. Again, it is clear that turning on several couplings can severely weaken the various constraints.

In summary, in the $tb$ sector isolating the right-handed current is complicated by the degeneracy with the $C_{Wb}$ dipole operator. Nevertheless, as shown in Table~\ref{sec:topanomalous}, the marginalized constraints on most of the anomalous $Wtb$ operators are very stringent.

\begin{figure}
\center
\includegraphics[width=6cm]{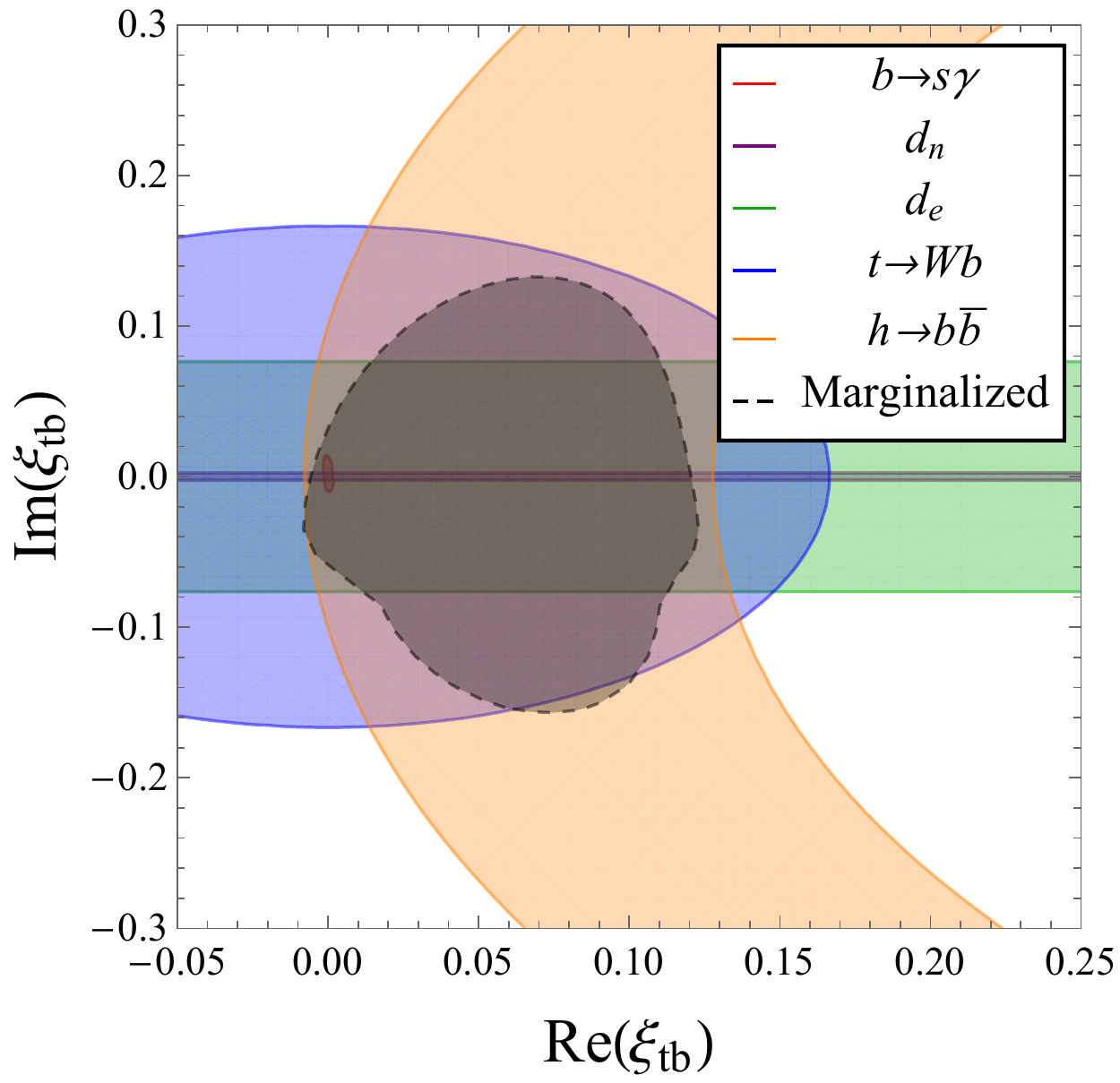}
\caption{\small  The figure shows the constraints in the ${\rm Re}\,\xi_{tb}-{\rm Im}\,\xi_{tb}$ plane. The constraints from $b\to s\g$, $d_n$, $d_e$, the helicity fractions, and $h\to b\bar b$ are shown in red, purple, green, blue, and orange, respectively, and assume that only $\xi_{tb}$ is generated at the scale $\Lambda$. The dashed black line is the resulting constraint when marginalizing over the other $tb$ couplings. }\label{Fig:xitb}
\end{figure}

\section{Conclusion}\label{sec:concl}

Motivated by the attractive possibility of parity restoration at high energies, we investigated possible footprints left behind by a right-handed extension of the SM. 
We studied in detail the right-handed charged-current couplings $\xi_{ij}$ defined in Eq.~\eqref{dim6edms}, 
looking at  their  manifestations in collider experiments, flavor physics, and low-energy precision tests. 

Our work provides  a case study of the complementarity of low-energy and collider experiments in probing 
heavy new physics which is not within direct reach of the LHC, and therefore can be  analyzed in the framework of the 
SMEFT.

In a first major thrust of our work,   assuming that at the scale $\Lambda$  the SM is modified  dominantly by the RHCC  operator, 
we  have worked out the  bounds on the $\xi_{ij}$ from a broad range of probes.   
The resulting  90\% C.L.\ limits are summarized in  Eqs.~\eqref{sumcoll} (collider) 
and  \eqref{boundsreal},  \eqref{boundsimag}  (global fit).   A graphical summary is presented in Fig.~\ref{fig:summary_chart}.
Note that in this setup one can straightforwardly compare the sensitivity of various direct and indirect probes both 
at high and low energy.  Such a comparison reveals that low-energy probes  
provide the strongest constraints, putting most of the $\xi_{ij}$  out of LHC sensitivity reach.

The above  results,  however,  should be put in a broader context. 
Since  most explicit models of new physics generate more than one class of operators at the 
UV-matching  scale $\Lambda$,  many of the observables used in our analysis 
would receive  contributions  form  several other dimension-six operators in the SMEFT.  
Therefore,  in a more general setting,  the low-energy constraints 
require  that certain linear combinations  of dimension-six Wilson coefficients (including the $\xi_{ij}$)  be highly constrained,  
which in turn imposes non-trivial constraints on new physics scenarios.  
Realizing this,  in a second thrust of our work  we have  explored: 
(i) the impact of  degeneracies  on the $\xi_{ij}$,  finding that  the  low-energy bounds can be  weakened to a level comparable to the collider sensitivity by turning on additional operators; 
(ii)  ways to remove this degeneracy,  identifying observables that would uniquely point to RHCC, both at collider and low-energy. 
Details can be found in Section~\ref{Disentangling},  with  focus on  the couplings $\xi_{ud}$, $\xi_{us}$, and $\xi_{tb}$.
Our analysis shows the importance of pursuing improved searches of $\xi_{ij}$ manifestations  
at both the energy and precision frontiers, and suggests new handles on RHCC at colliders. 

We conclude by  listing the main highlights of our analysis:

\begin{itemize}

\item Keeping in mind the significant theoretical uncertainties, we note that the introduction of appropriate $\xi_{ij}$ elements can help resolve some  tensions between data and SM predictions  in flavor physics, 
such as $\ep^\prime/\ep$~\cite{Cirigliano:2016yhc} and the inclusive-exclusive discrepancy in $V_{ub}$ and $V_{cb}$~\cite{Crivellin:2009sd,Buras:2010pz}.

\item In the framework of the linearly realized SMEFT, the most stringent collider constraint on the light elements $\xi_{ij}$, with $i \in \{u,c\}$ and $j \in \{d,s,b\}$, come from
the associated production of a Higgs and a $W$ boson, followed by $W$ production and Higgs production via vector boson fusion.
The right-handed charged-current operator $\xi$ also affects $WZ$ production, which, at the moment, provides somewhat weaker bounds.

\item Nucleon beta decay, and leptonic and semileptonic decays of pion, kaons, and $D$ mesons allow one to obtain strong bounds on the elements $\xi_{ud}$, $\xi_{us}$, 
$\xi_{cd}$, and $\xi_{cs}$.
Under the assumption that the SM is modified solely by the RHCC operator $\xi$, the low-energy bounds  put on $\xi_{ud}$ and $\xi_{us}$ are out of the collider reach, while in the case   of $\xi_{cd}$ and $\xi_{cs}$ improved constraints from the LHC Run II can compete with low-energy bounds.
If we allow for modifications to the couplings of left-handed quarks to the $W$ boson, or for additional semileptonic operators, 
the single coupling bounds on $\xi_{ud}$ can be weakened  to the percent level, see Fig.\ \ref{Fig:HWvsLE}, making it important to look for collider constraints on this coupling.

\item To this end, we identified differential distributions in $WH$ production  which are very sensitive to the Lorentz structure of the coupling of the light quarks to the $W$ boson. 
In the presence of a deviation from the SM expectations, these distributions could help identify the possible origin of the correction, disentangling RHCC interactions from other possible modifications  of the $WH$ process (see for example Fig.~\ref{Fig:phi}).

\item At colliders, it is hard to probe  $\xi_{ub}$ and $\xi_{cb}$ at a level comparable to the one achievable in exclusive and inclusive $B$ decays. 				   
Possible strategies might involve tagging $b$ jets in $WH$ and $VBF$, but in both cases it would remain hard to access values of $\xi_{ub}$ and $\xi_{cb}$ smaller than the corresponding CKM elements. 
We observe that, even including flavor observables, the bounds on $\xi_{ub}$ and $\xi_{cb}$ are not extremely strong. In  light of this, 
a more detailed study of RH contributions  to inclusive $B$-meson  decays might be appropriate. Such a study might also resolve whether RHCCs can explain the current tension between the determinations from exclusive and inclusive $B$ decays. 

\item The collider observables that are needed to constrain the third row of the $\xi$ matrix  are single-top production, top decays, with particular attention to the $W$ polarization in the decay, and $h \rightarrow b \bar b$. 
It is quite interesting that the loop process  $h \rightarrow b \bar b$ already probes $\xi_{tb}$ at a level comparable to top decays.

\item Right-handed currents in the top sector are, however, severely constrained by $B \rightarrow X_{s,d} \gamma$ and EDMs.  In a single-coupling analysis, the limits are two to three orders of magnitude stronger than collider limits,
and are not severely weakened by turning on $\xi_{td}$, $\xi_{ts}$, and $\xi_{tb}$ at the same time.
The collider limits become relevant only if we allow for more general modifications of the $Wtb$ vertex. In Section~\ref{sec:topanomalous} we therefore performed a global analysis, including all relevant low- and high-energy experiments, of the most general modification of the $Wtb$ interactions. In such a scenario, $\xi_{tb}$ and $C_{Wb}$ are strongly correlated (see Fig.~\ref{Fig:Cwb}) and the resulting limits are significantly softened. Our analysis extends those based on subsets of the available data, see e.g. Refs.~\cite{Cao:2015doa, Hioki:2015env, Castro:2016jjv}, and reflects the relevance of low-energy precision experiments.

\item Despite the large theoretical uncertainties, limits on EDMs provide strong constraints on the imaginary parts of many $\xi$ elements and even on some of the real parts because of the interplay with imaginary parts of certain CKM elements. Improvements of hadronic and nuclear theory could further strengthen the constraining power of EDM experiments as can be seen from Fig.~\ref{fig:summary_chart}. In the ideal case of negligible theoretical uncertainties, EDM experiments would set the strongest constraints on all imaginary parts (except for $\mathrm{Im}\,\xi_{bc}$), reaching the $\mathcal O(10^3\,\mathrm{TeV})$ scale for $\mathrm{Im}\,\xi_{ud}$, and the real parts of $\xi_{ub}$, $\xi_{cd}$, and $\xi_{td}$.

\end{itemize}

\section*{Acknowledgements}
SA acknowledges support
by the COFUND Fellowship under grant agreement PCOFUND-GA-2012-600377.
VC and EM  acknowledge support by the US DOE Office of Nuclear Physics and by the LDRD program at Los Alamos National Laboratory.
WD and JdV  acknowledge  support by the Dutch Organization for Scientific Research (NWO) 
through a RUBICON  and VENI grant, respectively.
We thank Paolo Gambino for discussion on the inclusive decays $B \rightarrow X_c l \nu$ and for providing us his code for the evaluation of the leptonic moments of the inclusive decay distribution. 
We thank Mikolaj Misiak for providing us the generalization of the anomalous dimension \eqref{eq:dipole3} to generic values of the charges $Q_u$ and $Q_d$.
We acknowledge several useful discussions with Zoltan Ligeti, Giuseppe Cerati, Andreas Crivellin, Ivo van Vulpen, and Marc de Beurs. We thank U.~Egede, M.~Gonz\'{a}lez-Alonso, and N.~Yamanaka for comments on the manuscript. 

\appendix

\section{Additional input for CKM fits}\label{AppB}
In this Appendix we discuss several processes that get small contributions from RHCCs, but that were used in the fits discussed in Section \ref{Single}
to constrain the SM CKM elements. In Appendix  \ref{sec:Beta} we discuss $B\to J/\psi K$, which determines the CKM angle $\beta$. RHCC in the $Wbc$ vertex contribute to this observable at tree level, but, as we will argue, the contribution is suppressed with respect to the SM. The remaining elements of the $\xi_{ij}$ matrix generate no or small contributions, after taking into account the limits from other observables.

In Appendix \ref{sec:Bmumu} we discuss the FCNC decays  $B_q \rightarrow \mu^+ \mu^-$, $K_L \rightarrow \pi^0 \nu \nu$, and the penguin contributions to $K_L \rightarrow \pi^0 e^+ e^-$. These contributions are quadratic in $\xi$ and necessarily involve two different $\xi$ elements. Thus, they do not play any role in a single coupling analysis. Finally, in Appendix \ref{sec:DeltaF2} we discuss $\Delta F = 2$ processes. Also in this case, contributions are quadratic in $\xi$. Both types of processes might play a more important role in a global analysis involving all $\xi_{ij}$ couplings, as possible cancellations may allow for larger values of the $\xi_{ij}$ couplings. However, this is beyond the scope of the current work, and we neglect all dimension-eight effects discussed in these appendices.

\subsection{$B\to J/\psi K$}\label{sec:Beta}
The CKM angle $\bt = {\rm arg} \big(-\frac{V_{cd}V_{cb}^*}{V_{td}V_{tb}^*}\big)$ is determined from $S_{J/\psi K}$ which appears in the CP asymmetry,
\bea
\frac{\Gamma(\bar B\to J/\psi K)-\Gamma(B\to J/\psi K)}{\Gamma(\bar B\to J/\psi K)+\Gamma(B\to J/\psi K)} = S_{J/\psi K}\sin (\Delta m_d t)+C_{J/\psi K}\cos (\Delta m_d t)\, .
\eea
Here 
\bea
S_{J/\psi K}=\frac{2{\rm Im}\lambda_{J/\psi K}}{1+|\lambda_{J/\psi K}|\sq},\qquad \lambda_{J/\psi K} = \frac{q}{p}\frac{\bar A_{J/\psi K}}{A_{J/\psi K}}=\frac{V_{tb}^*V_{td}}{V_{tb}V_{td}^*}\frac{\bar A_{J/\psi K}}{A_{J/\psi K}}\, ,
\eea
where the ratio $\frac{q}{p}=\frac{V_{tb}^*V_{td}}{V_{tb}V_{td}^*}$ is due to the $\bar B- B$ mixing and $A_{J/\psi K}$ ($\bar A_{J/\psi K}$) is the amplitude for the decay $B\to J/\psi K$ ($\bar B\to J/\psi K$). In the SM, this amplitude is due to a tree-level decay (proportional to $V_{cb}V_{cs}^*$) followed by $\bar K-K$ mixing (the real part of which is dominated by a term proportional to $V_{cs}V_{cd}^*$), such that to good approximation we have,
\bea
\frac{\bar A_{J/\psi K}}{A_{J/\psi K}}=\frac{V_{cb}V_{cd}^*}{V_{cb}^*V_{cd}} \, ,\qquad S_{J/\psi K}=\sin 2\bt  \, .
\eea
The experimental value is given by \cite{Olive:2016xmw},
\bea
S_{J/\psi K} = 0.682\pm0.019\, .
\eea
RHCCs can contribute to this observable through the amplitude, $A_{J/\psi K}$, or through $B$ or $K$ mixing. As discussed in Appendix \ref{sec:DeltaF2}, the RHCC contributions to meson mixing are quadratic in $\xi$, and we neglect them here. The RHCC contributions to the amplitude can arise from $\xi_{cs}$ and $\xi_{cb}$. The constraints on $\xi_{cs}$ from semileptonic $D$ and $D_s$ decays imply that the contribution to $ S_{J/\psi K}$ is at the percent level and therefore negligible. As for $\xi_{cb}$, the corrections to $S_{J/\psi K}$ are proportional to  $\xi_{cb}/V_{cb}$ and to ratios of the matrix elements of the left-right and SM operators, $\mathcal O^{cb\, cs}_{1,2\, LR}$ and $\mathcal O^{cb\, cs}_{1,2\, LL}$. In this case, inclusive and exclusive $B$ decays into charmed final states  and $B_q - \bar B_q$ oscillations allow for relatively large values of $\textrm{Im} ( \xi_{cb})/V_{cb}$,  $\textrm{Im} ( \xi_{cb})/V_{cb} \sim 0.3$. For exclusive $B$ decays  into $s$-wave charmonia, it was shown that the matrix elements factorize   \cite{Beneke:2000ry,Chay:2000xn,Beneke:2015wfa}. 
In particular, in the case of the left-right operators, one has to estimate the  matrix element 
$\langle J/\psi | \bar c_L c_R | 0 \rangle$.  While a precise evaluation is difficult, we notice that both in the nonrelativistic and in the $m_c\rightarrow 0$ limit, the matrix element vanishes at leading order. We therefore expect the corrections to  $A_{J/\psi K}$ to be relatively small.
Since the suppression factors in the two limits, the relative velocity of the charm quarks in the $J/\psi$  or the ratio $m_{J/\psi}/m_B$, are not extremely small, it might nonetheless be worthwhile to more rigorously investigate RHCC  contributions to  $B \rightarrow J/\psi K$.

\subsection{$\Delta F=1$ neutral current decays}\label{sec:Bmumu}

The effective Hamiltonian for $B^0_{s,d} \rightarrow \mu^+ \mu^-$, $K_L \rightarrow \mu^+ \mu^-$, $K_L \rightarrow \pi^0 e^+ e^-$ and $K_L \rightarrow \pi^0 \nu \nu$
contains the semileptonic operators  
\begin{eqnarray}
\mathcal H &=& - \frac{G_F^2 m_W^2}{16\pi^2}   \left\{ C^{i j}_{L\, V} \, \bar d^j \gamma^\mu P_L d^i\, \bar{l} \gamma_\mu l  + C^{i j}_{R\, V} \, \bar d^j \gamma^\mu P_R d^i\, \bar{l} \gamma_\mu l  
+ C^{i j}_{L \, ll}\, \bar d^j \gamma^\mu P_L d^i\, \bar{l} \gamma_\mu P_L l  \right. \nonumber \\ & & \left.
+ C^{i j}_{R\, ll}\, \bar d^j \gamma^\mu P_R d^i \, \bar{l} \gamma_\mu P_L l 
+ C^{i j}_{L\, \nu\nu}\, \bar d^j \gamma^\mu P_L d^i\, \bar{\nu} \gamma_\mu P_L \nu 
+ C^{i j}_{R\, \nu\nu}\, \bar d^j \gamma^\mu P_R d^i\, \bar{\nu} \gamma_\mu P_L \nu
\right\} .
\end{eqnarray}
The matching coefficients at the scale $\mu = m_W$ are obtained by computing penguin and box diagrams, and, for $i \neq j$, we find in the $\overline{\rm MS}$ scheme
\begin{eqnarray}\label{Eq:Penguins}
C^{i j}_{L\, V} &=& V^*_{t j} V_{t i} s_w^2 \left\{ \frac{8 (8 - 50 x_t + 63 x_t^2 + 6 x_t^3 - 24 x_t^4)}{9 (x_t-1)^4} \log x_t  
+ \frac{4 x_t (108 - 259 x_t + 163 x_t^2 - 18 x_t^3)}{9 (x_t -1)^3}\right\} \, ,\nn \\
C^{i j}_{R\, V} &=& \xi^*_{t j} \xi_{t i} s_w^2 \left\{  \frac{32}{3} (2 - 3 x_t)  \log \frac{\mu^2}{m_W^2}
+ \frac{8 (8 - 14 x_t - 81 x_t^2 + 222 x_t^3 - 168 x_t^4 + 36 x_t^5)}{9 (x_t-1)^4} \log x_t   \right.  \nonumber \\ & & 
\left.
- \frac{4 (-320 + 528 x_t + 141 x_t^2 - 493 x_t^3 + 162 x_t^4)}{27 (x_t -1)^3}\right\}  
\nn\\&&- \left(\frac{1472}{27}-\frac{256}{9}\log \frac{\mu\sq}{m_W\sq}\right) s^2_w ( \xi^*_{cj} \xi_{ci} +  \xi^*_{uj} \xi_{ui}) \, ,
\nn  \\
C^{i j}_{L\,ll} &=& V^*_{t j} V_{t i}  \left\{ \frac{12 x_t^2}{(x_t-1)^2}\log x_t + \frac{4 x_t (x_t -4)}{x_t -1} \right\} \, , \nn \\
C^{i j}_{R\,ll} &=& \xi^*_{t j} \xi_{t i}  \left\{ 4 (-3 + 4 x_t) \log \frac{\mu^2}{m_W^2}- \frac{4 x_t (10 - 11 x_t + 4 x_t^2)}{(x_t-1)^2} \log x_t + 4 \frac{4 -4x_t + 3 x_t^2}{x_t -1} \right\}  \nn \\
& & -\left( 16 +12\log \frac{\mu\sq}{m_W\sq}\right)( \xi^*_{c j} \xi_{c i} + \xi^*_{u j} \xi_{u i} ) \, , \nn 
\\
C^{i j}_{L\, \nu \nu} & =&  V^*_{t j} V_{t i} \left\{ - 12 \frac{x_t (x_t-2)}{(1-x_t)^2} \log x_t - 4 x_t \frac{2 + x_t}{x_t -1}  \right\} \, , \nn \\
C^{i j}_{R\, \nu \nu} & =&  \xi^*_{t j} \xi_{t i} \left\{ 4 (3 - 4 x_t)  \log\frac{\mu^2}{m_W^2} +  4 x_t \frac{4 - 11 x_t + 4 x_t^2 }{(1-x_t)^2} \log x_t +  4  \frac{2 + 4 x_t - 3 x_t^2}{x_t -1}  \right\} \nn \\
& & - \left(8 -12\log \frac{\mu\sq}{m_W\sq}\right)( \xi^*_{c j} \xi_{c i} + \xi^*_{u j} \xi_{u i} )\, ,
\end{eqnarray}
where we neglected powers of $x_c$ and $x_u$ and we used unitarity for the SM contributions.

Of the above operators, $C^{bq}_{L\, ll}$ and $C^{bq}_{R\, ll}$ contribute to $B^{0}_{s\,, d} \rightarrow l^+ l^-$ 
and $K_L \rightarrow \mu^+ \mu^-$. The photon penguins $C^{i j}_{L\, V}$ and $C^{ij}_{R\, V}$ do not contribute due to vector current conservation. 
The decay rate is
\begin{eqnarray}
\Gamma(B^0_{q} \rightarrow l^+ l^-) = \frac{1}{32 \pi }\sqrt{1 - \frac{4 m_l^2}{m_{B_q}^2}} m_{B_q} f_{B_q}^2 m_l^2 \, \left( \frac{G_F^2 m_W^2}{16 \pi^2} \right)^2 \, 
 \left| C^{bq}_{L\, ll} - C^{bq}_{R\, ll} \right|^2 \, ,
\end{eqnarray}
where the minus sign between the Wilson coefficients  is due to the fact that  only the axial part of the quark current contributes.
The observed branching ratios for $B_{d,s} \rightarrow \mu^{+} \mu^-$ are \cite{Amhis:2016xyh}
\begin{eqnarray}
\textrm{BR} (B^0_d \rightarrow \mu^+ \mu^-) = \left(3.9 ^{+1.6}_{-1.4}\right) \cdot 10^{-10}, \qquad \textrm{BR} (B_s \rightarrow \mu^+ \mu^-) = \left(2.8 ^{+0.7}_{-0.6}\right) \cdot 10^{-9}.
\end{eqnarray}
From Eq.\ \eqref{Eq:Penguins} one sees that RHCC contributions to these processes are relevant if $\xi^*_{tb} \xi_{t q} \sim V^*_{tb} V_{t q}$. As discussed in Section \ref{Single}, this possibility is ruled out by 
$B \rightarrow X_{s,d} \gamma$, even when all $\xi_{tj}$ are turned on at the same time. Similarly, the RHCC contributions to $K_L \rightarrow \mu^+ \mu^-$ are small after taking into account limits from semileptonic decays and $B\to X_q\g$.
In our analysis we therefore only use $\textrm{BR} (B^0_{s,d} \rightarrow \mu^+ \mu^-)$  to constrain the CKM elements $V_{ts}$ and $V_{td}$.

$C^{sd}_{L\, \nu \nu}$ and $C^{sd}_{R\, \nu \nu}$ contribute to the decay $K_L \rightarrow \pi^0 \nu \bar \nu$. The contributions of the SM and RHCC are of similar size if $\textrm{Im}\, (\xi^*_{ts} \xi_{td}) \sim
\textrm{Im} (V^*_{ts} V_{td}) \sim 10^{-4}$, which is ruled out by EDMs, and by the branching ratio and CP asymmetry in $B \rightarrow X_{s,d} \gamma$. Therefore, this channel might  become interesting  only
in scenarios in which multiple operators are turned on at the same time.    

The operators $C^{sd}_{L\,V}$ and $C^{sd}_{L\, \mu \mu}$ give the leading SM contribution to $K_L \rightarrow \pi^0 e^+ e^-$. They are related to the operators $C_{7V}$
and $C_{7A}$ defined in Ref. \cite{Buchalla:1995vs} by
\begin{equation}
C_{7V} = \frac{\alpha_{\textrm{em}}}{32 \pi s_w^2} \left( C^{ds}_{L\, V} + \frac{C^{ds}_{L\, \mu\mu}}{2} \right), \qquad C_{7A} = - \frac{\alpha_{\textrm{em}}}{64 \pi s_w^2}  \, C^{ds}_{L\, \mu\mu}\, .
\end{equation}
$C_{7A}$ does not run, while $C_{7V}$ mixes with tree-level charged currents. Factoring out a factor of $\alpha_{\textrm{em}}/2\pi$, the authors of Ref. \cite{Buchalla:1995vs} define the couplings
\begin{eqnarray}
\tilde y_{7V}(\mu) &=& P_0(\mu)  - 4 \left( C_0(x_t) + \frac{1}{4} D_0(x_t) \right)  + \frac{Y_0(x_t)}{s_w^2}, \qquad \tilde y_{7 A} = - \frac{Y_0(x_t) }{s_w^2}  \, ,
\end{eqnarray}
with 
\begin{eqnarray}
Y_0(x_t) &=& \frac{x_t}{8} \left( \frac{4-x_t}{1-x_t} + \frac{3 x_t}{(1-x_t)^2} \log x_t \right)\, , \nonumber \\
C_0(x_t) &=& \frac{x_t}{8} \left( \frac{x_t-6}{x_t-1} + \frac{3 x_t + 2}{(1-x_t)^2} \log x_t \right)\, , \nonumber \\
D_0(x_t) &=& -\frac{4}{9} \log x_t +  \frac{-19 x_t^3 + 25 x^2_t}{36 (x_t-1)^3} + \frac{x_t^2 (5 x_t^2 - 2 x_t - 6)}{18 (1-x_t)^4} \log x_t\, ,
\end{eqnarray}
where $x_t = m_t^2/m_W^2$. Without resummation, $P_0 = -4/9 \log x_c$. The value of $P_0(\mu)$ at different scales is given in  Ref. \cite{Buchalla:1995vs}.

\subsection{$\Delta$F = 2 processes}\label{sec:DeltaF2}

The effective Hamiltonian for $\Delta S =2$ processes in the presence of a RHCC  is given by \cite{Buras:2001ra}
\begin{equation}\label{HamiltonianDF2}
\mathcal H^{eff} = \frac{G_F^2 m_W^2}{16\pi^2} \sum C_i(\mu) \mathcal O_i(\mu),
\end{equation}
with 
\begin{eqnarray}\label{DeltaF2basis}
\mathcal O_1^{\textrm{VLL}} &=&  (\bar s \gamma_\mu \, P_L d)\, (\bar s \gamma^\mu \, P_L d), \qquad \mathcal O_1^{\textrm{VRR}} =  (\bar s \gamma_\mu \, P_R d)\, (\bar s \gamma^\mu \, P_R d), \nonumber \\
\mathcal O_1^{\textrm{LR}} &=&  (\bar s \gamma_\mu \, P_L d)\, (\bar s \gamma^\mu \, P_R d), \qquad \mathcal O_2^{\textrm{LR}} =  (\bar s  P_L d)\, (\bar s  P_R d), \nonumber \\
\mathcal O_1^{\textrm{SLL}} &=&  (\bar s P_L d)\, (\bar s P_L d),  \qquad \qquad \; \, \mathcal O_1^{\textrm{SRR}} = (\bar s P_R d)\, (\bar s P_R d), \nonumber \\
\mathcal O_2^{\textrm{SLL}} &=&  (\bar s \sigma^{\mu\nu} P_L d)\, (\bar s \sigma_{\mu\nu} P_L d)  \qquad \mathcal O_2^{\textrm{SRR}} =  (\bar s \sigma^{\mu\nu} P_R d)\, (\bar s \sigma_{\mu\nu}P_R d) \, .
\end{eqnarray}
An analogous Hamiltonian can be written for $\Delta B =2$ and $\Delta C = 2$ processes.

The coefficients $C_i$ are obtained by computing the box diagrams with two $W$ exchanges, for which we find in the $\overline{\rm MS}$ scheme
\begin{eqnarray}\label{matchDF2}
C^{\textrm{VLL}}_1 &=& V_{i s}^* V_{i d} V_{j s}^* V_{j d}  \left(  \left(6 - x_i - x_j\right)\log \frac{\mu^2}{m_W^2}+ f_1(x_i,x_j) \right), \nonumber \\
C^{\textrm{LR}}_1 &=& 2 V_{i s}^* V_{i d} \xi_{j s}^* \xi_{j d}  \left(  \left(6 - x_i - x_j\right)  \log \frac{\mu^2}{m_W^2}+ f_2(x_i,x_j) \right) ,\nonumber \\
C^{\textrm{LR}}_2 &=& 2  \frac{m_i  m_j}{m_W^2}\, \xi_{i s}^* V_{i d} V_{j s}^* \xi_{j d}  \left(  - 4 \log \frac{\mu^2}{m_W^2} + f_3(x_i,x_j) -4\delta^{i}_{u,c}\delta^{j}_{u,c}\,g(x_i,x_j)\right) ,\nonumber \\
C^{\textrm{SLL}}_1 &=&   \frac{m_i  m_j}{m_W^2}\, \xi_{i s}^* V_{i d} \xi_{j s}^* V_{j d}  \left(  - 4 \log \frac{\mu^2}{m_W^2}+ f_3(x_i,x_j) -4\delta^{i}_{u,c}\delta^{j}_{u,c}\,g(x_i,x_j)\right), \nonumber \\
C^{\textrm{SRR}}_1 &=&   \frac{m_i  m_j}{m_W^2}\, V_{i s}^* \xi_{i d} V_{j s}^* \xi_{j d}  \left(  - 4 \log \frac{\mu^2}{m_W^2}+ f_3(x_i,x_j) -4\delta^{i}_{u,c}\delta^{j}_{u,c}\,g(x_i,x_j)\right), \nonumber \\
C^{\textrm{SLL}}_2 &=&   \frac{m_i  m_j}{m_W^2}\, \xi_{i s}^* V_{i d} \xi_{j s}^* V_{j d}  \,  \left(f_4(x_i,x_j) -\delta^{i}_{u,c}\delta^{j}_{u,c}\,g(x_i,x_j)\right),\nonumber \\
C^{\textrm{SRR}}_2 &=&   \frac{m_i  m_j}{m_W^2}\, V_{i s}^* \xi_{i d} V_{j s}^* \xi_{j d}  \, \left( f_4(x_i,x_j)  -\delta^{i}_{u,c}\delta^{j}_{u,c}\,g(x_i,x_j)\right)\, ,
\end{eqnarray}
where $i=u,c,t$ and $j=u,c,t$ label the internal up-type quark, and a summation over $i$, $j$ is understood. $m_i$, $m_j$ are the masses of the internal up-type quarks, and $x_i = m_i^2/m_W^2$.
The loop functions are  
\begin{eqnarray}
f_1(x_i,x_j) &=& = - \frac{x_j^2 (4  -8 x_j  + x^2_j)}{ (x_i - x_j) (-1 + x_j)^2  } \log(x_j)  + \frac{x_i^2 (4  -8 x_i  + x^2_i) }{ (-1 + x_i)^2 (x_i - x_j)} \log x_i \nonumber \\
& & +2 - \frac{3}{2} (x_i + x_j) - \frac{3 (x_i + x_j - x_i x_j)}{(1-x_i)(1-x_j)},\nonumber \\
f_2(x_i,x_j) &=&  - \frac{(-4 + x_j)^2 x_j^2}{ (x_i - x_j) (x_j-1 )^2} \log(x_j) - \frac{(-4 + x_i)^2 x_i^2}{ (-1 + x_i)^2 (-x_i + x_j)}\log x_i \nonumber \\ & & + 14 - \frac{3}{2}(x_i + x_j) + 9 \frac{x_i  + x_j - x_i x_j}{(1-x_i)(1-x_j)} ,\nonumber \\
f_3(x_i,x_j) & = & -\frac{4 x_j (4 -2 x_j + x_j^2 )}{(x_i - x_j) (-1 + x_j)^2} \log x_j  + \frac{4 x_i (4  -2 x_i  + x_i^2)}{(-1 + x_i)^2 (x_i - x_j)} \log x_i ,\nonumber \\
& & + \frac{4 ( 2 + x_i + x_j - x_i x_j)}{(-1 + x_i) (-1 + x_j)}, \nonumber \\
f_4(x_i,x_j) & = & \frac{2 (-2 + x_j) x_j}{(x_i - x_j) (-1 + x_j)^2} \log x_j + \frac{2 (-2 + x_i) x_i}{(-1 + x_i)^2 (-x_i + x_j)} \log x_i + \frac{2}{(-1 + x_i) (-1 + x_j)}\, . \nonumber\\
\end{eqnarray}
The remaining function,  $g(x_i,x_j)$, arises from the matching contributions in the theory below $\mu=m_W$, which is why it does not receive contributions from diagrams involving the top quark. Up to $\Or(x_u,x_c)$ corrections, it is given by,
\bea
g(x_i,x_j) &=& -4\left(1+\log \frac{\mu\sq}{m_W\sq} -\frac{x_i\log x_i-x_j\log x_j}{x_i-x_j}\right)\, .
\eea

We verified that the expressions in Eq.\ \eqref{matchDF2} are gauge independent. Our results are in agreement with Ref. \cite{He:2009rz}, except that we find 
matching contributions to $C_{2}^{\textrm{LR}}$, $C_{2}^{\textrm{SLL}}$ and $C_{2}^{\textrm{SRR}}$, which are not given in Ref. \cite{He:2009rz}, 
and we do not assume unitarity of the $\xi_{ij}$ matrix. Most expressions in Eq. \eqref{matchDF2} are  UV divergent. For the SM coefficient $C_{1}^{\textrm{VLL}}$ the unitarity of the CKM guarantees that after summing over $i,j$ the divergence cancels. In all other cases, the divergence indicates mixing of two insertions of RHCC onto $\Delta F=2$ four-fermion operators between the high-energy scale, $\Lambda$, and $m_W$. 
The QCD running of the operators in Eq. \eqref{DeltaF2basis} below the scale $m_W$  is discussed in detail in Ref. \cite{Buras:2001ra}.

\subsubsection{$B - \bar B$ oscillations}

The Hamiltonian  in Eq.\ \eqref{HamiltonianDF2} can be used to compute the mass difference between mass eigenstates in the $B^0_{d,s}$ -- $\overline{B}^0_{d,s}$ systems
\begin{eqnarray}
\Delta m_{q} = 2|M_{12}^{(q)}|=\frac{\left| \langle \bar{B}_q^0 | \mathcal H^{eff} | B_q^0 \rangle \right|}{m_{B_q}}  =  \left(\frac{G_F^2 m_W^2}{16 \pi^2}\right) \frac{1}{m_{B_q}}  \left| \sum_i C_i(\mu)   \,\langle \bar{B}^0_q | \mathcal O_i | B^0_q \rangle \right|. 
\end{eqnarray}
The matrix elements for the operators in the basis \eqref{DeltaF2basis} have been computed on the lattice in Ref. \cite{Carrasco:2013zta}, and we have
\begin{eqnarray}
\langle \bar{B}^0_q | \mathcal O^{\textrm{VLL}}_1 | B^0_q \rangle &=& \langle \bar{B}^0_q | \mathcal O^{\textrm{VRR}}_1 | B^0_q \rangle =  \frac{8}{3} B^q_1(\mu)  m^2_{B_q} f^2_{B_q}\, , \\
\langle \bar{B}^0_q | \mathcal O^{\textrm{SLL}}_1 | B^0_q \rangle &=& \langle \bar{B}^0_q | \mathcal O^{\textrm{SRR}}_1 | B^0_q \rangle = - \frac{5}{12} B^q_2(\mu) R(\mu)\,  m^2_{B_q} f^2_{B_q}
\, , \\
\langle \bar{B}^0_q | \mathcal O^{\textrm{SLL}}_2 | B^0_q \rangle &=& \langle \bar{B}^0_q | \mathcal O^{\textrm{SRR}}_2 | B^0_q \rangle =  \left( \frac{5}{3} B^q_2(\mu)
- \frac{2}{3} B_3^q \right)\, R(\mu) m^2_{B_q} f^2_{B_q} \, , \\
\langle \bar{B}^0_q | \mathcal O^{\textrm{LR}}_1 | B^0_q \rangle &=&   -\frac{1}{3} B^q_5(\mu)\, R(\mu) m^2_{B_q} f^2_{B_q}\, , \\
\langle \bar{B}^0_q | \mathcal O^{\textrm{LR}}_2 | B^0_q \rangle &=&   \frac{1}{2} B^q_4(\mu) \, R(\mu) m^2_{B_q} f^2_{B_q}\, ,
\end{eqnarray}
where $R(\mu) = m^2_{B_q}/(m_b(\mu) + m_q(\mu))^2$.
The bag parameters, in the $\overline{\textrm{MS}}$ scheme at the scale $\mu = m_b = 4.2$ GeV are summarized in Table \ref{TabBag}.
The FLAG average for the $B_d$ and $B_s$ decay constants is given in Table \ref{LQCDinput}. 
The  RGE factors to run the coefficients in Eq. \eqref{matchDF2} to the scale $\mu = m_t$ to $\mu = m_b$ are given in Ref. \cite{Buras:2001ra}.

Neglecting the RHCC contributions which are quadratic in $\xi$, one has to good approximation in the SM,
\bea
\Delta m_{q} = 2|M_{12}^{(q)}|= \frac{G_F\sq m_W\sq}{6\pi\sq}|V_{tq}V_{tb}^*|\sq f_{B_{q}}\sq \hat B_{B_q}\eta_B S_0(x_t,x_t)\, ,
\eea
 in which $x_t$ should be evaluated at $\mu=m_t$ and $\eta_B = 0.55\pm0.01$ \cite{Buras:2013ooa}, $S_0(x_i,x_j) = \frac{1}{4}(f_1(x_i,x_j)-f_1(0,x_j)-f_1(x_i,0)+f_1(0,0))$.
In place of $B_1^{s,d}$, it is convenient to introduce the renormalization-group-independent bag parameters $\hat B_{B_{d,s}}$, for which we use   \cite{Aoki:2016frl}
\bea
f_{B_d}\sqrt{\hat B_{B_d}} = 219\pm 14\, {\rm MeV}\, ,\qquad f_{B_s}\sqrt{\hat B_{B_s}} = 270\pm 16\, {\rm MeV}\, .
\eea 
The experimental values of $\Delta m_s$ and $\Delta m_d$ are
\begin{eqnarray}
\Delta m_d = \left(0.5064 \pm 0.0019\right)  \, \textrm{ps}^{-1}\, , \qquad  \Delta m_s = \left(17.757 \pm 0.021\right)  \, \textrm{ps}^{-1}\, . 
\end{eqnarray}

\begin{table}
\center
\begin{tabular}{||c|ccccc||}
\hline
& $B_1$ & $B_2$ & $B_3$ & $B_4$ & $B_5$ \\
\hline 
$B^0_d - \bar{B}^0_d$ \cite{Carrasco:2013zta}  & $0.85  \pm 0.04$ & 0.72 $\pm$ 0.03 & 0.88 $\pm$ 0.13 & 0.95 $\pm$ 0.05 & 1.47 $\pm$ 0.12 \\
$B^0_s - \bar{B}^0_s$ \cite{Carrasco:2013zta} & $0.86  \pm 0.03$ & 0.73 $\pm$ 0.03 & 0.89 $\pm$ 0.12 & 0.93 $\pm$ 0.04 & 1.57 $\pm$ 0.11 \\
$K^0 - \bar{K}^0$ \cite{Aoki:2016frl} & 0.56 $\pm$ 0.01 & 0.50 $\pm$ 0.01 & 0.77 $\pm$ 0.03  & 0.93 $\pm$ 0.02 & 0.72 $\pm$ 0.04 \\ 
\hline 
\end{tabular}
\caption{Bag parameters for $B_q - \bar{B}_q$ and $K^0 - \bar K^0$ oscillations, in the $\overline{\textrm{MS}}$ scheme. For $B_q - \bar B_q$ oscillations, we use the results of Ref. \cite{Carrasco:2013zta},
and the bags parameters are given at the renormalization scale $\mu = m_b$. 
For $K^0 - \bar K^0$ oscillation, we quote the FLAG averages of simulations performed with $n_f = 2+1$ flavors \cite{Aoki:2016frl}.
In this case, $B_1 = B_K$ is given at the renormalization scale $\mu = 2$ GeV, while $B_{2, \ldots, 5}$ are given at $\mu = 3$ GeV.}\label{TabBag}
\end{table}

\subsubsection{$\varepsilon_K$}

In the case of $K^0 - \bar K^0$ oscillations, the mass difference $ \Delta m_K = m_{K_L} - m_{K_S}$ receives sizable long-distance contributions  \cite{Buchalla:1995vs},
whose uncertainties prevent the use of $\Delta m_K$ for a  precise extraction of the CKM elements. On the other hand, CPV in $K_0 - \bar K_0$  mixing is dominated by short-distance effects.     
The indirect CP violation in $K \rightarrow \pi \pi$ decays is parametrized by the parameter $\varepsilon_K$, which, up to $\Or(\xi\sq)$ corrections,  is given by \cite{Buras:1998raa}
\bea
\varepsilon_K &=& \frac{G_F\sq m_W\sq }{12\pi\sq}\frac{m_K f_K\sq \hat{B}_K}{\sqrt{2} \Delta m_K}\kappa_\varepsilon{\rm Im}\bigg(\eta_{cc}(V_{cs}^*V_{cd})\sq S_0(x_c)+2\eta_{ct}V_{cs}^*V_{cd}V_{ts}^*V_{td} S_0(x_c,x_t)
\nn \\ && 
+\eta_{tt}(V_{ts}^*V_{td})\sq S_0(x_t) \bigg)\, .\nn
\eea
Here  $x_t$ should be evaluated at $\mu=m_t$ and $x_c$ at $\mu=m_c$, furthermore from FLAG and Ref.\ \cite{Buras:2013ooa}
\bea
\hat{B}_K = 0.717\pm 0.018\pm0.016\, ,\qquad \kappa_\varepsilon = 0.94\pm0.02\,  ,\nn\\
\eta_{cc}=1.87\pm0.76,\qquad \eta_{ct}=0.496\pm0.047\, ,\qquad \eta_{tt}= 0.5765\pm 0.065\, .
\eea
$\hat{B}_K$ is the renormalization-group-invariant bag factor. 
In Ref. \cite{Cirigliano:2016yhc}, we considered long-range contributions to $\varepsilon_K$, linear in $\xi_{us}$
and $\xi_{ud}$. Since the ensuing constraints on these couplings are much weaker than the one from $\ep^\prime/\ep$, we do not include these corrections here.

\section{Two-loop contributions to the electron EDM}\label{app:eEDM}
As the $\xi$ operator only couples the $W$ boson to quarks it mainly induces hadronic EDMs. However, the $\xi_{tb}$ coupling also generates a (fairly small) electron EDM at the two-loop level. Here we briefly describe this contribution.

In the relevant diagram two $W$ bosons connect an electron line with a top-bottom loop which emits a photon. Neglecting the lepton masses, this produces the following contribution to the electron EDM,
\bea
v\sq\, \tilde c_{\g l}^{(ee)}|_{\rm 2loop} = -16N_c\frac{y_t y_b}{(4\pi)^4}\text{Im}\, \big(\xi_{tb}V_{tb}^*\big)\bigg[\frac{Q_t}{Q_e} F(x_t,x_b)+(t\leftrightarrow b)\bigg]\, ,\label{Xi2loopBZ}
\eea
where $x_i\equiv m_i\sq/m_W\sq$, and
\bea
F(x_i,x_j) &=& \frac{1}{2}\int_0^1 dx\frac{x-1}{x\sq+x(x_j-x_i-1)+x_i}\ln\frac{x(1-x)}{x(x_j-x_i)+x_i}\, .
\eea
In the approximation of small $x_b$, the loop function becomes,
\bea
F(x_t,0) &=& \frac{1}{2}\bigg[{\rm Li}_2(1-1/x_t)-\frac{\pi\sq}{6}\bigg],\nn\\
F(x_b,x_t) &\simeq  & \frac{1}{2}\frac{1}{x_t-1}\bigg[\ln x_t\ln \tfrac{x_b}{x_t}-(x_t+1){\rm Li}_2(1-1/x_t)\bigg]+\frac{\pi\sq}{12}\, .
\eea
Below the scale $\mu=m_W$, a second matching contribution arises from an operator of the form, $\vL = 
C^{(b,e)} \bar b \simu b\, \bar e \,i\sigma_{\mu\nu}\g_5 e$. This operator is generated at one loop and, in turn, induces the electron EDM through an additional loop. All combined, the matching conditions at the scales $\mu=m_W$ and $\mu=m_b$  become
\bea
v\sq \,C^{(b,e)}(\mu_W)&=&\frac{y_t y_e}{(4\pi)\sq}\frac{\ln x_t}{1-x_t}\text{Im}\, \big(\xi_{tb}V_{tb}^*\big)\, ,\nn\\
\tilde c_{\g l}^{(ee)}(\mu_W) &=& \tilde c_{\g l}^{(ee)}|_{\rm 2loop} - \frac{8N_c}{(4\pi)\sq}\frac{m_b Q_b}{m_e Q_e}\ln\frac{m_b\sq}{\mu_W\sq}\, C^{(b,e)}(\mu_W)\, ,\nn\\
\tilde c_{\g l}^{(ee)}(\mu_b^-)&=&\tilde c_{\g l}^{(ee)}(\mu_b^+)+\frac{8N_c}{(4\pi)\sq}\frac{m_b Q_b}{m_e Q_e}\ln\frac{m_b\sq}{\mu_b\sq}\, C^{(b,e)}(\mu_b)\, ,
\eea
where $\mu_{W}\simeq m_W$ indicates a scale around $\mu=m_{W}$, while $\mu_b^+$ ($\mu_b^-$) refers to a scale just above (below) the $b$-quark threshold. Finally, the RG evolution between $m_W$ and $m_b$, which determines $ C^{(b,e)}(\mu_b^+)$ and $\tilde c_{\g l}^{(ee)}(\mu_b^+)$, is given by
\bea
\frac{d}{d\ln\mu} \tilde c_{\g l}^{(ee)}(\mu) = 16 N_c\frac{1}{(4\pi)\sq}\frac{m_b Q_b}{m_e Q_e} C^{(b,e)}(\mu)\, ,\quad
\frac{d}{d\ln\mu}C^{(b,e)}(\mu) =2C_F \frac{\al_s}{4\pi} C^{(b,e)}(\mu)\, .
\eea
The electron EDM does not evolve under RG $\mu=m_b$ (apart from small QED corrections), we use $\tilde c_{\g l}^{(ee)}(2\, {\rm GeV}) = \tilde c_{\g l}^{(ee)}(\mu_b^-)$ which is given in Table \ref{MQCDXi}.

\bibliographystyle{JHEP} 
\bibliography{bibliography}

\end{document}